\pdfoutput=1
\PassOptionsToPackage{pdfpagelabels=false}{hyperref}
\documentclass[fleqn, usenatbib]{mnras}
\usepackage[utf8]{inputenc}
\usepackage[english]{babel}
\usepackage{lmodern}

\usepackage[T1]{fontenc}
\usepackage{ae, aecompl}

\usepackage{layouts}
\usepackage{graphicx}	% Including figure files
\usepackage{amsmath}	% Advanced maths commands
\usepackage{amssymb}	% Extra maths symbols
\usepackage{stackengine}
\usepackage[usenames, dvipsnames]{xcolor}

\hypersetup{
colorlinks,
pdfborder = {0 0 0},
breaklinks,
linkcolor=black,
urlcolor=NavyBlue,
citecolor=NavyBlue,
}

\def\figrelpath{./}%./SupplementaryMaterialGGLRescalingPaper/

\renewcommand{\vec}[1]{\ensuremath{\boldsymbol{#1}}}
\newcommand{\dd}[1]{\,\mathrm{d} #1\,}

\newcommand{\expVal}[1]{\ensuremath{\left \langle #1 \right \rangle}}
\DeclareMathOperator\erfc{erfc}

\defcitealias{Angulo:2009rc}{AW10}
\defcitealias{Angulo:2014gza}{AH15}
\defcitealias{2016MNRAS.460.1214L}{L16}

\title[Halo profiles in rescaled simulations]{Halo mass and weak galaxy-galaxy lensing profiles in rescaled cosmological $N$-body simulations}

\author[Renneby, Hilbert, \& Angulo]{
Malin Renneby$^{1, \, 2}$\thanks{E-mail: \href{mailto:malin.renneby@physik.uni-muenchen.de}{malin.renneby@physik.uni-muenchen.de}},
Stefan Hilbert$^{1,\, 2}$
and Ra{\'u}l E. Angulo$^{3}$
\\
$^{1}$Excellence Cluster Universe, Boltzmannstraße 2, 85748 Garching bei M{\"u}nchen, Germany\\
$^{2}$Ludwig-Maximilians-Universit{\"a}t, Fakult{\"a}t f{\"u}r Physik, Universit{\"a}ts-Sternwarte, Scheinerstra{\ss}e 1, 81679 M{\"u}nchen, Germany\\
$^{3}$Centro de Estudios de F{\'i}sica del Cosmos de Arag{\'o}n, Plaza de San Juan 1, 44001 Teruel, Spain
}

\date{Accepted XXX. Received YYY; in original form ZZZ}

\pubyear{2018}

\begin{document}
%\selectlanguage{english}
\label{firstpage}
\pagerange{\pageref{firstpage}--\pageref{lastpage}}
\maketitle

\begin{abstract}
We investigate 3D density and weak lensing profiles of dark matter haloes predicted by a cosmology-rescaling algorithm for $N$-body simulations. We extend the rescaling method of Angulo \& White (2010) and Angulo \& Hilbert (2015) to improve its performance on intra-halo scales by using models for the concentration-mass-redshift relation based on excursion set theory. The accuracy of the method is tested with numerical simulations carried out with different cosmological parameters. We find that predictions for median density profiles are more accurate than $\sim 5\,\%$ for haloes with masses of $10^{12.0} - 10^{14.5} h^{-1}\,M_{\sun}$ for radii $0.05 < r/r_{200\text{m}} < 0.5$, and for cosmologies with $\Omega_\text{m} \in [0.15,\,0.40]$ and $\sigma_8 \in [0.6,\,1.0]$. For larger radii, $0.5 < r/r_{200\text{m}} < 5$, the accuracy degrades to $\sim20\,\%$, due to inaccurate modelling of the cosmological and redshift dependence of the splashback radius. For changes in cosmology allowed by current data, the residuals decrease to $\lesssim2\,\%$ up to scales twice the virial radius. We illustrate the usefulness of the method by estimating the mean halo mass of a mock galaxy group sample. We find that the algorithm's accuracy is sufficient for current data. Improvements in the algorithm, particularly in the modelling of baryons, are likely required for interpreting future (dark energy task force stage IV) experiments.
\end{abstract}


%%%%%%%%%%%%%%%%%%%%%%%%%%%%%%%%%%%%%%%%%%%%%%%%%%

%%%%%%%%%%%%%%%%% BODY OF PAPER %%%%%%%%%%%%%%%%%%

\section{Introduction}

The relation between galaxies and their dark matter haloes is of great interest not only for the study of galaxy evolution, but also for precision cosmology. To fully exploit future large-scale structure measurements requires a thorough quantitative understanding of the connection between galaxies as visible tracers of cosmic structure and the predominantly dark cosmic web. One of the most sensitive probes to constrain this relation is galaxy-galaxy lensing (GGL). 

GGL quantifies the relationship between galaxies and the dark matter density field through the cross-correlation of the observed shapes of distant galaxies and the positions of foreground galaxies. These foreground galaxies, together with their surrounding dark matter haloes, act as gravitational lenses since the associated gravity induces a differential deflection of light from the background sources \citep[e.g.][]{Bartelmann:1999yn}. Typically, the resulting image distortions are small. However, the effect can be measured statistically by considering a large number of systems. 

Since its first detection by \citet{1996ApJ...466..623B}, GGL has become well understood in terms of statistical and systematic uncertainties. Recent GGL observations report signal-to-noise ratios $\sim 120$  \citep{2015MNRAS.452.3529V}. The available data will increase substantially from ongoing and upcoming surveys such as the Dark Energy Survey (DES), the Kilo Degree Survey (KiDS), the Hyper Suprime-Cam Subaru Strategic Survey (HSC), the Large Synoptic Survey Telescope (LSST) survey, and the Euclid mission. This creates new challenges for GGL theoretical modelling.

Two of the most widely-used frameworks to interpret GGL measurements are halo-occupation distribution (HOD) models \citep[e.g.][]{2000MNRAS.318.1144P, 2000MNRAS.318..203S, 2002ApJ...575..587B, 2002PhR...372....1C,
2011ApJ...738...45L, 2012ApJ...744..159L, 2015MNRAS.454.1161Z} and (sub-)halo abundance matching (SHAM) techniques \citep{2004ApJ...609...35K, 2004ApJ...614..533T, 2006MNRAS.371.1173V, 2006ApJ...647..201C, 2009ApJ...696..620C, 2010ApJ...710..903M, 2010ApJ...717..379B}. There are however hints that there may be aspects poorly understood for certain galaxy samples \citep{2017MNRAS.467.3024L}. This might be a product of shortcomings of and/or simplifications in these models. For instance, effects such as assembly bias, the non-gravitational physics induced by baryons, and the overall dependence on cosmological parameters are difficult to incorporate accurately.

A more faithful description of GGL might be constructed from a joint numerical treatment of galaxy formation and the evolution of the density field. In recent years, elaborate modelling of the baryonic gas physics has become feasible in hydrodynamical simulations such as Illustris \citep{2014MNRAS.444.1518V, 2014Natur.509..177V, 2014MNRAS.445..175G} and Eagle \citep{2015MNRAS.446..521S, 2015MNRAS.450.1937C} in sufficiently large volumes to allow for a direct comparison with GGL observations \citep{2017MNRAS.467.3024L, 2017MNRAS.471.2856V}.

A complementary approach is to employ semi-analytical models (SAMs) of galaxy formation \citep{1991ApJ...379...52W,1999MNRAS.303..188K, 2001MNRAS.328..726S,2006MNRAS.370..645B,2007MNRAS.375....2D,Guo:2010ap, 2013MNRAS.431.3373H, 2015MNRAS.451.2663H} together with gravity-only simulations. In this approach, halo merger trees extracted from $N$-body simulations are populated with galaxies whose physical processes, such as cooling, star formation, and feedback, are tracked by a set of coupled differential equations. This allows for self-consistent and physically-motivated predictions for the galaxy population and the respective dark matter, which can then be used to compute the expected weak lensing signal for various lens galaxy samples \citep[e.g.][]{2009A&A...499...31H, 2010MNRAS.404..486H, 2011A&A...531A.169P, 2012A&A...547A..77S, 2013MNRAS.431.1439G, 2015MNRAS.454.1432S, 2016MNRAS.456.2301W, 2017A&A...601A..98S}. 

The computationally cost of carrying out numerical simulations over many different cosmological parameters is currently prohibitively expensive. A way to alleviate this challenge is to carry out a small number of high-quality simulations which could then be manipulated to mimic different background cosmologies. This idea was originally brought forth by \citet{Angulo:2009rc}, henceforth \citetalias{Angulo:2009rc}. Their method is to rescale the time and length units such that the variance of the linear matter field in the rescaled fiducial and target simulations match over a range of scales relevant for halo formation. In \citet{Angulo:2014gza}, hereafter \citetalias{Angulo:2014gza}, an additional requirement on a matched linear growth history was introduced, which improved the accuracy of predictions for shear correlations functions. 

Despite the improvements, the rescaling method still produced noticeable biases in the internal structure of dark matter haloes, owing to different formation times in the fiducial and target cosmologies. In this paper, we propose an enhancement to the original algorithm by taking advantage of recent theory developments in predicting the concentration-mass relation of dark matter haloes by \citet[][]{2016MNRAS.460.1214L}, henceforth \citetalias{2016MNRAS.460.1214L}. We then investigate if the updated rescaling algorithm can capture the \emph{small and intermediate scales of the cosmic web} interpretable by GGL. 

This paper is organised as follows: In Section~\ref{sec:theory}, we recap the key ingredients of our rescaling algorithm. Details on the simulations, halo samples, and summary statistics for testing the algorithm are described in Section~\ref{sec:observables}. We present the results using the original as well as our updated scaling predictions in Section~\ref{sec:results}. We discuss our results and their implications, e.g. for the estimation of lens masses and predictions for concentration biases, in Section~\ref{sec:discussion}. We summarise our main findings in Section~\ref{sec:conclusions}.

\section{Theory}
\label{sec:theory}

In this section we present the main aspects of our scaling algorithm. We briefly recap the \citetalias{Angulo:2009rc} and \citetalias{Angulo:2014gza} algorithm in Section~\ref{sec:cosmResc}. In Section~\ref{sec:haloProfiles} and \ref{sec:rescaledCMZ} we define halo concentrations and how they transform under rescaling. In Section~\ref{sec:cmz}, we summarise the model of \citetalias{2016MNRAS.460.1214L}, which will be employed later in the paper. Throughout the paper we use comoving coordinates and densities.

\subsection{Determining the rescaling coefficients}
\label{sec:cosmResc}

For the details of the rescaling algorithm, we refer to \citetalias{Angulo:2009rc} and \citetalias{Angulo:2014gza}. Here we note that it determines a length rescaling factor $\alpha$ and a redshift $z_\ast$ in the fiducial cosmology to match to a redshift $z_\ast^\prime$ in the target cosmology based on (i) the difference in the variance $\sigma$ of the linear matter field between two smoothing lengths determined by the range of halo masses one would like to emulate and (ii) the difference in growth history. Letting primed symbols denote quantities in the target cosmology, comoving positions $\vec{x}$ and simulation particle masses $m_\text{p}$ in the fiducial simulation are rescaled as
\begin{align}
\label{eq:position_rescaling}
\begin{aligned}
\vec{x} \left [ \text{Mpc}/h \right ] &\mapsto \vec{x^\prime} \left [ \text{Mpc}/h^\prime \right ]   = \alpha \vec{x} \left [ \text{Mpc}/h \right ],
\end{aligned}\\
\label{eq:mass_rescaling}
\begin{aligned}
m_{\text{p}} \left [ M_\odot/h \right  ] &\mapsto m_{\text{p}}^\prime \left [ M_\odot/h^\prime \right ] = \alpha^3 \frac{\Omega_\text{m}^\prime}{\Omega_\text{m}} \frac{h^{\prime^2}}{h^2} m_{\text{p}} \left [ M_\odot/h \right ] \\ &= \beta_\text{m} m_{\text{p}} \left [ M_\odot/h \right ].
\end{aligned}
\end{align}
Here, $\Omega_\text{m}$ denotes the cosmic mean matter density (in units of the critical density) and $H_0 = 100 \, h \, \text{km/s/Mpc}$ is the Hubble constant.
The comoving matter density $\rho_\text{m}$ then transforms as: 
\begin{equation}
\label{eq:comovingMatterDensityRescaling}
\rho_\text{m} \mapsto \rho_\text{m}^\prime = \alpha^{-3} \beta_\text{m} \rho_\text{m}.
\end{equation} 
The simulation box length and redshift change to:
\begin{align}
\label{eq:boxRescaling}
L &\rightarrow L^\prime  = \alpha L,\\
z &\rightarrow z^\prime, \, z \leqslant z_\ast, \, z^\prime \leqslant z_\ast^\prime,
\end{align}
where higher redshifts are acquired through the linear growth factor relation, 
\begin{equation}
\label{eq:linGrowthRelation}
D^\prime(z^\prime) = D(z)/D(z_\ast) \cdot D^\prime(z^\prime_\ast). 
\end{equation}

The growth constraint from \citetalias{Angulo:2014gza} is implemented through a comparison of a range of scale factors $a$ around the value $a_\ast$ in the (unscaled) fiducial cosmology corresponding to the best redshift fit $z_\ast$ of the target simulation at $z = 0$ for a range of proposed scaling options $( \alpha, \, z_\ast)$ with the growth history\footnote{The best relative weight on emulating the variance vs. the  growth for a given observable is still an open question.} of the target simulation. In \citetalias{Angulo:2009rc}, the last step of the algorithm involves a large-scale structure correction to account for the differences in the primordial linear power spectrum between the fiducial and target cosmologies, which amounts to moving the particles with respect to one another to reach a better agreement with the positions in the target simulation. Since this analysis focuses on the non-linear regime where this correction translates to an almost uniform displacement, we neglect this correction. As the snapshot output of an $N$-body simulation usually is discrete in time, the closest match to $(\alpha, \, z_\ast)$ is selected.
 
The chief advantage of the algorithm is that all quantities are calculated in the linear regime, wherein we either have explicit predictions or adequate fits for a range of different cosmologies. This allows for a fast evaluation ($\leqslant5\,\mathrm{s}$ on a contemporary laptop).

\subsection{Halo profiles}
\label{sec:haloProfiles}

As a model for comoving matter density profiles of haloes, we consider the NFW profile \citep{Navarro:1995iw, Navarro:1996gj}:
\begin{equation}
\label{eq:NFWprofile}
\rho_\text{NFW}(r) =
 \frac{\rho_\text{crit}(z)\delta_\text{c}}{(r/r_\text{s})( 1 + r/r_\text{s})^2}.
\end{equation}
Here, $\delta_\text{c}$ denotes the characteristic density of the halo, $r_\text{s}$ its scale radius, and $\rho_\text{crit}(z)$ the comoving critical density at halo redshift $z$. For a spatially flat universe with cold dark matter (CDM) and a cosmological constant $\Lambda$, $\rho_\text{crit} (z) = 3 H_0^2 (8\pi \mathrm{G})^{-1}  E(z)^2 (1 + z)^{-3}$, where $\mathrm{G}$ is the gravitational constant, and
$E(z)^2 = \Omega_\text{m}(1 + z)^3 + (1 - \Omega_\text{m})$.

For a given overdensity threshold $\Delta$, one may define the halo radius $r_{\Delta\text{c}}$ as the radius at which the mean interior density is $\Delta \times \rho_\text{crit}(z)$. The halo concentration $c_{\Delta\text{c}}$ is then defined by $c_{\Delta\text{c}} = r_{\Delta\text{c}}/ r_\text{s}$ with the associated halo mass $M_{\Delta\text{c}} = \Delta (4/3) \pi r_{\Delta\text{c}}^3 \rho_\text{crit}(z)$ and the characteristic density $\delta_\text{c}$ 
\begin{equation}
	 \delta_\text{c} = \frac{\Delta}{3}\frac{c_{\Delta\text{c}}^3}{\ln(1+c_{\Delta\text{c}}) - c_{\Delta\text{c}}/(1+c_{\Delta\text{c}})}.
\end{equation}

We also consider as halo radius $r_{\Delta\text{m}}$, at which the halo's mean interior density is $\Delta$ times the cosmic mean. The associated halo concentration $c_{\Delta\text{m}} = r_{\Delta\text{m}}/ r_\text{s}$, and the halo mass $M_{\Delta\text{m}} = \Delta (4/3) \pi r_{\Delta\text{m}}^3 \Omega_\text{m} \rho_\text{crit}(0)$. 

In addition, we also model the density field with Einasto profiles \citep{1965TrAlm...5...87E}:
\begin{equation}
\label{eq:einastoDensityProfile}
\rho_\text{Einasto}(r) = \rho_\text{s} \exp \left( -\frac{2}{\alpha} \left [ \left( \frac{r}{r_\text{s}} \right)^\alpha - 1 \right ]\right),
\end{equation}
where $\alpha$ denotes a profile shape parameter, $r_\text{s}$ the scale radius, and $\rho_\text{s}$ is a density normalisation parameter. The shape parameter is connected to the local average density in the initial field, encompassing the peak curvature \citep{2008MNRAS.387..536G, 2017MNRAS.465L..84L}. Following \citetalias{2016MNRAS.460.1214L}, we fix $\alpha = 0.18$.

\subsection{Rescaled concentrations}
\label{sec:rescaledCMZ}

The halo scale radii $r_\text{s}$ transform under rescaling as $r_\text{s} \mapsto r_\text{s}^\prime = \alpha r_\text{s}$. NFW halo radii $r_{\Delta\text{m}}$, masses $M_{\Delta\text{m}}$, and concentrations $c_{\Delta\text{m}}$ based on halo overdensities relative to the cosmic mean density also follow simple transformation rules:
$r_{\Delta\text{m}} \mapsto r_{\Delta\text{m}}^\prime = \alpha         r_{\Delta\text{m}}$,
$M_{\Delta\text{m}} \mapsto M_{\Delta\text{m}}^\prime = \beta_\text{m} r_{\Delta\text{m}}$, and
$c_{\Delta\text{m}} \mapsto c_{\Delta\text{m}}^\prime =                c_{\Delta\text{m}}$.

The rescaling transformation laws for NFW profile quantities based on overdensities relative to the \emph{critical} density are more involved. Applying Eq.~\eqref{eq:comovingMatterDensityRescaling} to the NFW profile definition Eq.~\eqref{eq:NFWprofile}, we find for the characteristic densities:
\begin{equation}
	\delta_\text{c}^\prime \rho_\text{crit}^\prime(z^\prime) = \frac{\Omega_\text{m}^\prime}{\Omega_\text{m}} \left(\frac{H_0^\prime}{H_0} \right)^2 \delta_\text{c} \rho_\text{crit}(z). 
\end{equation}
Thus, the concentration $c_{\Delta\text{c}}$ transforms as
\begin{equation}
	c_{\Delta\text{c}} \mapsto c_{\Delta\text{c}}^\prime, 
\end{equation}
with $c_{\Delta\text{c}}^\prime$ given by the (numerical) solution to 
\begin{equation} \label{eq:criticalConcentrationRescaling}
\delta_\text{c}^\prime(c_{\Delta\text{c}}^\prime)  =  \frac{\Omega_\text{m}^\prime}{\Omega_\text{m}} \frac{(1 + z^\prime)^3}{(1 + z)^3} \frac{E(z)^2}{E^\prime(z^\prime)^2} \delta_\text{c} (c_{\Delta\text{c}}).
\end{equation}
The halo mass $M_{\Delta\text{c}}$ then transforms according to
\begin{equation} \label{eq:criticalMassRescaling}
	M_{\Delta\text{c}} \mapsto M_{\Delta\text{c}}^\prime = \beta_\text{c}	M_{\Delta\text{c}},
\end{equation}
with
\begin{equation} \label{eq:critPrimeConcentrationFromFiducialConcentration}
\begin{split}
\beta_\text{c} &= 
\left(\frac{c_{\Delta\text{c}}^\prime}{c_{\Delta\text{c}}}  \right)^3 \cdot \alpha^3 \cdot 
\left(\frac{H_0^\prime}{H_0}\right)^{2}
\frac{ E(z^\prime)^2 }{ E(z)^2}
\frac{ (1 + z)^{3}}{(1 + z^\prime)^{3}}
,
\end{split}
\end{equation}
and $c_{\Delta\text{c}}^\prime$ as the numerical solution to Eq.~\eqref{eq:criticalConcentrationRescaling}. As a range of $c_{\Delta\text{c}}$ values could correspond to a given $M_{\Delta\text{c}}$, this means that the rank order of $M_{\Delta\text{c}}$ is not invariant under rescaling.

One may also use
\begin{equation} \label{eq:critToMeanHaloMass}
M_{\Delta\text{m}}  = \left( \frac{c_{\Delta\text{m}}}{c_{\Delta\text{c}}}\right)^3 \frac{\Omega_\text{m}(1 + z)^3}{E(z)^2} M_{\Delta\text{c}},
\end{equation}
to first convert $M_{\Delta\text{c}}$ to $M_{\Delta\text{m}}$, then rescale $M_{\Delta\text{m}}$ to $M_{\Delta\text{m}}^\prime$, and then convert back to $M_{\Delta\text{c}}^\prime$. We show how to rescale Einasto concentrations in Appendix~\ref{sec:einastoConcentrations}.

\subsection{Concentration-mass-redshift relation}
\label{sec:cmz}

We focus on what excursion sets \citep{1974ApJ...187..425P, 1991ApJ...379..440B} predict for the concentration of haloes \citep{1993MNRAS.262..627L}. One approach for CDM has been to tie the concentration to the mass accretion history of the halo \citep[e.g.][]{2014MNRAS.441..378L, 2015MNRAS.452.1217C}. However, this is not suitable for warm dark matter (WDM) models where the concentration-mass relation is non-monotonic despite the different accretion histories of low and high mass haloes. Revisiting the original NFW argument \citep{Navarro:1995iw, Navarro:1996gj}, it was proposed that the characteristic density of the halo $\delta_\text{c}$ is an imprint of the critical density of the Universe at an appropriate collapse redshift, when progenitors exceeding a fraction $f$ of the final virial halo mass constituted half of this mass. \citetalias{2016MNRAS.460.1214L} argued that choosing the mean density $\expVal{\rho_\text{s}}$ inside the scale radius $r_\text{s}$ to be proportional to the critical density of the Universe at the collapse redshift (instead of $\delta_\text{c}$) and letting the mass inside the scale radius $M_\text{s}$ define the characteristic collapsed mass (instead of the virial mass) yields a better agreement for CDM \emph{and} WDM. This relation then takes the form
\begin{equation}
\label{eq:critMassScaleMassRelation}
 M_\text{s}
 = \frac{4 \pi}{3} r_\text{s}^3 \expVal{\rho_\text{s}}
 = \frac{4 \pi}{3} r_\text{s}^3 \cdot C \cdot \rho_{\text{crit}}(z_\text{s}),
\end{equation}
where $C$ is a proportionality constant and $z_\text{s}$ the collapse redshift.
According to excursion sets \citep{1993MNRAS.262..627L}, the collapsed mass fraction is given by
\begin{equation}
\label{eq:collapsedMassHistory}
\frac{M_\text{s}(f,  z)}{M_{\Delta \text{c}}}
  = \erfc \left( \frac{\delta_\text{sc} (z_\text{s}) - \delta_\text{sc}(z_0)}{\sqrt{2} \cdot \sqrt{\sigma^2\left(f M_{\Delta \text{c}}  \right) - \sigma^2\left(M_{\Delta \text{c}}  \right)}}\right), \\
\end{equation}
where $M_{\Delta\text{c}}$ is the final mass at $z_0$, $\sigma^2 (M)$ the variance of the linear density field on scales equivalent to the mass $M$, and $\delta_{\text{sc}}(z)$ a linear barrier height $\delta_\text{sc} (z) = \delta_\text{sc} (z_0) /D(z)$, where the linear growth is normalised such that $D(z_0) = 1$, and the linear density threshold satisfies $\delta_{\text{sc}}(z_0) = \delta_{\text{sc}}(z = 0) \approx 1.686$ corresponding to spherical collapse at redshift $z = 0$. Combining this with Eq.~\eqref{eq:critMassScaleMassRelation} and an assumed density profile, this system of three equations yields numerical fits for the $c(M, \, z)$-relation. The best-fits for the two constants were determined\footnote{To achieve internal consistency for a spherical collapse model, $C = 400$ would have been the preferred value, but $C= 650$ produced better fits. This inconsistency primarily affects high mass haloes, which are rare in our simulations. Moreover, we limit the possible length scale factors to $\alpha \in \left [ 0.5, \, 2 \right ]$ in Eq.~\eqref{eq:position_rescaling}. For the cosmological parameters in this study, this ensures that $\beta_\text{m} M_{\Delta \text{m}}$ remains in the range of validity.} to be $f = 0.02$ and $C = 650$. We neglect the mild cosmological and redshift dependences of $\delta_{\text{sc}}(z_0)$ in this study.

In \citetalias{2016MNRAS.460.1214L} this relation was found to fit the median $c(M, \, z)$-relation estimated with Einasto profiles for relaxed haloes (see Section~\ref{sec:haloSamples}) for the same simulations that we are using in this paper (see Section~\ref{sec:simulations}) with the $M_{\Delta\text{c}}$ mass definition with $\Delta = 200$. We thus calculate the $c(M, \, z)$-relation with Eq.~\eqref{eq:critMassScaleMassRelation} and Eq.~\eqref{eq:collapsedMassHistory}, assuming an NFW profile Eq.~\eqref{eq:NFWprofile}, with $z_0 = z_\ast$ and $z_\ast^\prime$ in the fiducial and target simulations, respectively, then adapt the relations for $M_{\Delta\text{m}}$ and $c_{\Delta \text{m}}$.

\section{Methodology}
\label{sec:observables}

In this section we present details of our adopted methodology to test the performance of the scaling algorithm. In Section~\ref{sec:simulations}, we describe our fiducial simulation along with five others carried out adopting significantly different cosmologies. We discuss the construction of halo samples in Section~\ref{sec:haloSamples}. In Section~\ref{sec:ggl}, we define the differential excess surface mass
density profiles and provide details about how to measure them, as well as halo concentrations in our simulations.

\subsection{Numerical simulations}
\label{sec:simulations}

This study is conducted with several $N$-body simulations employing \textsc{GADGET-2} \citep{2005MNRAS.364.1105S} with $1080^3$ particles. The fiducial simulation spans a $( 250 \, h^{-1} \,\text{Mpc} )^3$ comoving volume, uses a softening length of $l_\text{s} = 5\,h^{-1}\, \text{kpc}$, and has particle masses $m_\text{p} = 8.61\times 10^{8} h^{-1}\, M_{\sun}$. It assumes a flat $\Lambda$CDM cosmology with a cosmological constant energy density parameter $\Omega_\Lambda = 1 - \Omega_\text{m} = 0.75$, a matter density parameter $\Omega_\text{m} =  \Omega_\text{cdm} + \Omega_\text{b} = 0.25$, baryon density parameter $\Omega_\text{b} = 0.045$, Hubble constant $H_0 = 100 \,h\, \text{km} \, \text{s}^{-1} \, \text{Mpc}^{-1}$ with $h = 0.73$, matter power spectrum normalisation $\sigma_8 = 0.90$, and spectral index $n_\text{s} = 1$. The cosmological parameters and force and mass resolution are identical to those of the Millennium simulation \citep{Springel:2005nw}.

We rescale the fiducial simulation to cosmologies with different values for $\Omega_\text{m}$ and $\sigma_8$. We then compare these rescaled simulations to simulations carried out directly assuming the target cosmologies. These \lq{}direct\rq{} and \lq{}rescaled\rq{} simulations have initial conditions with identical phases. The softening lengths, box sizes, and particle masses in these direct simulations have been chosen to match those in the rescaled simulations. Details are provided in Table~\ref{tab:simulationTable} (the other configurations and parameters are the same as in the fiducial run).

Though the rescaling algorithm captures non-linear structure evolution, it cannot arbitrarily adapt to different growth histories. As dark energy becomes more important at lower redshifts, the growth and expansion histories of different $\Lambda$CDM cosmologies deviate in different manners from an Einstein-de-Sitter evolution. Thus, we expect the inaccuracy of the scaling to grow with cosmic time. For this reason, we focus on structures at redshift $z = 0$ to obtain a conservative estimate on the accuracy of the scaling method. Finally, note that the rescaling parameters $(\alpha, \, z_\ast)$ are identified following \citetalias{Angulo:2009rc} and \citetalias{Angulo:2014gza} for scales corresponding to halo masses in the range $10^8 - 10^{15} \, h^{-1} M_{\sun}$.

\begin{table}
\caption{
\label{tab:simulationTable}
Simulation configurations (fiducial cosmology in the first row) with their values of $\Omega_\text{m}$ and $\sigma_8$ listed. The scale factors $\alpha$ from Eq.~\eqref{eq:boxRescaling} are obtained by dividing the box lengths $L$ with the first column entry. The softening lengths are set as $\alpha \times l_\text{s}$ for the direct simulations with $\alpha = 1$ for the fiducial run. The particle masses $m_\text{p}^\prime$ are calculated using Eq.~\eqref{eq:mass_rescaling}. The rescaling redshifts $z_\ast$ of the fiducial cosmology's snapshots are listed in the last column.
}
\begin{tabular}{c c c c c}
\hline
		$\Omega_\text{m}$ & $\sigma_8$ & $L \, \left[ h^{-1}\, \text{Mpc}\right]$ & $m_\text{p} \, \left[ 10^8 h^{-1} \, M_{\sun} \right]$ & $z_\ast$ \\
\hline		
		0.25 &	0.90 &	250.0 & 8.61 &  -    \\  
		0.15 &	1.00 &	373.3 & 17.2 & 	0.32 \\ 
		0.25 &	0.60 &	205.3 & 4.77 & 	0.56 \\
		0.29 &	0.81 &	224.4 & 7.22 & 	0.06 \\
		0.40 &	0.70 &	176.4 & 4.84 & 	0    \\
		0.80 &	0.40 &	 88.2 & 1.21 & 	0    \\
\hline
\end{tabular}
\end{table}

\subsection{Halo samples}
\label{sec:haloSamples}

Haloes in the simulations are first identified using a friends-of-friends (FOF) algorithm \citep{1985ApJ...292..371D} with a linking length of 0.2 times the mean particle separation. The FOF haloes are then processed with \textsc{SUBFIND} \citep{2001MNRAS.328..726S}, employing the same settings as for the MXXL simulation \citep{2012MNRAS.426.2046A}, to identify self-bound structures, possibly returning a main subhalo and further self-bound subhaloes.

We will mostly consider halo samples defined by their (rescaled) $M_{200\text{m}}$ mass. However, in some cases we will also consider halo samples that only include matched haloes in direct-rescaled pairs of simulations. Following \citetalias{Angulo:2009rc}, we identify as match candidate for each halo in the direct simulation the halo in the rescaled simulation with the most particles with ids matching those of the direct simulation's halo. We repeat the process with the simulations' roles swapped, and consider a haloes matched if they are each others match candidates.

Note that the most accurate rescaling approach would be to transform individual simulation particles and then re-run the group finding algorithm. However, this is computationally expensive, and similarly accurate results can be obtained by directly rescaling the halo catalogue, as shown by \cite{Ruiz2011} \citep[see also][]{Mead2014a, Mead2014b}, which is the procedure we adopt here; we rescale the position and mass of each snapshot particle but keep the fiducial halo catalogue and rescale it accordingly.

Unrelaxed haloes are poorly described by NFW profiles, and their best fit concentrations tend to be lower than those of relaxed systems \citep{2007MNRAS.381.1450N}. To test for this in our results, in some cases we will consider samples of haloes that satisfy two criteria. The first criterion is based on the offset between the centre-of-mass $\vec{r}_\text{CM}$ and the gravitational potential minimum $ \vec{r}_\text{pot}$ relative to the halo radius $r_{200}$ \citep{2001MNRAS.324..450T, 2007MNRAS.378...55M, 2007MNRAS.381.1450N} $d_\text{off} = \left \vert \vec{r}_\text{pot} - \vec{r}_\text{CM} \right \vert / r_{200}$. We consider haloes relaxed if $d_\text{off} < 0.1$. The second criterion is a substructure threshold \citep{2007MNRAS.381.1450N, 2012MNRAS.427.1322L}, $f_\text{sub} = M_\text{sub}/M_{200} < 0.1$, where $M_\text{sub}$ is the mass of all bound particles in the subhaloes apart from the main halo identified by the substructure finder. 

These criteria lead to similar results as imposing the $d_\text{off}$ cut and a dynamical age criterion, $t_{50} \geqslant 1.25 \, t_\text{cross}$ \citep{2016MNRAS.458.2848J, 2016MNRAS.460.1214L} curtailing the allowed accretion of the main progenitor w.r.t. its crossing time $t_\text{cross} = 2 \, r_{200}/V_{200}$, as they exclude recent mergers of structures with similar mass.\footnote{However, a dynamical timescale cut also discriminates against haloes at maximum contraction following a massive merger, which are still present in our subsample.} With the $M_{200\text{m}}$ mass definition\footnote{Given $\beta_\text{c}$ in Eq.~\eqref{eq:critPrimeConcentrationFromFiducialConcentration}, the cuts w.r.t. $M_{200\text{c}}$ are not rescaling invariant. Since the measured concentrations are influenced by these cuts \citep{2007MNRAS.381.1450N}, a recursive rescaling fitting scheme is required to find the passing haloes in the target cosmology.}, the geometric cuts on $f_\text{sub}$ and $d_\text{off}$ are trivially invariant under the rescaling mapping\footnote{provided we ignore implicit relations, e.g. redshift evolution which affects $f_\text{sub}$ \citep[e.g.][]{2005MNRAS.359.1029V}}. This invariance does not hold for other dynamical relaxation criteria such as bounds on the virial ratio\footnote{If the simulation's softening length $l_\text{s} \mapsto \alpha \text{s}$ and $\alpha_\text{vel} \approx \alpha$ for the velocities whose transform is given in \citetalias{Angulo:2009rc} then $\eta \mapsto \eta^\prime \simeq \Omega_\text{m}/\Omega_\text{m}^\prime (H_0/H_0^\prime)^2 \eta$ with the potential $U$ given in \citet{Springel:2005nw}. Since $U$ and $T$ have different transform prefactors, mapping $\lambda \mapsto \lambda^\prime$ is non-trivial.} $\eta = 2K/|U|$ \citep[e.g.][]{1996MNRAS.281..716C} or the spin parameter\footnote{In \citetalias{Angulo:2009rc}, $\lambda$ was comparable for the haloes in the direct and rescaled simulation snapshots, hinting at similar internal dynamical states, whereas the halo concentrations estimated from velocities displayed a systematic bias.} $\lambda$ \citep[e.g.][]{2007MNRAS.376..215B}.

\subsection{Halo density and weak-lensing profiles}
\label{sec:ggl}

We measure the spatial cross-correlation between the halo and matter fields in our simulations to obtain mass profiles in 3D and 2D. In 3D, we consider spherically averaged radial matter density profiles for haloes as a function of halo mass. As analytic approximations to these profiles we consider NFW profiles Eq.~\eqref{eq:NFWprofile} and Einasto profiles Eq.~\eqref{eq:einastoDensityProfile}.

The 3D density field is not readily available in the real Universe. However, galaxy-galaxy lensing can be used to probe the cross-correlation between galaxies and matter. Assuming statistical isotropy, this cross-correlation $\xi_{\text{gm}}(\left \vert \vec{r}^\prime - \vec{r} \right \vert) = \expVal{\delta_\text{g}(\vec{r}) \delta_\text{m}(\vec{r}- \vec{r}^\prime)}$ between the total overdensity of matter $\delta_\text{m}$ and the overdensity of lens galaxies $\delta_\text{g}$ at comoving positions $\vec{r}$ and $\vec{r^\prime}$, respectively, is related to the mean projected surface mass overdensity $\Sigma$ at projected comoving transverse distance $r$ through
\begin{equation}
\label{eq:projectedSurfaceMassDensity2PT}
\Sigma(r) = \bar{\rho}\int \text{d} l \,\xi_{\text{gm}}(\sqrt{r^2 + l^2}),
\end{equation}
with $\bar{\rho}$ as the mean comoving density. The differential excess surface mass density $\Delta \Sigma (r)$ then reads
\begin{equation}
\label{eq:excessSurfaceMassDensity}
\Delta \Sigma (r) =  \bar{\Sigma}(\leqslant r) - \Sigma(r),
\end{equation}
where
\begin{equation}
\bar{\Sigma}(\leqslant r) = \frac{1}{\pi r^2} \int_0^{r} \dd{r'} 2\pi r' \Sigma(r'),
\end{equation}
denotes the mean projected surface mass overdensity inside a circular aperture with radius $r$. $\Delta \Sigma$ can be estimated from the tangential shear $\bar{\gamma}_\text{t} = \Delta \Sigma / \Sigma_\text{crit}$ induced by lens galaxies at redshift $z_\text{d}$ in images of source galaxies at redshift $z_\text{s} > z_\text{d}$ \citep{1991ApJ...370....1M, 1996ApJ...473...65S, 2001ApJ...555..572W}, where $\Sigma_\text{crit}=\Sigma_\text{crit}(z_\text{d},z_\text{s})$ denotes the comoving critical surface mass density for lenses at redshift $z_\text{d}$ and sources at redshift $z_\text{s}$. Hence, the tangential shear of background galaxies provides information on the matter distribution around foreground galaxies.

As analytical models, we consider NFW lenses \citep{2000ApJ...534...34W, 2009JCAP...01..015B}. The lensing expressions are acquired by integrating the NFW density profile Eq.~\eqref{eq:NFWprofile} along the line-of-sight. Expressed in terms of the dimensionless ratio $x = r/r_\text{s}$, the projected surface mass density at a radius $x$ is then acquired through\footnote{We ignore differences between halo density and overdensity profiles, since these do not affect the differential excess surface mass density $\Delta\Sigma$.}
\begin{equation}
\label{eq:projectedSurfaceMassDensityAnalyticalImplicit}
\Sigma(x) = 2 r_\text{s} \int^{\infty}_0 \dd{l} \rho_\text{NFW}(\sqrt{l^2 + x^2}),
\end{equation}
whereas $\Delta \Sigma$ is given by Eq.~\eqref{eq:excessSurfaceMassDensity}. We restrict the comparison to scales $\lesssim$ the halo virial radii and leave modelling of the large scales for future studies. We do not model the lenses with Einasto profiles as those are similar to NFW lenses \citep{2012A&A...540A..70R, 2016JCAP...01..042S}.

Operationally, we compute 3D radial halo profiles $\rho$ and projected radial profiles $\Sigma$ by binning all particles in spherical and cylindrical shells, respectively, around the recorded halo centres given by the positions of their most bound particles. To moderate triaxiality \citep[e.g.][]{2002ApJ...574..538J} and other deviations from azimuthal symmetry, we project the cylinders along the three principal simulation box axes and let the mean signal describe the halo sample, effectively tripling our sample size. For the rescaled simulation, the profiles are computed after applying the adequate rescaling to ensure matching bin boundaries.

In order to assess the errors due to the limited volume, we bootstrap resample \citep[e.g.][]{Efron:1979} the haloes in each mass bin with 100 realisations to estimate the variance. For $\Delta \Sigma$ we calculate 100 realisations per axis.

We consider halo samples selected by mass with 0.1 dex width above $10^{12} \, h^{-1} \, M_{\sun}$ to approximately $10^{14.5} \, h^{-1} \, M_{\sun}$ where we record twenty haloes per bin. For the halo mass function we show the result in 0.05 dex bins. For the 3D profiles, we follow \citet{2007MNRAS.381.1450N}, where the matter density profiles were estimated using 32 $\log$-equidistant bins between $r_{200\text{c}}$ and $\log_{10} (r/r_{200\text{c}}) = - 2.5$ where we replace $r_{200\text{c}}$ with $r_{200\text{m}}$. To suppress the impact of outliers on the 3D profile fits, we use the median particle count per spherical shell as input, unless otherwise specified. We then minimise the difference in $\ln \rho$ between the measured median profile and the analytic profile to determine the best fit parameters. We also present concentration estimates for individual haloes from the separate particle counts. To investigate the transition regime between the 1-halo and 2-halo terms, we bin the particles in 64 $\log$-equidistant bins for $0.05 \, r_{200\text{m}} < r < 5 \, r_{200\text{m}}$.

GGL profiles for each mass-selected halo sample are obtained through Eq.~\eqref{eq:excessSurfaceMassDensity}, with the projected profiles computed by binning the particles in 40 $\log$-equidistant bins in the $30 \, h^{-1} \, \text{kpc} - 3 \,h^{-1} \, \text{Mpc}$ range. The average GGL profiles are fitted by analytical profiles Eq.~\eqref{eq:projectedSurfaceMassDensityAnalyticalImplicit} minimising
\begin{equation}
\label{eq:fom}
\chi^2 = \sum^{N_r}_{i=1} r_i^2 \left[ \Delta\Sigma_\text{data}(r_i) - \Delta\Sigma_\text{NFW}(r_i; r_{200\text{m}}, c_{200\text{m}}) \right]^2,
\end{equation}
w.r.t. $r_{200\text{m}}$ and $c_{200\text{m}}$. The radial weights $\propto r^2$ are observationally motivated, as the shape noise error on the signal scales with the number density of background galaxies, which is proportional to the area of the projected cylinder assuming a constant source density. In observations, masking and blending of background galaxies by foreground galaxies becomes a major systematic as one approaches the central galaxy \citep{2015MNRAS.452.3529V}, which motivates the lower cutoff.

\section{Results}
\label{sec:results}

In this section we quantify the performance of the scaling algorithm and present alternatives to further improve it. We first focus on the halo mass functions (Section~\ref{sec:haloMassFunctionResults}), the 3D density profiles (Section~\ref{sec:results3DDensityProfiles}), and the differential excess surface mass density profiles (Section~\ref{sec:resultsTangentialShearAndProjectedSurfaceMassDensity}) for
the original algorithm. The accuracy of the rescaling for the concentration-mass relation is quantified and compared to the theoretical prediction of \citetalias{2016MNRAS.460.1214L} in Section~\ref{sec:concentrationFits}. In Section~\ref{sec:correctedProfiles} we use this model to correct the rescaled profiles and show the resulting improvements.
Attempts at further ameliorations for the halo outskirts based on models for the position of the splashback radius are discussed in Section~\ref{sec:outerProfileBias}. We will focus on representative cases using one of the cosmologies studied where the others manifest similar trends and primarily report on the findings for $(0.80, \, 0.40)$ in Appendix~\ref{sec:resultsForAlmostEinsteinDeSitter} as these parameters strongly deviate from current observational constraints.

\subsection{Halo mass function}
\label{sec:haloMassFunctionResults}

\begin{figure}
	\includegraphics[width=1.04\columnwidth]{\figrelpath 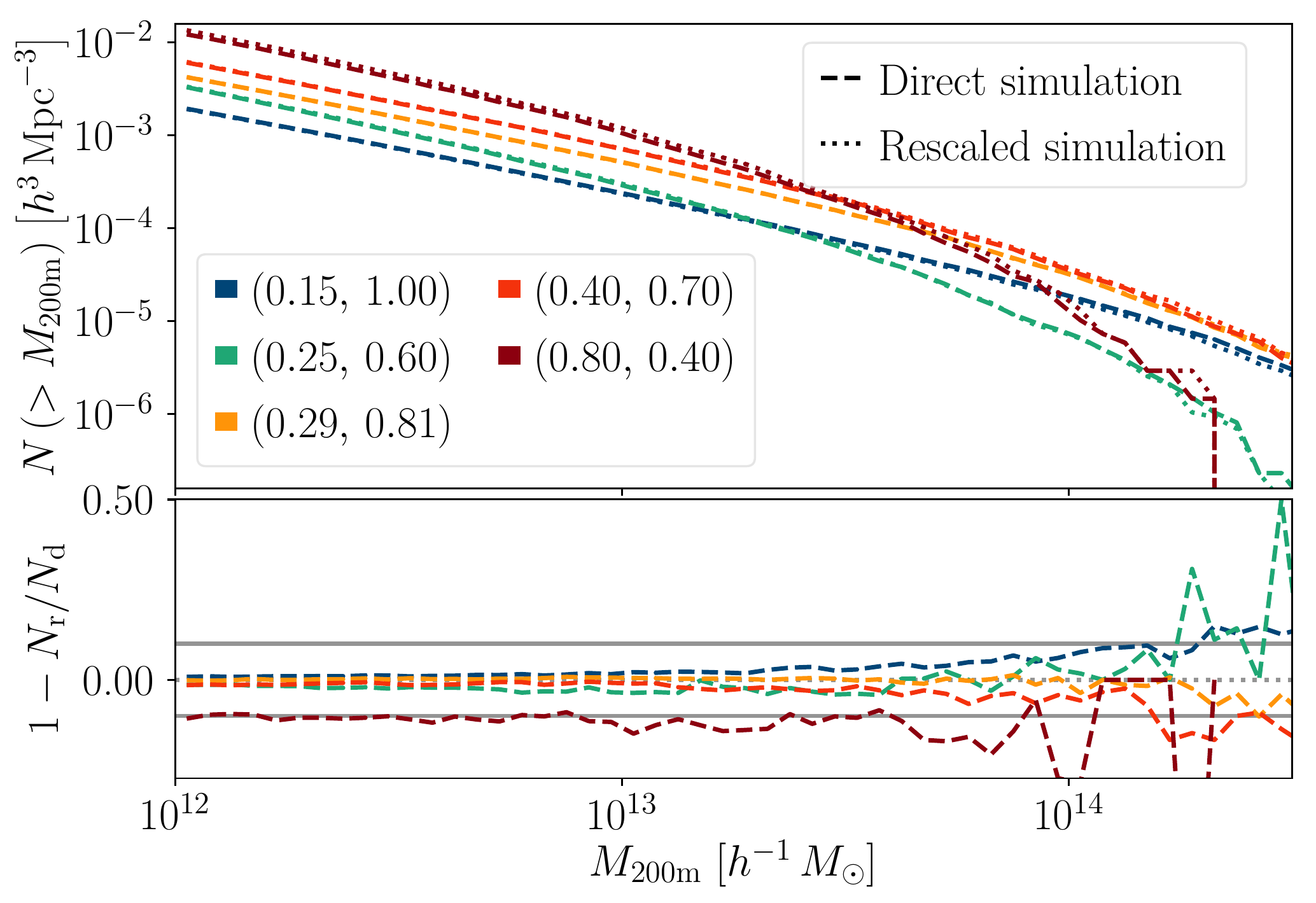}
    \caption{
    \label{fig:hmfRawFull1012}
		Cumulative halo mass function at $z=0$ (in 0.05 dex bins) for simulations with different values for ($\Omega_\text{m}$, $\sigma_8$) as indicated by the legend. For each cosmology, we display results for direct and rescaled simulations. The fractional differences between these two cases are shown in the bottom panel where solid lines mark $\pm \,10\,\%$.}
\end{figure}

One of the most basic quantities predicted by simulations is the halo mass function. The cumulative halo mass function (HMF) $N(> M)$ defines the number of haloes above a certain mass $M$ per comoving volume. In \citetalias{Angulo:2009rc}, the number densities were properly matched with a bias of order $\lesssim$ 10\,\%. To avoid numerical artefacts, we only compare HMFs for haloes with (rescaled) masses exceeding $10^{12} \, h^{-1} M_{\sun}$ (i.e. objects resolved with $>1000$ particles).

In Fig.~\ref{fig:hmfRawFull1012}, we show $N(>M)$ for all haloes in the direct and rescaled cosmologies with the fractional difference in the bottom panel. In numbers, there are 100\,154, 28\,427, 47\,519, 33\,123 and 8\,325 haloes with $M_{200\text{m}} > 10^{12} \, h^{-1} M_{\sun}$ in the direct simulations (listed according to increasing $\Omega_\text{m}$), and 97\,232, 28\,145, 46\,620, 32\,888 and 8\,999 haloes in the rescaled snapshots. As seen in Fig.~\ref{fig:hmfRawFull1012}, the error in the number counts is in the range $\pm\,10\,\%$ for all simulations except for $(0.80, \, 0.40)$ and for masses $<10^{14} \, h^{-1} M_{\sun}$. At higher masses, Poisson noise is significant. In addition, these clusters are the last structures to have collapsed and thus are most sensitive to changes in the growth rate governed by the background cosmology. Since we opt for a minimisation scheme covering a large range of halo masses, the rescaling parameters are not necessarily the best ones for cluster-size haloes. This could then bias the predicted masses. The best matches are found for the $(0.29, \, 0.81)$ and $(0.40, \, 0.70)$ cosmologies, with fractional differences $\lesssim 3\,\%$. Overall, this performance is similar to that stated in \citetalias{Angulo:2009rc}.

\begin{figure}
	\includegraphics[width=1.04\columnwidth]{\figrelpath 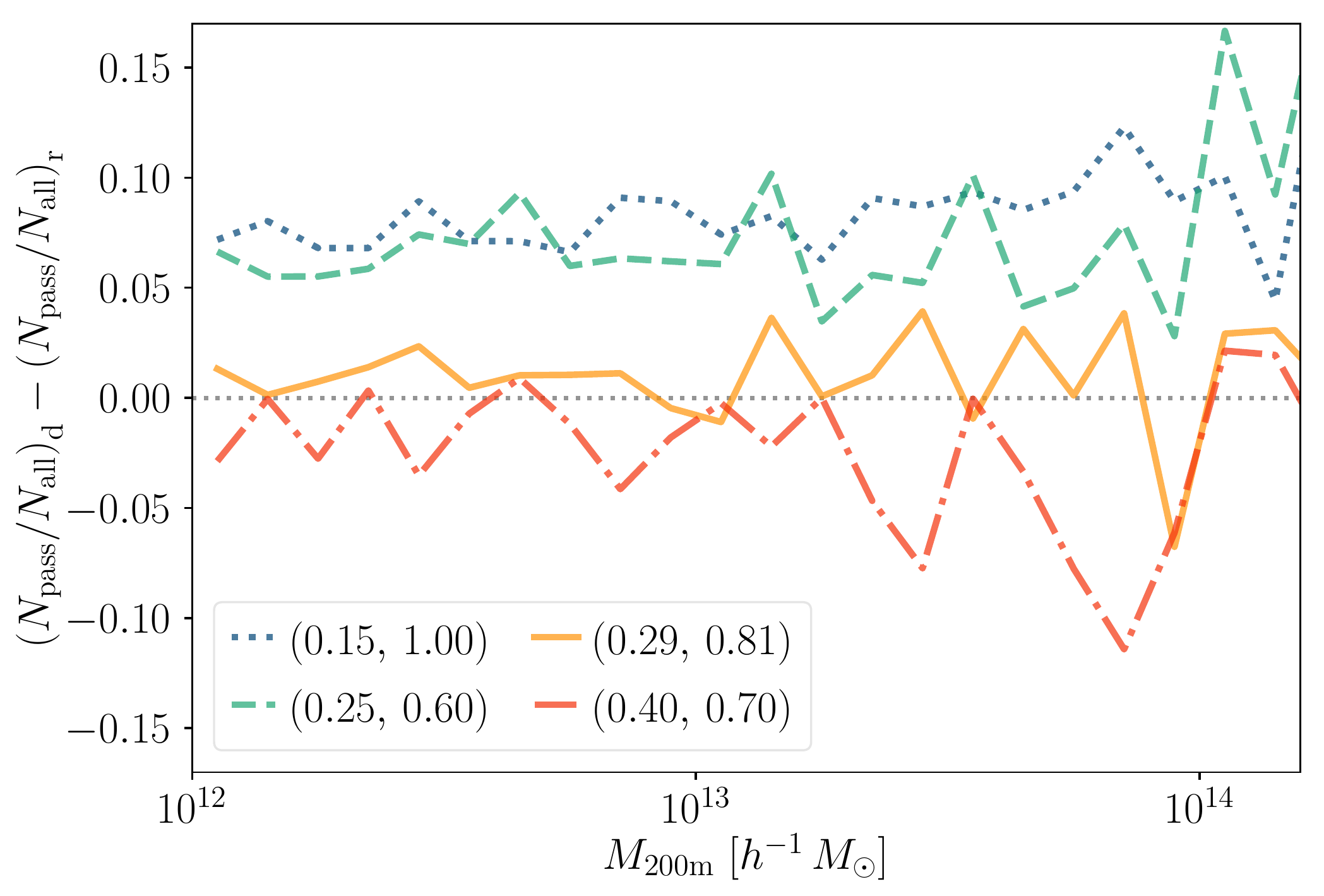}
    \caption{
    \label{fig:hmfDoffCutFsubCutWrtAll}
		Difference in the fraction of relaxed haloes between the direct and rescaled simulations per 0.1 dex mass bin with the $d_\text{off} + f_\text{sub}$ cuts enforced (the results are similar if only the $d_\text{off}$ cut is applied).
		}
\end{figure}

Trends for passing the relaxation cuts are similar in the direct and rescaled simulations, with cuts more effective at the high mass end, and peak passing rates between 54 and 73\,\% for the $10^{12.0} - 10^{12.1} h^{-1}\,M_{\sun}$ mass bin. As Fig.~\ref{fig:hmfDoffCutFsubCutWrtAll} illustrates, there are however some differences between the direct and rescaled simulation in the fraction of haloes per mass bin which satisfy the relaxation criteria. For $(0.15, \, 1.00)$ and $(0.25, \, 0.60)$, fewer haloes per mass bin survive the cuts, which may indicate a possible redshift dependence of the cut efficiency, as the rescaled signals come from fiducial snapshots at higher redshifts. This implies that we do not only have a slight scatter in the number of haloes but also in the properties of the haloes which pass the relaxation cuts.

\begin{figure}
	\includegraphics[width=1.04\columnwidth]{\figrelpath 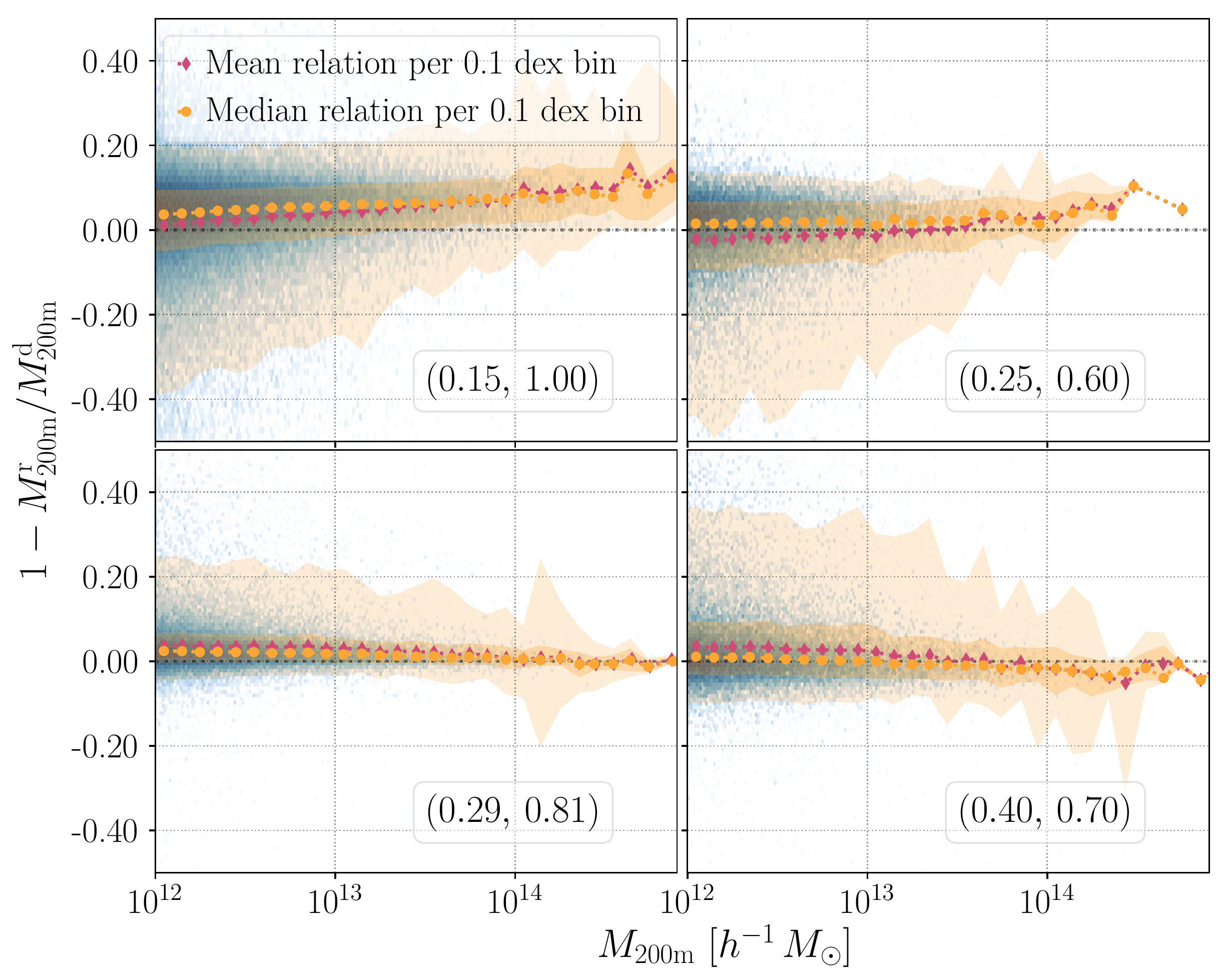}
    \caption{
    \label{fig:hmfMatchedMassDifference}
		Fractional difference in the mass of matched haloes identified in direct and rescaled simulations. Each panel shows results for a different combination of $\Omega_\text{m}$ and $\sigma_8$ indicated in the legend. Contours enclose 68\,\% and 95\,\% of the distributions, and symbols mark the mean and median values per mass bin.
		}
\end{figure}

Almost all haloes ($\sim99\,\%$) with $M_{200\text{m}} \geq 10^{12} h^{-1} \, M_{\sun}$ in the direct simulations have matches in the rescaled simulation (and the few non-matches have no significant impact on the profile statistics considered here). However, properties of matching haloes are usually not identical. The fractional difference in recorded $M_{200\text{m}}$ between the matched haloes in the direct simulation and their matched rescaled counterparts is shown in Fig.~\ref{fig:hmfMatchedMassDifference}. Both a scatter and a systematic trend with mass and cosmology are discernible. For example, haloes in the rescaled simulation tend to be less massive than their counterparts for $(0.15, \, 1.00)$. These trends are in part responsible for differences in the halo profiles between the direct and rescaled simulations discussed in the following sections.

\subsection{3D density profiles}
\label{sec:results3DDensityProfiles}

\begin{figure}
	\includegraphics[width=1.04\columnwidth]{\figrelpath 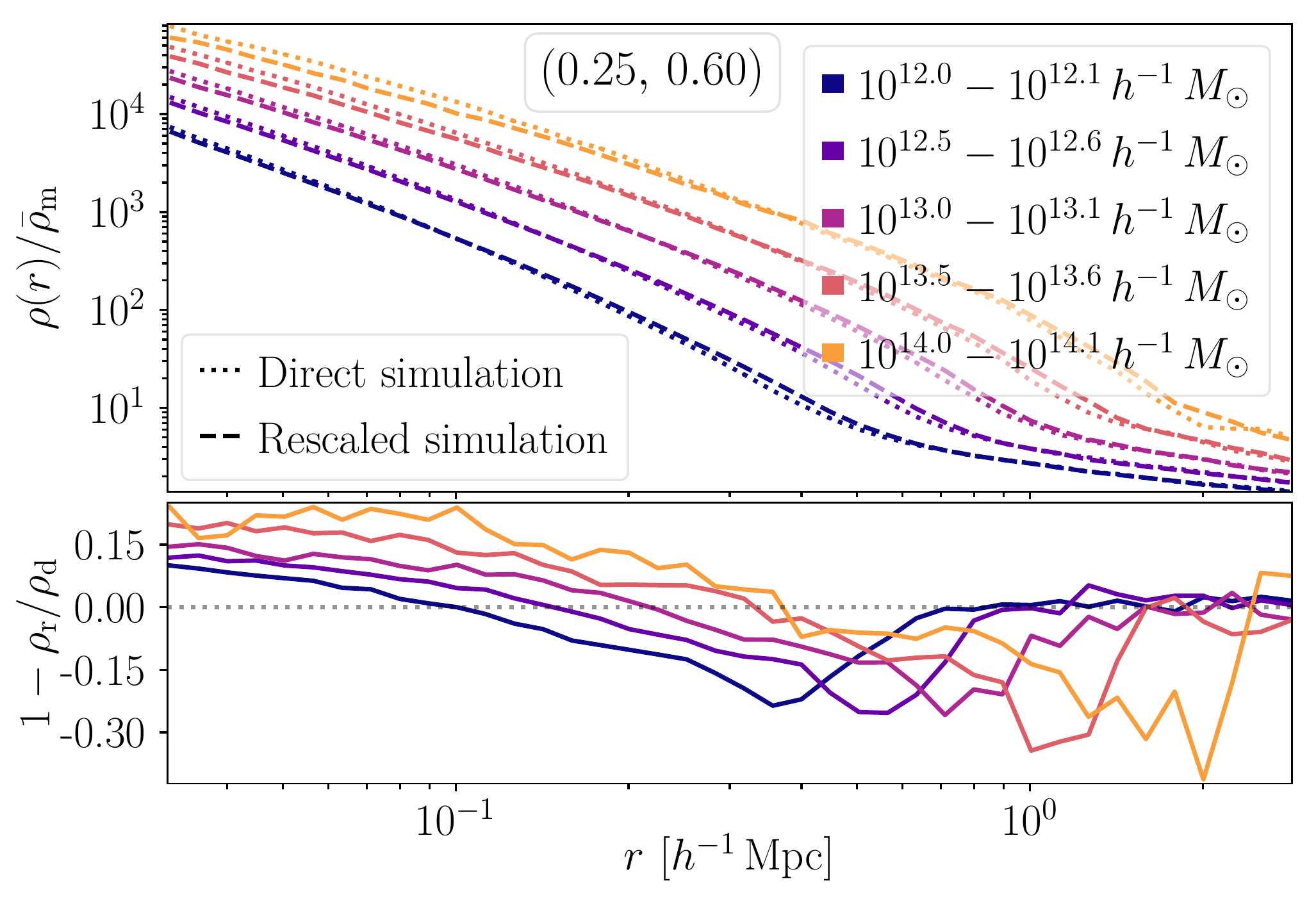}
  \caption{
  \label{fig:3DDensityProfileAllHaloBiasIllustration}
	  3D comoving matter density profiles $\rho(r)$ in units of the cosmic mean density $\bar{\rho}_\text{m}$ as function of radius $r$ for all haloes in direct and rescaled simulations of the $\Omega_\text{m}=0.25,\,\sigma_8=0.60$ cosmology for five different mass bins (see legend). Fractional differences between the results from the two simulations are shown in the bottom panel.
		}
\end{figure}

In Fig.~\ref{fig:3DDensityProfileAllHaloBiasIllustration} we plot the median density profiles for five mass bins in the $(0.25, \, 0.60)$ cosmology in 40 $\log$-equidistant bins between $0.03 - 3 \, h^{-1} \, \text{Mpc}$. The halo profiles in the direct and rescaled simulations display remarkable agreement, with differences of at most $20\,\%$ over two orders of magnitude in density and scale. The differences likely reflect different mass accretion histories and formation times for the direct and rescaled haloes. They are characterised by two features: (i) an underestimation (overestimation) of the density near the halo centre, and (ii) an overestimation (underestimation) of the density near the transition scale between the 1-halo and 2-halo terms for the $(0.15, \, 1.00)$, $(0.25, \, 0.60)$ and $(0.29, \, 0.81)$ cosmologies, with the opposite signs for $(0.40, \, 0.70)$ and $(0.80, \, 0.40)$. 

\begin{figure*}
\begin{centering}
	\includegraphics[width=2.1\columnwidth]{\figrelpath 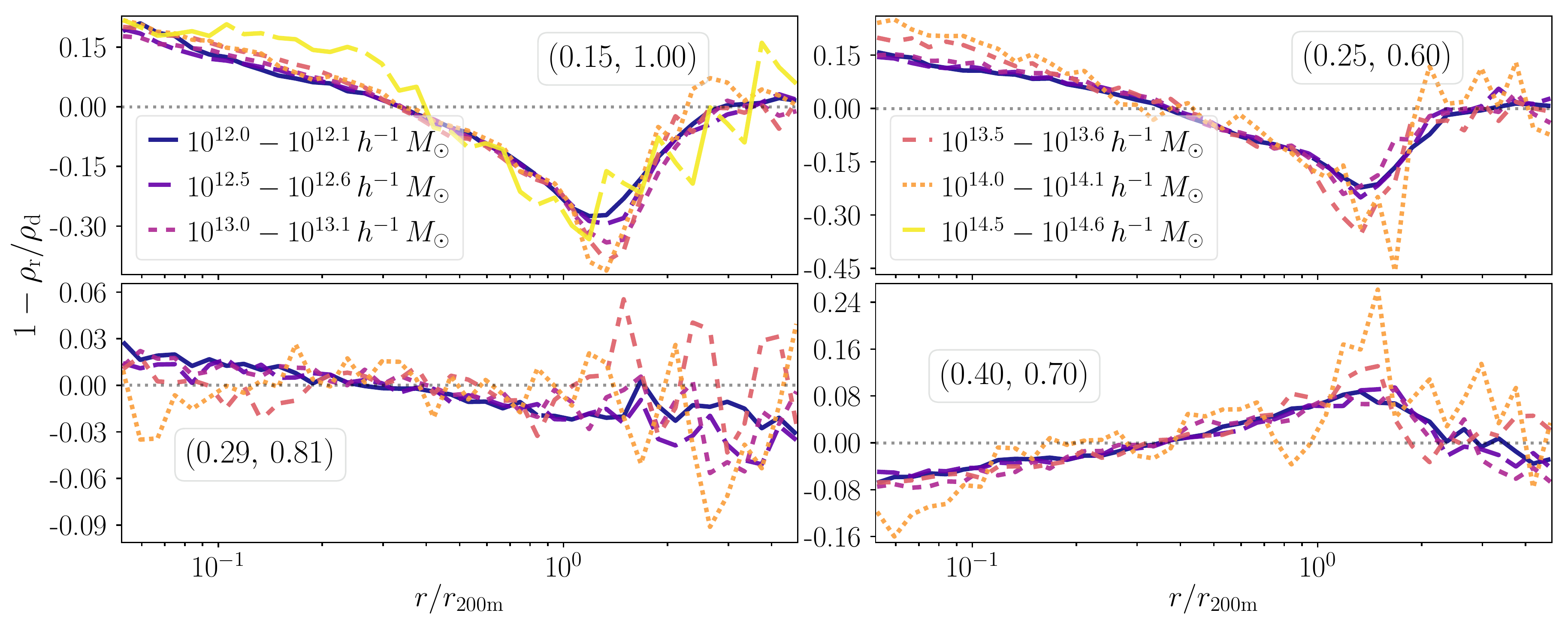}
  \caption{
  \label{fig:som0Rho}
		Fractional differences in the 3D density profiles of haloes in the direct and rescaled simulation snapshots. Each panel displays results for a different background cosmology and for five to six disjoint halo mass bins, as indicated by the legend. The $x$-axis is in units of the halo $r_{200\text{m}}$ radius, which highlights that the differences are almost independent of mass. 
	}
\end{centering}
\end{figure*}

Fig.~\ref{fig:som0Rho} shows the fractional difference for four of our test simulations for haloes in four to six mass bins, where more than twenty haloes have been recorded in the direct and rescaled simulations. The magnitude (though not always the sign) of the differences is similar to that for the $(0.25, \, 0.60)$ cosmology. From approx. $0.3$ to $3 \, r_{200\text{m}}$, the rescaled profiles have an outer bias with the opposite sign to the inner ($r \lesssim 0.3 \,r_{200\text{m}}$) profile bias, until they reach better agreement at larger scales ($r > 3 \, r_{200\text{m}}$). This suggests that the simulations have a similar halo bias. Fewer haloes in the higher mass bins lead to a larger scatter, predominantly in the outskirts where the active evolution takes place. Performing the same tests with just haloes passing the relaxation cuts or matched haloes yield similar results as for the whole population, indicating that the biases are universal features. We show the corresponding fractional differences for matched haloes only in Appendix~\ref{sec:matchedHaloes}.

\subsection{Weak lensing profiles}
\label{sec:resultsTangentialShearAndProjectedSurfaceMassDensity}

\begin{figure*}
\begin{centering}
	\includegraphics[width=2.1\columnwidth]{\figrelpath 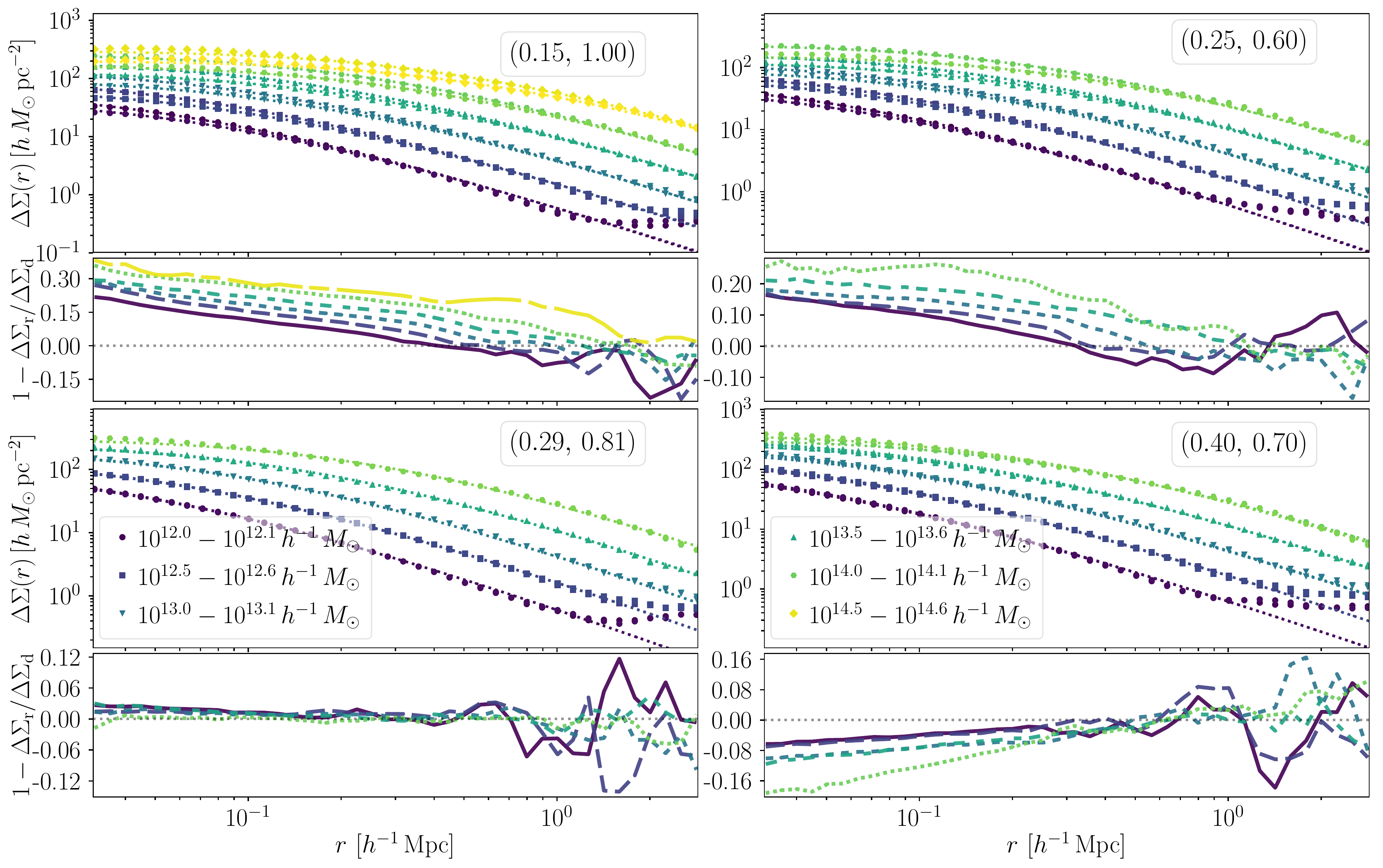}
    \caption{
    \label{fig:som0DeltaSigma}
		Differential excess surface mass density profiles $\Delta \Sigma(r)$ for stacks of haloes in the direct and rescaled simulations. Different colours indicate the different halo mass bins displayed whereas different panels show results for different cosmologies, where the bottom sub-panels show fractional differences with the same mass bin line styles as in Fig.~\ref{fig:som0Rho}. In the upper sub-panels, the best fit NFW profiles are indicated by dotted lines. The results for the $(0.80, \, 0.40)$ cosmology are presented in Appendix~\ref{sec:resultsForAlmostEinsteinDeSitter}.
		}
    \end{centering}
\end{figure*}

As shown in Fig.~\ref{fig:som0DeltaSigma}, the small differences in the 3D density profiles propagate to small differences in the weak lensing profiles. The best agreement between the profiles of the rescaled and direct simulations is reached for $(0.29, \, 0.81)$. The other cosmologies show larger differences, in particular in the inner profiles. In contrast, the outer profile bias is barely discernible except for the low mass bins for $(0.25, \, 0.60)$, implying that it is washed out by taking the mean and calculating the projection. If we increase the mass bin width to 0.2 dex and recompute the profiles, the outer profile bias almost completely vanishes in 2D but it is still discernible in 3D for median profiles. The transition regime scatter does not necessarily dampen at larger scales\footnote{We calculated the large scale $\Delta \Sigma$ for $(0.29, \, 0.81)$ for the same mass bins for $3 - 30 \, h^{-1} \, \text{Mpc}$ and there are small differences at the level of the scatter over this range.}.
For the total maximum and median values of the residuals below $r_{200\text{m}}$, we refer to Table~\ref{tab:deltaRhoSigmaBeforeAndAfterCorrection}.

As for the 3D density profiles, we find negligible differences between all haloes and all matched haloes. However, the scatter in the 2-halo transition regime is dampened, and the inner and outer profile biases are accentuated, especially for $(0.29, \, 0.81)$. In addition, there are no conspicuous differences between the profiles for all haloes, for those which pass the $d_\text{off}$ relaxation cut and for those which pass both $d_\text{off} + f_\text{sub}$ relaxation cuts.

\begin{figure*}
\begin{centering}
	\includegraphics[width=1.5\columnwidth]{\figrelpath 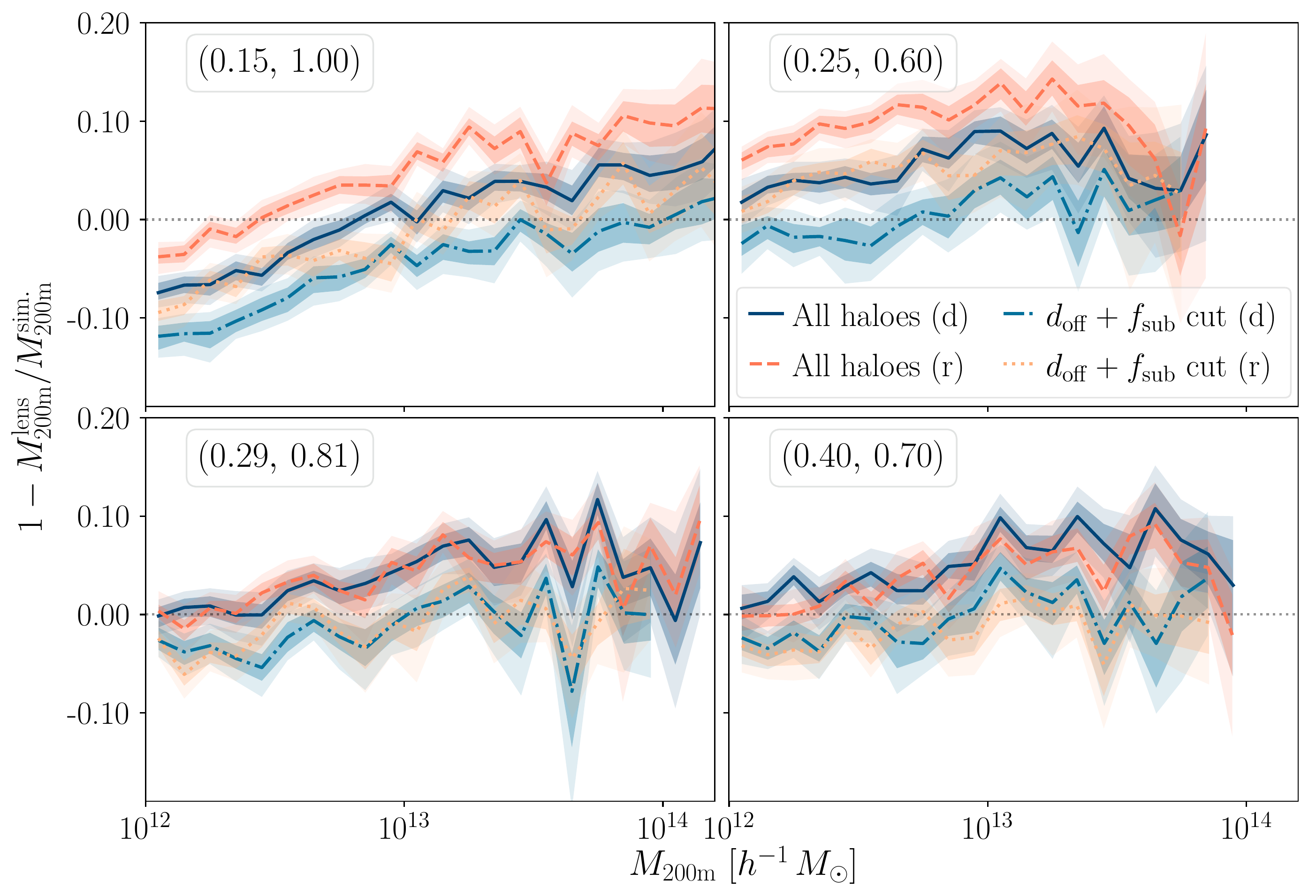}
    \caption{
    \label{fig:massBiasFitVsSimMass}
		Fractional differences between the true mean mass of haloes in our simulations, $M_{200\text{m}}^{\text{sim.}}$, and that inferred from their $\Delta \Sigma$ profiles, $M_{200\text{m}}^{\text{lens}}$. Each panel focuses on a particular cosmology, and it shows results from the direct and rescaled simulations for all haloes and only for those relaxed according to two different criteria. The coloured regions mark the $68 \, \%$ and $95 \, \%$ percentiles, estimated from the bootstrap resample.
		}
      \end{centering}
\end{figure*}

Fig.~\ref{fig:som0DeltaSigma} also illustrates that the $\Delta \Sigma$ profiles for $r \lesssim  r_{200\text{m}}$ are well described by NFW lens profiles. We fit the measured mean profiles by minimising Eq.~\eqref{eq:fom} with both $c_{200\text{m}}$ and $r_{200\text{m}}$ as free parameters. Fig.~\ref{fig:massBiasFitVsSimMass} shows the relative difference between the mean $M_{200\text{m}}$ recorded by the halo finder and the value fitted from the $\Delta \Sigma$ profiles. For the simulations with rescaled fiducial snapshots close to $z = 0$, the rescaled and direct simulation mass biases have similar amplitudes and show a similar evolution in mass with additional scatter at the high mass end. Introducing relaxation cuts shifts the amplitude consistently in the direct and rescaled simulation towards zero and for some high mass bins the bias changes signs, presumably due to scatter. The results with only the $d_\text{off}$ cut enforced are similar to the ones where both cuts are imposed.

The negative bias for low mass haloes, particularly for $(0.15, \, 1.00)$, is likely due to a lack of spatial resolution, which causes the measured lensing profiles to fall below the analytic profiles in the innermost regions. Moreover, for $(0.15, \, 1.00)$ and $(0.25, \, 0.60)$, there is a visible systematic offset between fit masses of the rescaled and direct simulations, which is preserved with the introduction of cuts. Small but significant cosmology-dependent deviations from the analytic NFW lens profiles even for relaxed haloes might cause this offset. This requires further investigation in future work.

\subsection{Concentration-mass relations}
\label{sec:concentrationFits}

\begin{figure*}
\begin{centering}
\large
\stackon[5pt]{\includegraphics[width=1.03\columnwidth]{\figrelpath 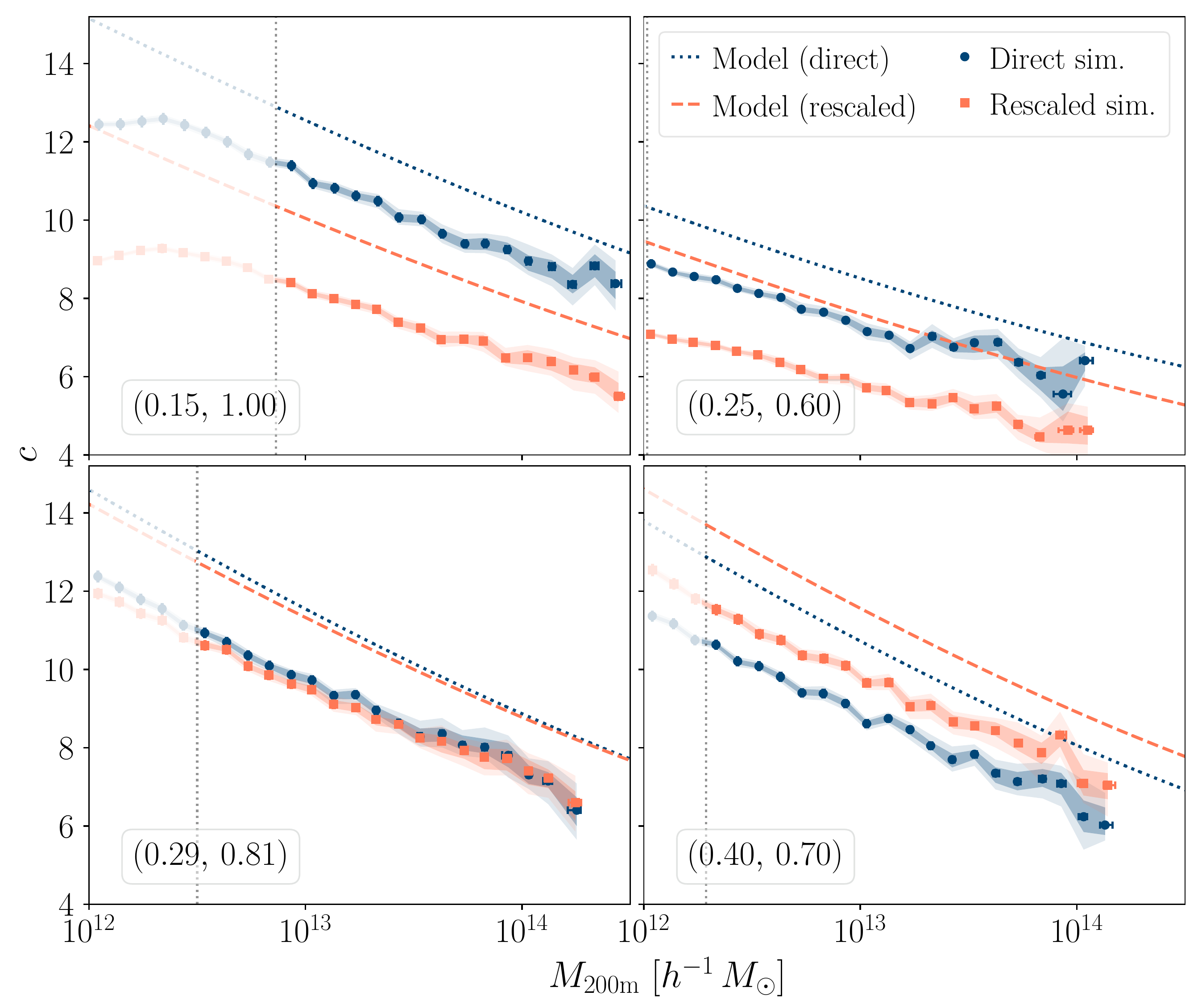}}{{\bf Density profiles}}
\stackon[5pt]{\includegraphics[width=1.03\columnwidth]{\figrelpath 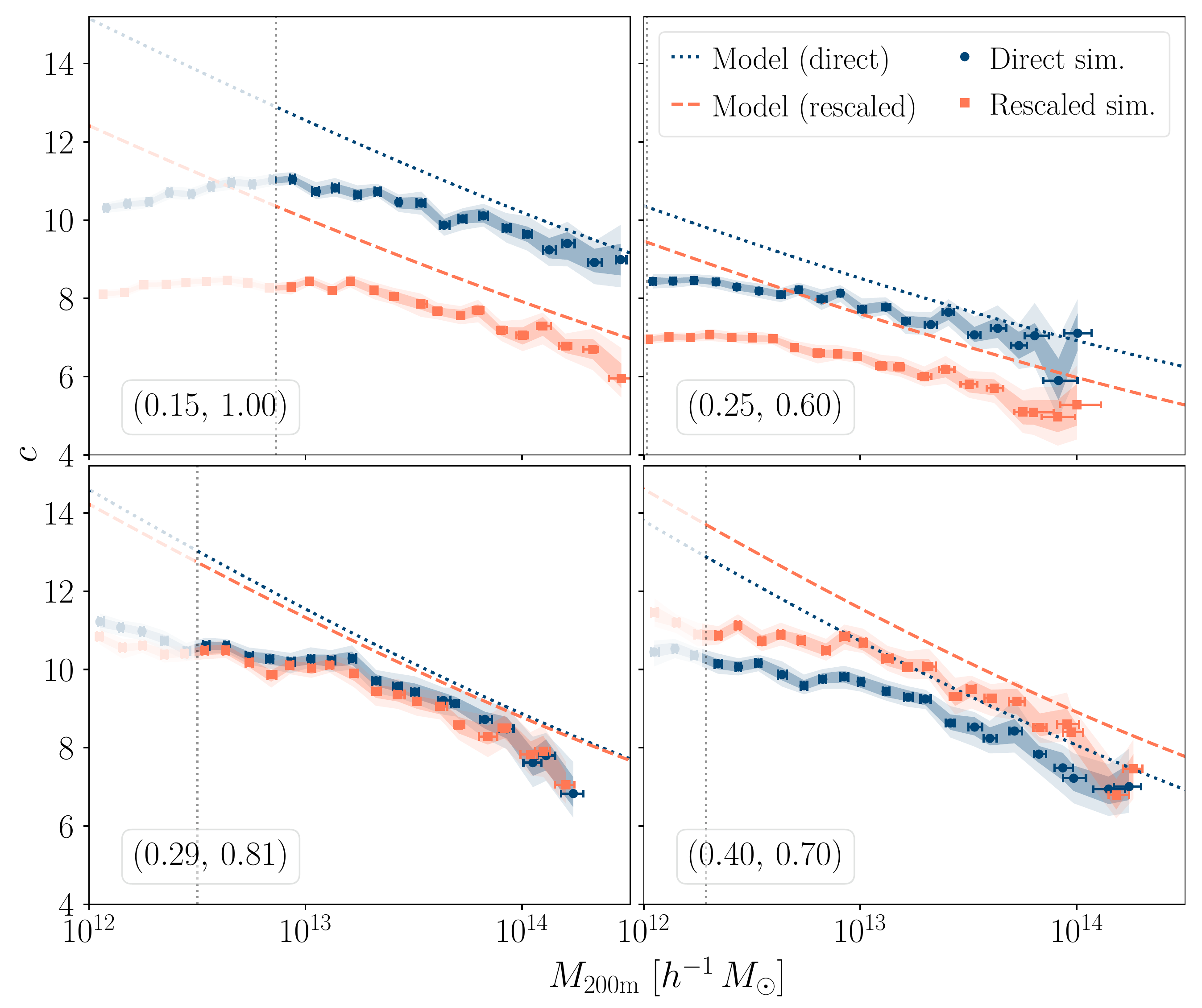}}{{\bf $\Delta \Sigma$ profiles}}
\caption{
\label{fig:cMM200MeanMean3D2D}
The concentration-mass relation of haloes in rescaled and direct simulations. Concentrations were estimated from NFW fits with the halo mass as a free parameter. The left and right plots show results from employing 3D density profiles and $\Delta \Sigma$, respectively, and each sub-panel focuses on a different cosmology. Dotted and dashed lines show the predictions of the model by \citetalias{2016MNRAS.460.1214L}. Symbols mark the mean relations, and shaded regions show the $68\, \%$ and $95\, \%$ of the distribution at a fixed mass. Horizontal error bars indicate the spread in the fitted $M_{200\text{m}}$ masses. The vertical dotted lines denote the mass limit below which the finite force resolution affects the concentration estimates. Note that the disagreement between the model and the measurements originates mostly from unrelaxed haloes (cf. Fig.~\ref{fig:som015CMWithCuts}). 
For an analogous plot using concentrations obtained with Einasto profiles, see Appendix~\ref{sec:einastoConcentrations}.
}
\end{centering}
\end{figure*}

\begin{figure}
	\includegraphics[width=1.04\columnwidth]{\figrelpath 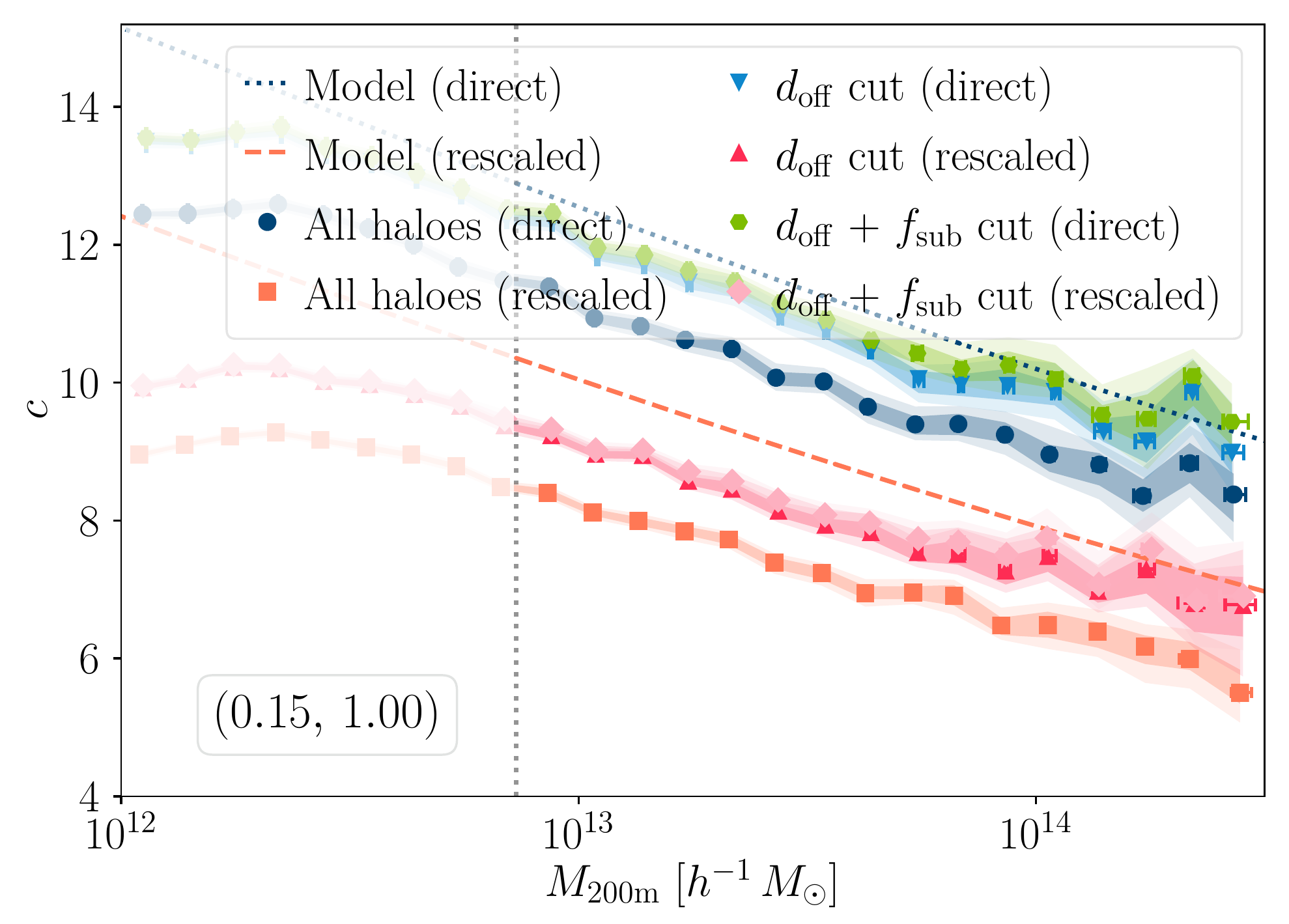}
    \caption{
    \label{fig:som015CMWithCuts}
		The impact of unrelaxed haloes in the concentration-mass relation. Different lines show the results for direct and rescaled halo catalogues after different cuts were applied to eliminate unrelaxed systems. Note that applying relaxation cuts increases the amplitude of the relation and produces a better agreement with the theoretical models.
		}
\end{figure}
 
In Fig.~\ref{fig:cMM200MeanMean3D2D}, we compare the values of the concentration parameter from the 3D and 2D NFW fits to the predictions of the model described in Section~\ref{sec:rescaledCMZ}. 
At the low mass end, the finite force resolution of the simulations affects the inner halo profiles and thus the concentrations estimates noticeably, in particular for $(0.15, \, 1.00)$ due to its larger softening scale. The vertical dotted lines in Fig.~\ref{fig:cMM200MeanMean3D2D} and \ref{fig:som015CMWithCuts} mark the halo mass above which the scale radius exceeds $r_\text{s} > 6\, l_\text{s}$ for the theory predictions, and thus the concentrations estimates are less affected by the finite force resolution.

In 3D, the model fails to predict the concentration-mass relation within the statistical errors for the general population. Additional cuts remove the tension, as Fig.~\ref{fig:som015CMWithCuts} shows for $(0.15, \, 1.00)$. For low mass haloes, the Einasto fits favour higher $c$-values than the NFW fits (see Appendix~\ref{sec:einastoConcentrations}) and have the best agreement with the \citetalias{2016MNRAS.460.1214L} model with the cuts enforced (which is encouraging since the model is supposed to match such relations). We are able to reach a complete agreement with the model with the cuts enforced with the Einasto parameterisation for all cosmologies where we use snapshots close to $z = 0$ in the fiducial run.

Yet, the model cannot describe the measured rescaled $c(M)$-relation for $(0.25, \, 0.60)$. This is caused by a failure to model the signal at $z = 0.56$ in the fiducial cosmology. We have also computed the unscaled $M_{200\text{c}}$ concentration-mass relations for median Einasto $c(M, \, z)$ relations with the corresponding cuts implemented\footnote{Since $M_{\Delta\text{m}} > M_{\Delta\text{c}}$ generally holds, the cuts are more conservative with a $M_{\Delta\text{c}}$ mass definition as neither the centre-of-mass, the position of the most bound particle nor the mass contained in substructure are altered for the same halo.}, which yield the highest available concentrations per mass bin. Even in this case, the model predicts higher than observed concentrations. This could be due to the neglect of the redshift evolution of the collapse threshold.

In 2D, the model fits the measured values well at high masses, particularly for the relaxed subpopulations. Due to limited resolution, we cannot discern the expected monotonous $c(M)$-relation in 2D below $ \approx 10^{12.7} \, h^{-1} M_{\sun}$ for $(0.15, \, 1.00)$. This effect is present in the low mass bins for $(0.25, \, 0.60)$ as well. The relations in 2D and 3D mainly differ due to different binning choices; in 3D we follow the approach in \citetalias{2016MNRAS.460.1214L} whereas we opt for an observation conforming choice in 2D. Fewer bins in the inner projected regions of the stacked haloes combined with the down-weighting of these bins result in less sensitivity to the concentration, which explains the flat relations for low mass haloes. On the other hand, the masses are still determined well which is reflected in the small horizontal error bars.
\begin{figure*}
\begin{centering}
\large
\stackon[5pt]{\includegraphics[width=1.02\columnwidth]{\figrelpath 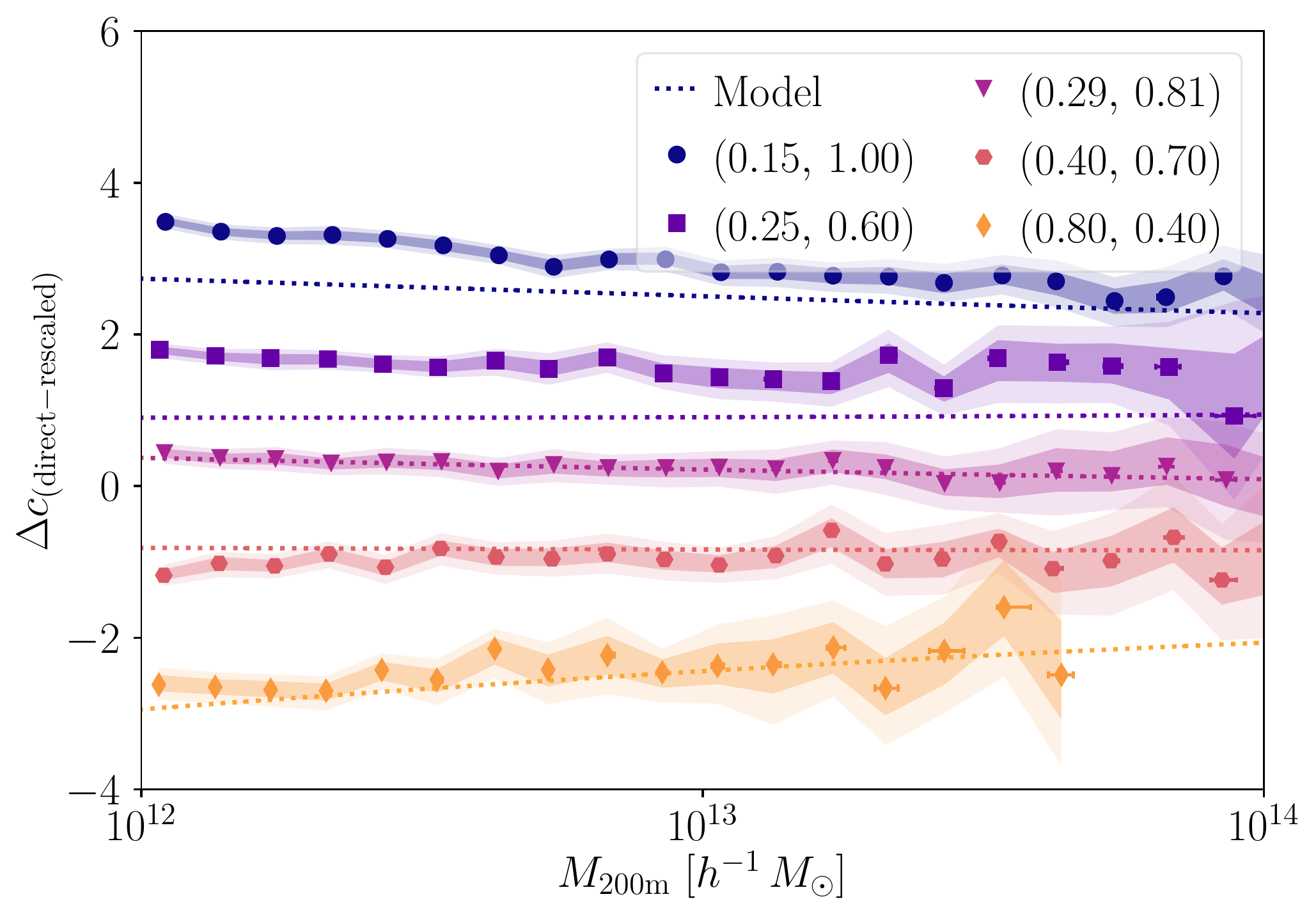}}{{\bf Concentration bias 3D}}
\stackon[5pt]{\includegraphics[width=1.02\columnwidth]{\figrelpath 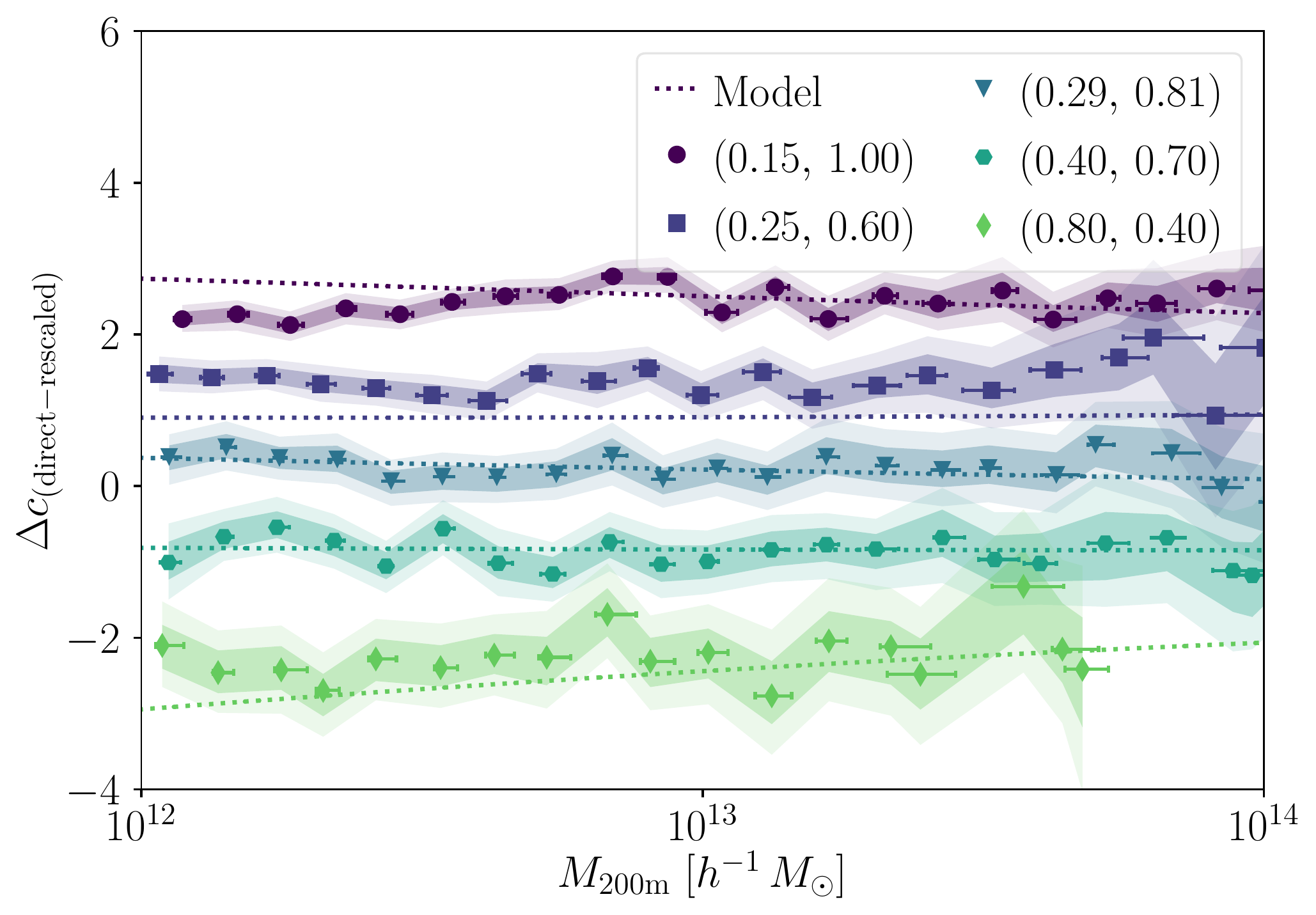}}{{\bf Concentration bias $\Delta \Sigma$}}
\caption{
\label{fig:concentrationDifference}
The difference in concentrations measured in the direct and rescaled simulations, $\Delta c_{(\rm direct - rescaled)}$, as a function halo mass at $z = 0$. Concentrations were measured by fitting NFW profiles to 3D density profiles (left panel), and to $\Delta \Sigma$ profiles (right panel). Different colours indicate results for different combinations of ($\Omega_\text{m}, \, \sigma_8$). Dotted lines correspond to the predictions for this quantity based on the \citetalias{2016MNRAS.460.1214L} model. The shaded regions mark the $68\, \%$ and $95 \, \%$ percentiles, and horizontal error bars the range of fitted halo masses.
}
\end{centering}
\end{figure*}

As Fig.~\ref{fig:concentrationDifference} illustrates, the difference in concentration $\Delta c$ between the direct and rescaled simulations is approximately constant for haloes in the mass range $10^{12} - 10^{14} \, h^{-1} M_{\sun}$, and moreover roughly consistent with the model predictions. The deviation for $(0.25, \, 0.60)$ results in a discrepancy between the model and the measured difference relation, but for $(0.29, \, 0.81)$, $(0.40, \, 0.70)$ and partly for $(0.15, \, 1.00)$ at the high mass end, there is consistency both in 3D and for the lensing profiles. For low mass haloes, resolution effects and the relatively higher amplitude of the (not modelled) 2-halo term obscure the results. At the high mass end, the low number of haloes cause a larger scatter.

The constant $\Delta c$ relations hold for the relaxed populations as well, especially for $\Delta \Sigma$, though the variance increases. The small changes for the 3D density profiles are quantified by comparing $\Delta c_\text{relaxed} /\Delta c_\text{all haloes}$ for haloes with $10^{12} - 10^{14} \, h^{-1} \, M_{\sun}$ masses and record the median differences in the mass bins where we have more than twenty haloes for each imposed cut. This produces variations of the order of $5 \, \%$ but there are no consistent trends present for both the NFW and Einasto parameterisations. This means that whereas the \citetalias{2016MNRAS.460.1214L} model fails to accurately predict the concentration-mass relations for halo samples containing both relaxed and unrelaxed systems, it can predict the difference in this relation between two simulations for such a mixed population very well both for 3D density and $\Delta \Sigma$ profiles. Hence, it is suitable for modern surveys.

\subsection{Concentration corrected profiles}
\label{sec:correctedProfiles}

\begin{figure*}
\begin{centering}
	\includegraphics[width=2.1\columnwidth]{\figrelpath 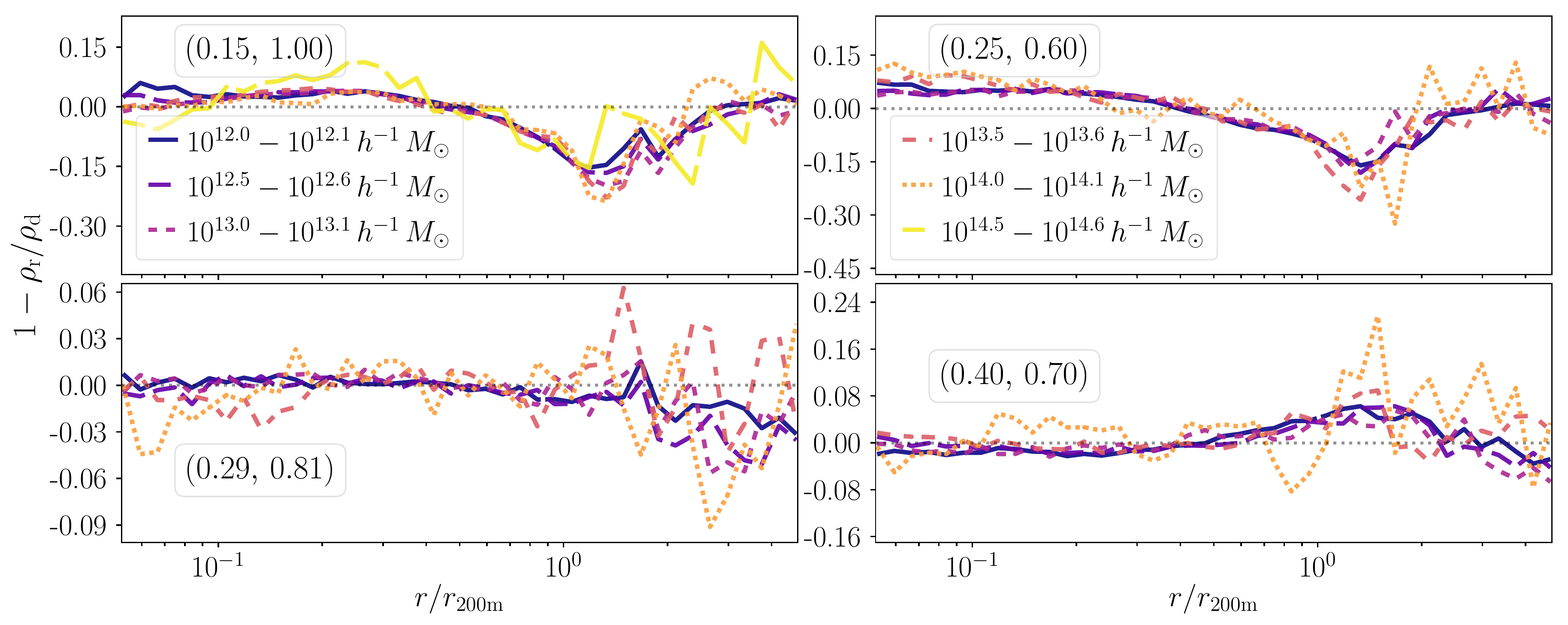}
    \caption{
    \label{fig:rhoNFWCorrected}
		Same as Fig. \ref{fig:som0Rho} but after correcting the inner profiles of rescaled haloes.
		}
    \end{centering}
\end{figure*}

\begin{figure*}
\begin{centering}
	\includegraphics[width=2.1\columnwidth]{\figrelpath 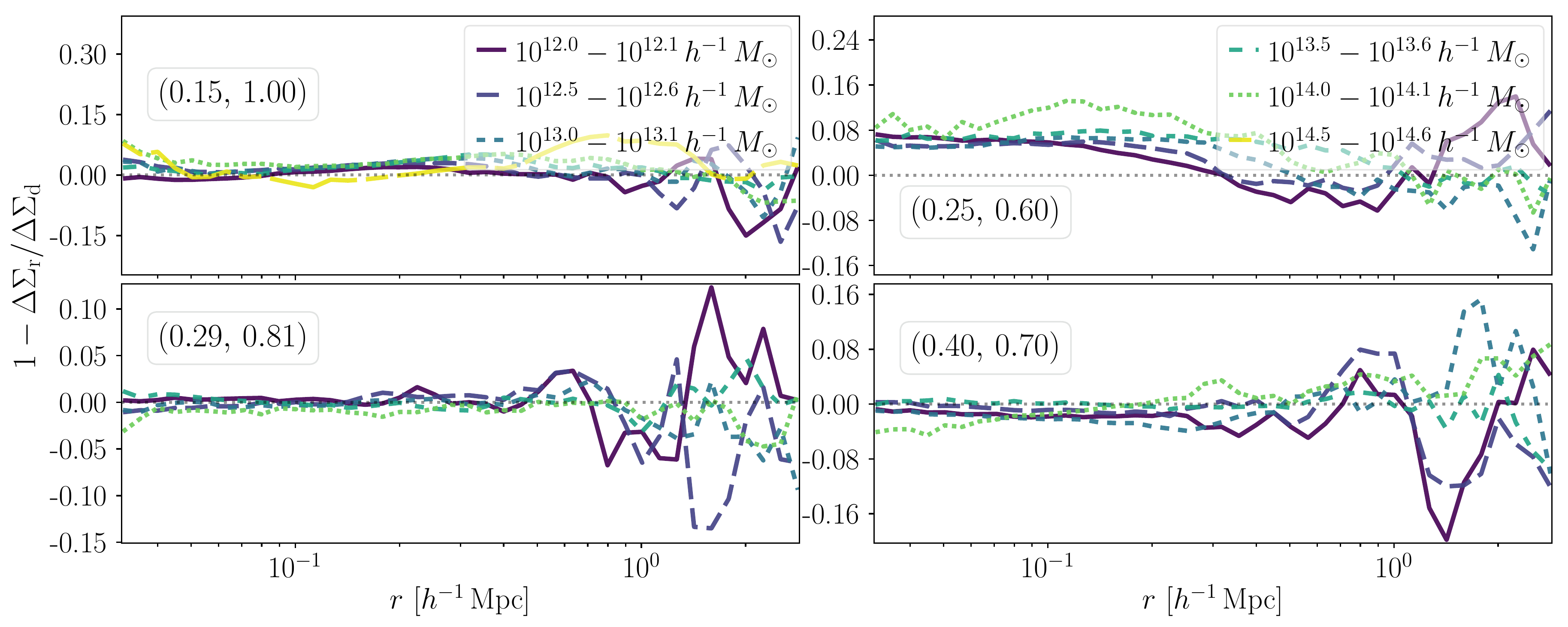}
    \caption{
    \label{fig:som0DeltaSigmaCorrected}
		Same as Fig. \ref{fig:som0DeltaSigma} but after correcting the inner profiles of rescaled haloes.
		}
   \end{centering}
\end{figure*}

\begin{table*}
	\centering
	\caption{
  \label{tab:deltaRhoSigmaBeforeAndAfterCorrection}
	  Total and median maximum deviation between the direct and rescaled simulation, $1 - \rho_\text{r}/\rho_\text{d}$ and $1 - \Delta \Sigma_\text{r}/\Delta \Sigma_\text{d}$, for 3D median and for 2D mean profiles per mass bin for radial bins in the given range before and after the concentration correction.
	}	
	\begin{tabular}{l | l | c c | c c | c c | c c |}
	\hline
		\multicolumn{2}{ |c| }{Residuals:} & \multicolumn{4}{ |c| }{$\rho(r), \, 30\, h^{-1} \, \text{kpc}  < r < r_{200\text{m}}$} & \multicolumn{4}{ |c| }{$\Delta \Sigma (r), \, 30\, h^{-1} \, \text{kpc}  < r < r_{200\text{m}}$} \\
		& & \multicolumn{2}{ c }{Pre-correction} & \multicolumn{2}{ c }{Post-correction} & \multicolumn{2}{ c }{Pre-correction} & \multicolumn{2}{ c }{Post-correction}\\
		Simulation & Halo mass range & Max & Median & Max & Median & Max & Median & Max & Median \\
  	\hline
		$(0.15, \, 1.00)$ & $10^{12.0} - 10^{14.8} h^{-1}\, M_{\sun}$ & $35 \, \%$ & $22 \, \%$ & $-17 \, \%$ & $-9.1\, \%$ & $39 \, \%$ & $30 \, \%$ & $10 \, \%$ & $5.2\, \%$ \\
		$(0.25, \, 0.60)$ & $10^{12.0} - 10^{14.2} h^{-1}\, M_{\sun}$ & $25 \, \%$ & $15 \, \%$ & $-17 \, \%$ & $-7.4\, \%$ & $36 \, \%$ & $18 \, \%$ & $19 \, \%$ & $7.3\, \%$ \\
		$(0.29, \, 0.81)$ & $10^{12.0} - 10^{14.5} h^{-1}\, M_{\sun}$ & $16\, \%$ & $2.5\, \%$ & $-16\, \%$ & $1.3\, \%$ & $6.1\, \%$ & $2.4\, \%$ & $5.1\, \%$ & $-1.0\, \%$ \\
		$(0.40, \, 0.70)$ & $10^{12.0} - 10^{14.4} h^{-1}\, M_{\sun}$ & $25 \, \%$ & $7.4\, \%$ & $-18 \, \%$ & $3.6\, \%$ & $-26 \, \%$ & $-9.6\, \%$ & $-15 \, \%$ & $-2.1\, \%$ \\
		$(0.80, \, 0.40)$ & $10^{12.0} - 10^{13.8} h^{-1}\, M_{\sun}$ & $43 \, \%$ & $22 \, \%$ & $11 \, \%$ & $6.5\, \%$ & $-42 \, \%$ & $-29 \, \%$ & $13 \, \%$ & $-4.3\, \%$ \\
	\hline
	\end{tabular}
\end{table*}

\begin{figure*}
\begin{centering}
\large
\stackon[5pt]{
\includegraphics[width=1.02\columnwidth]{\figrelpath 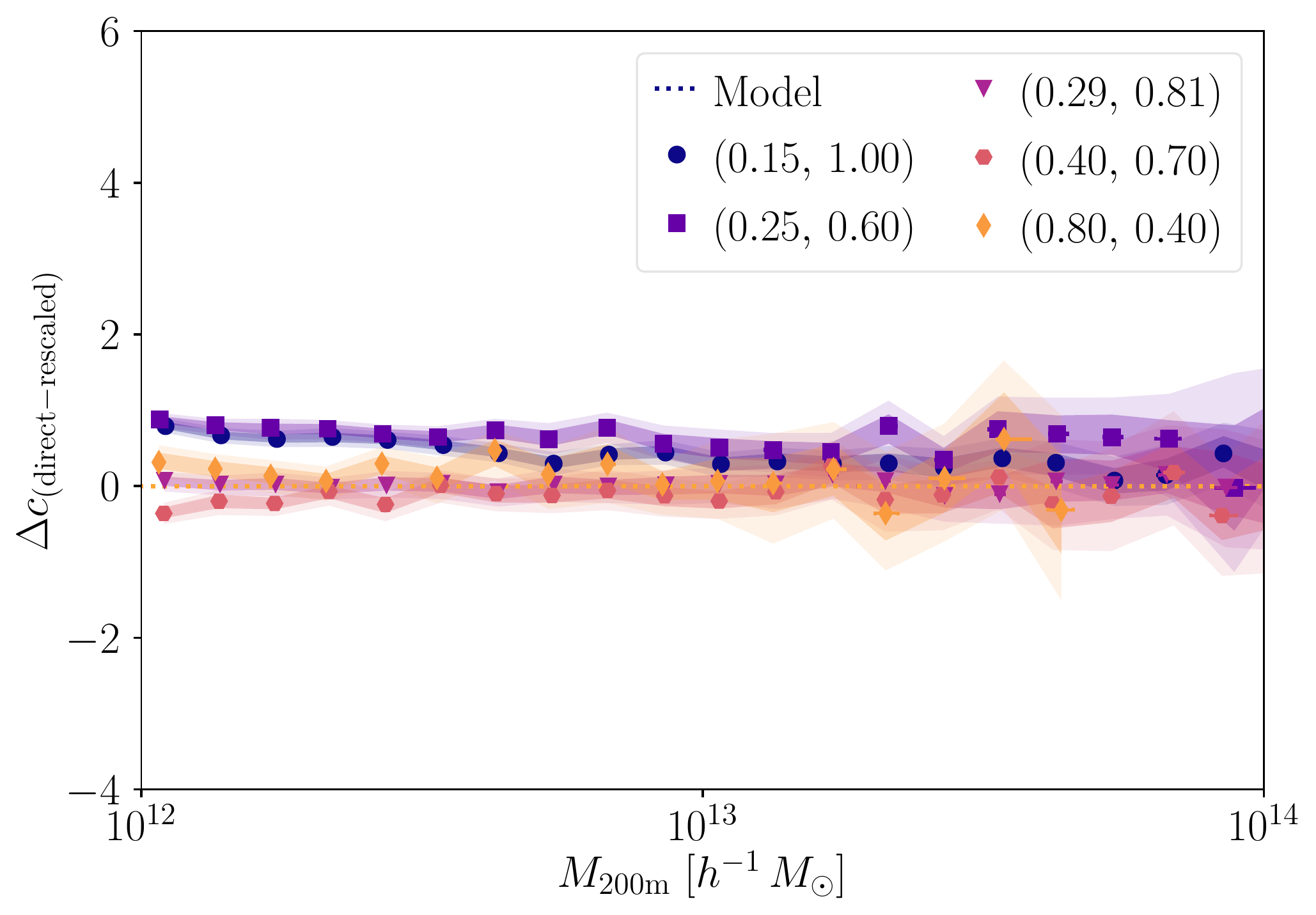}}{{\bf Concentration bias 3D post-correction}}
\stackon[5pt]{
\includegraphics[width=1.02\columnwidth]{\figrelpath 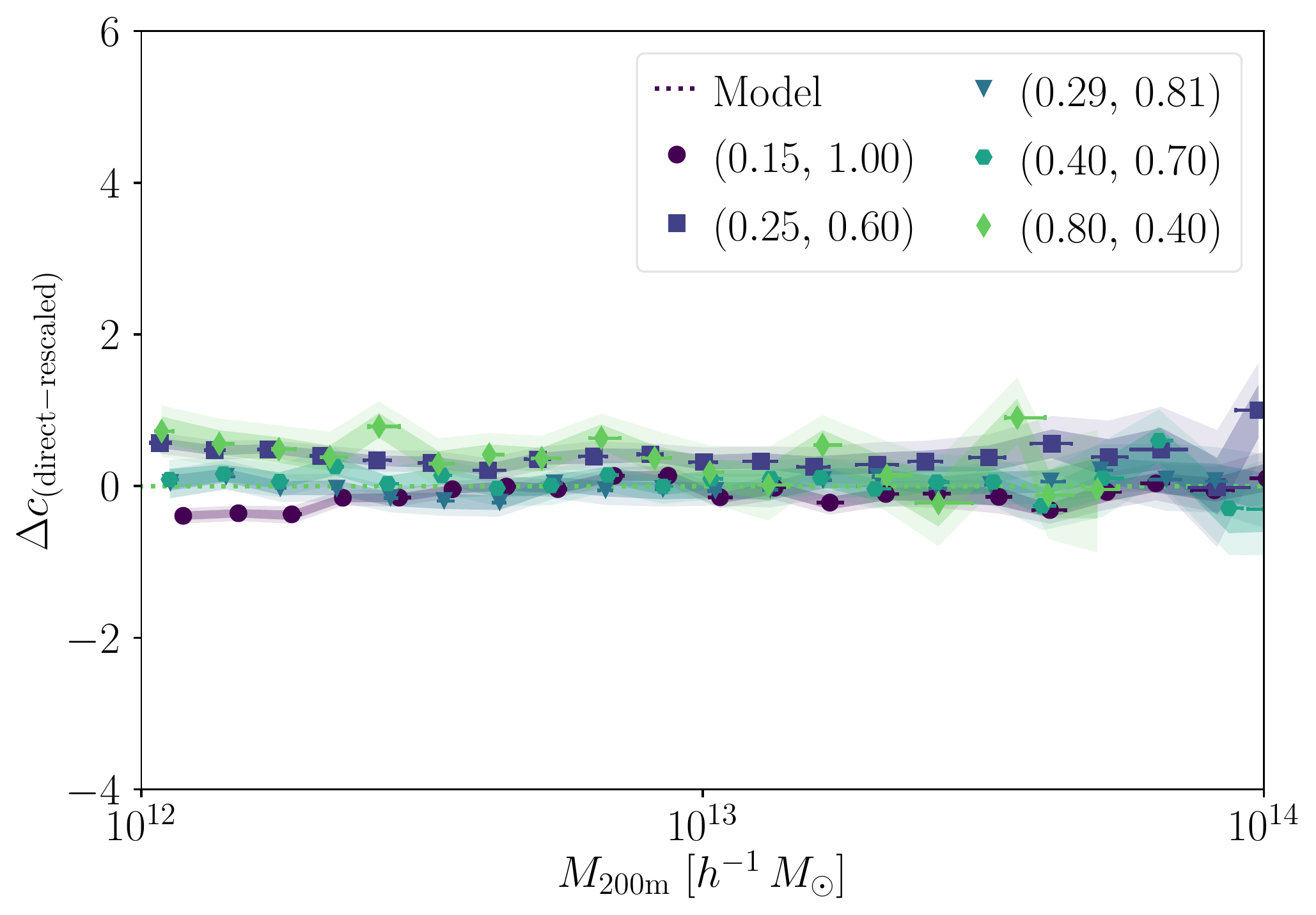}}{{\bf Concentration bias $\Delta \Sigma$ post-correction}}
\caption{Same as Fig. \ref{fig:concentrationDifference} but after applying our corrections in Eqs.~\eqref{eq:densityProfileCorrection} and~\eqref{eq:lensingProfileCorrection} to the rescaled profiles. The concentration bias for the corrected profiles is reduced considerably.}
\label{fig:correctedConcentrationDifference}
\end{centering}
\end{figure*}

Motivated by the good agreement in Fig.~\ref{fig:concentrationDifference}, we correct the rescaled profiles by multiplying the measured values with the ratio between the fitted profile to the rescaled simulation data and a modified profile with the concentration bias from the model, $\Delta c( r_{200\text{m}})$:
\begin{align}
\label{eq:densityProfileCorrection}
\rho'(r) &\mapsto \frac{\rho_\text{NFW}\left(r, \, c + \Delta c\left( r_{200\text{m}} \right), \, r_{200\text{m}} \right)}{\rho_\text{NFW}\left(r, \, c, \, r_{200\text{m}} \right)} \rho'(r)
, \\
\label{eq:lensingProfileCorrection}
\Delta \Sigma'(r) &\mapsto \frac{\Delta \Sigma_\text{NFW}\left(r, \, c + \Delta c\left( r_{200\text{m}} \right), \, r_{200\text{m}} \right)}{\Delta \Sigma_\text{NFW}\left(r, \, c, \, r_{200\text{m}} \right)} \Delta \Sigma'(r),
\end{align}
for all radii $r \lesssim r_{200\text{m}}$. We will refer to these correction factors as $\gamma(r_i)$. The Einasto correction is calculated in the same manner (see Appendix~\ref{sec:einastoConcentrations}). Since $\Delta c(M)$ only weakly depends on $M$, there are no significant differences between using the fitted $M_{200\text{m}}$ or halo finder value.

Correcting the profiles up to $3\, h^{-1} \, \text{Mpc}$ does not significantly affect the lensing signal, but jeopardises the agreement for the 2-halo term in 3D (see Fig.~\ref{fig:effectOfDensityFieldCorrections}). We find that restricting the correction to $r < 1.8 \, r_{200\text{m}}$ reduces differences in the 1-to-2-halo transition region without compromising the agreement on larger scales.

The concentration correction could be additive instead of multiplicative. This gives a slightly better performance on scales $ r > r_{200\text{m}}$, since the field differences are small, but this correction also induces a small bias and should thus be applied below a cutoff radius. The multiplicative correction preserves the shape of the residual throughout the transition regime slightly better. Otherwise, we have checked that there are no significant differences between the two for all halo mass bins and cosmologies with NFW or Einasto parametrizations for matched haloes, in bootstrapped stacks or individually. Both largely preserve the width and shape of the $\Delta c$ distribution around the median or the mean concentration, with no obvious advantages, and yield $\Delta c = 0$ if we correct the rescaled profiles with the measured direct concentrations.

The residuals for the corrected 3D density profiles are shown in Fig.~\ref{fig:rhoNFWCorrected} and for the corrected $\Delta \Sigma$ profiles in Fig.~\ref{fig:som0DeltaSigmaCorrected}. The maximum and median pre- and post-correction profile differences are listed in Table~\ref{tab:deltaRhoSigmaBeforeAndAfterCorrection} for the 40 radial bins setup. Typically, the largest differences occur in the most or second most massive halo mass bin. In most cases, the correction reduces the differences by factors of two to five. For $(0.25, \, 0.60)$, both the residual profiles and residual concentration differences indicate that a larger concentration correction than predicted by the \citetalias{2016MNRAS.460.1214L} model could improve the agreement between direct and rescaled profiles.

However, when comparing the measured halo concentrations pre- and post-correction, we find significant improvement in the concentration mismatch between rescaled and direct simulations for all considered cosmologies, as Fig.~\ref{fig:correctedConcentrationDifference} illustrates for all haloes (see Appendix~\ref{sec:matchedHaloes} for the result for matched populations).

\subsection{Correcting individual halo profiles}
\label{sec:individualCorrections}

\begin{figure}
	\includegraphics[width=1.04\columnwidth]{\figrelpath 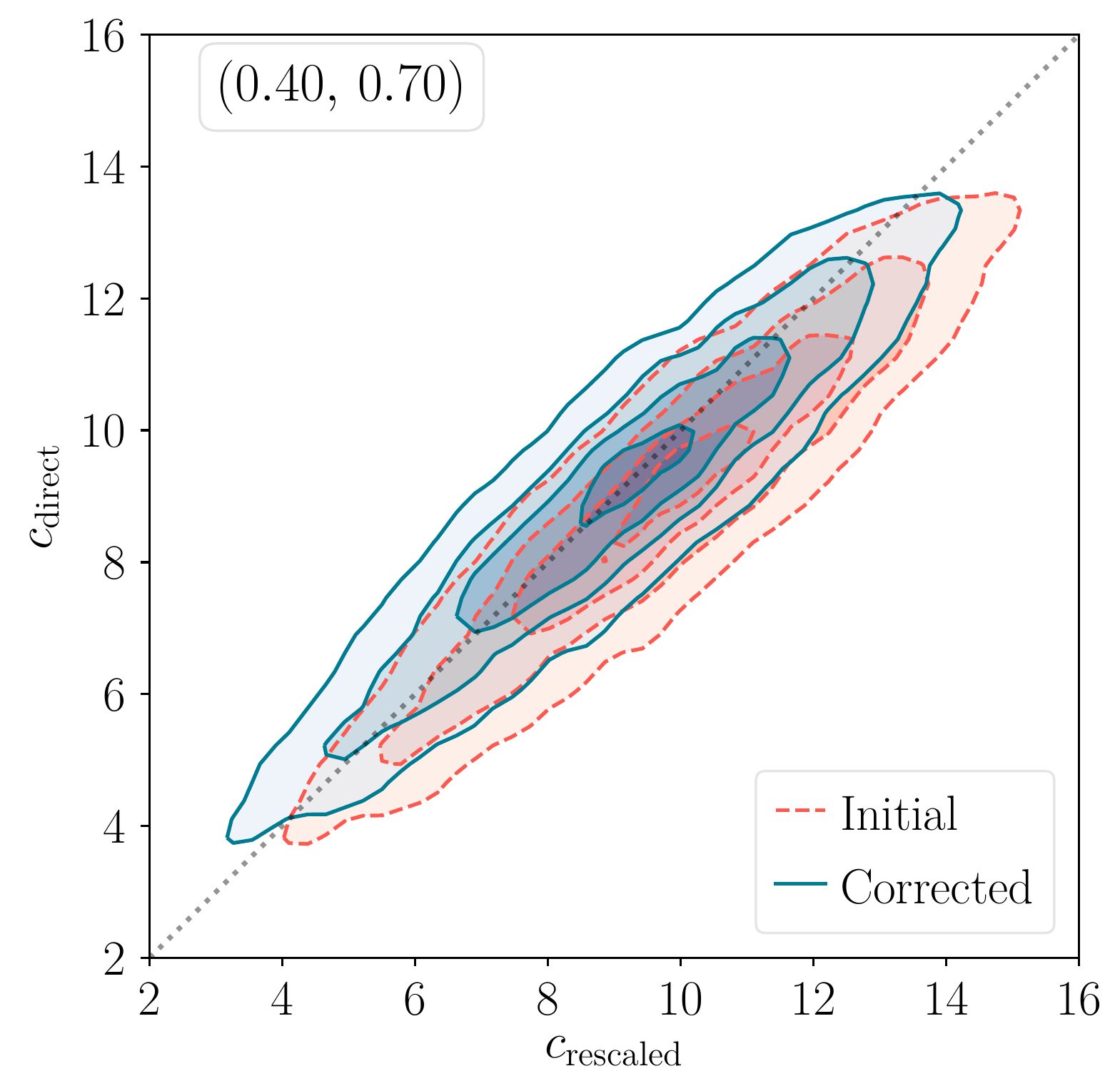}
    \caption{
    \label{fig:som04IndHaloDensityFieldCEstimatesPrePostcorrection}
		Effect of the density field correction on the NFW estimated concentration distribution for individual matched haloes in the direct and rescaled simulation with $(0.40, \, 0.70)$ where the haloes in the direct simulation have $ M_{200\text{m}} > 10^{12.5} \, h^{-1} \, M_{\sun}$. For smoother contours, the distributions have been convolved with a Gaussian filter with $\sigma = 1$.}
\end{figure}

We also examine how the correction in Eq.~\eqref{eq:densityProfileCorrection} affect the concentrations from 3D profile fits to individual haloes. The joint distribution of concentrations for haloes above $10^{12.5} \, h^{-1} M_{\sun}$ in the $(0.40, \, 0.70)$-simulation and their rescaled counterparts is shown in Fig.~\ref{fig:som04IndHaloDensityFieldCEstimatesPrePostcorrection}. Applying the concentration correction translates the distribution towards the diagonal in a similar manner for high and low concentration haloes. This is a consequence of the modest mass evolution of the concentration bias for the cosmologies in this study. However, the correction cannot account for a slight tilt between the two simulations, with low-$c$ (high-$c$) haloes having higher (lower) concentrations in the direct simulation than in the rescaled simulation.\footnote{This tilt persists when relaxation cuts are enforced, regardless of whether $r_{200\text{m}}$ is fixed or a free parameter, and is also present with Einasto parameterisations (Appendix~\ref{sec:einastoConcentrations}).}

The tilt is stronger for cosmologies with $\Delta \Omega_\text{m} > 0$ away from the fiducial simulation with a clockwise tilt relative to the diagonal (see Appendix~\ref{sec:matchedHaloes}). For $(0.15, \, 1.00)$ and $(0.25, \, 0.60)$, there is a slight counter-clockwise tilt. The results are robust to changes in the fitting scheme.\footnote{For all profile fits we use the Levenberg-Marquardt algorithm with $( c = 4, \, r_{200\text{m}} = r_{200\text{m}, \, \text{sim.}})$ as a starting point. We have checked that the results are insensitive to the starting point choice for physically viable parameter values. In addition we have computed the parameters with the limited-memory BFGS algorithm with bounds $c \in [1, \, 30]$ and $r_{200} \in [0.5 \, r_{200\text{m}, \, \text{sim.}}, \, 2 \, r_{200\text{m}, \, \text{sim.}}]$ and obtain consistent results.} We have checked that there are negligible differences for all cosmologies between the $c(M)$ relations computed from the median profiles and the median $c(M)$ relations from fits to individual haloes, and that the tilt in the distributions persist when one corrects the individual halo concentrations with the median measured relations.

The tilt in the joint distribution is also present for halo samples selected in narrower mass ranges. The asymmetry is partly washed out in the results for the median profiles, as both high $c$ and low $c$ haloes contribute to the effective density field per mass bin. However, this secondary rescaling concentration bias could influence analyses where the halo population is split into different concentration samples at fixed mass, such as assembly bias studies. Further studies with larger simulation volumes are required to accurately quantify this effect.

\subsection{Halo outskirts}
\label{sec:outerProfileBias}

\begin{figure}
\includegraphics[width=1.04\columnwidth]{\figrelpath 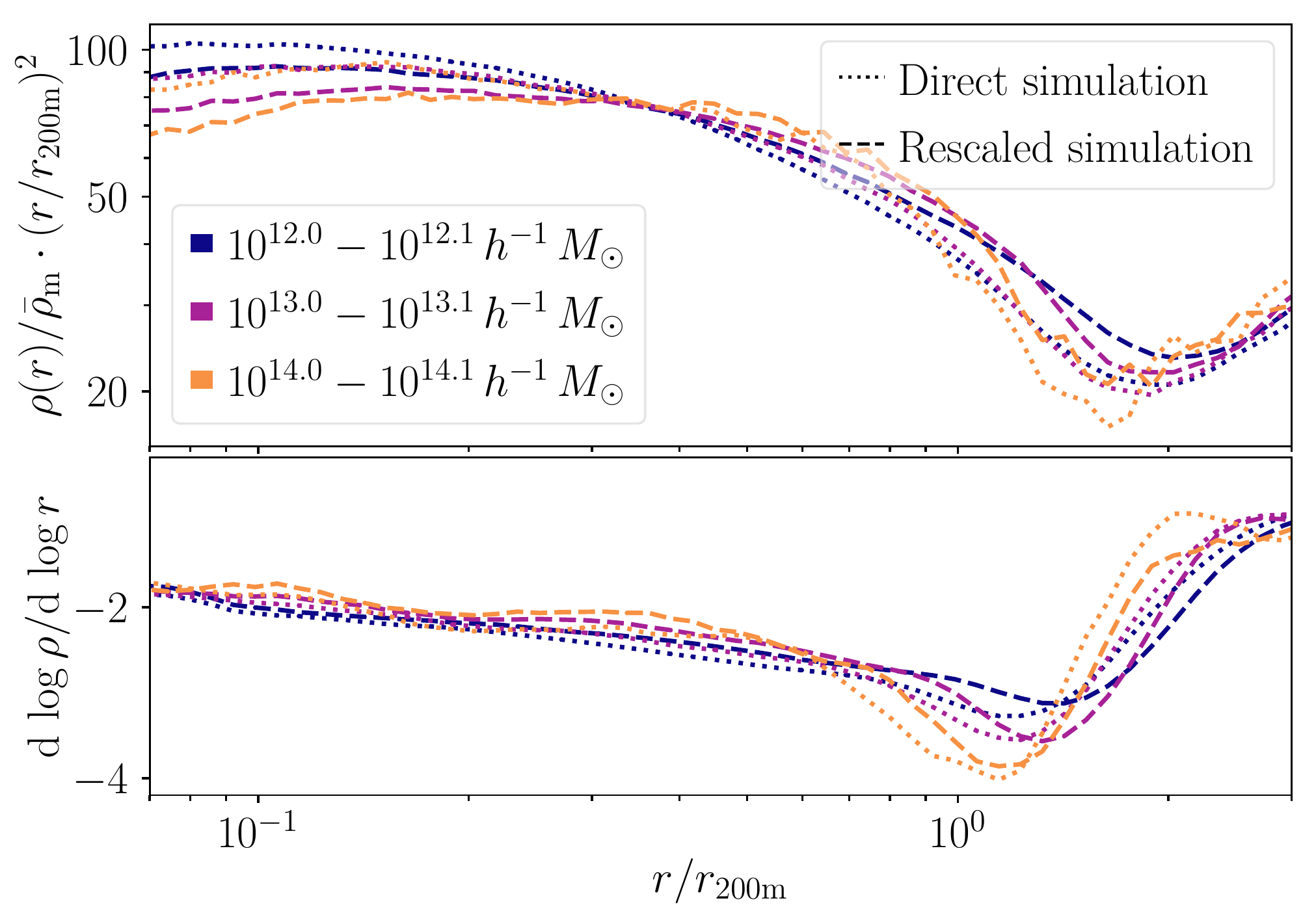}
    \caption{
    \label{fig:profilesOuterBiasCorrected}
     Comparison between direct and rescaled profiles and their radial derivatives for matched haloes for $(0.25, \, 0.60)$ for three mass bins. The concentration bias is visible as an amplitude offset close to the halo centres whereas the outer profile bias corresponds to a radial shift of the profiles at the halo boundaries. This shift is visible in the radial derivatives of the field (computed with a fourth-order Savitzky-Golay filter with a window length of 15 bins) in the lower panel as well, where there are offsets in the positions of the steepest slope between the direct and rescaled profiles.
   }
\end{figure}

The concentration correction does not fully account for differences in the halo outskirts, as it focuses on rearranging material within the halo. Subsequent outer corrections could redefine the halo boundary and potentially improve agreement in the halo mass function. 
Fig.~\ref{fig:profilesOuterBiasCorrected} highlights that the profile bias in the inner halo regions is mostly an amplitude offset, whereas the bias in the halo outskirts is rather a radial offset. Hence, correcting the rescaled profiles by shifting them radially in the outskirts can mitigate the outer profile bias.

\begin{figure}
	\includegraphics[width=1.04\columnwidth]{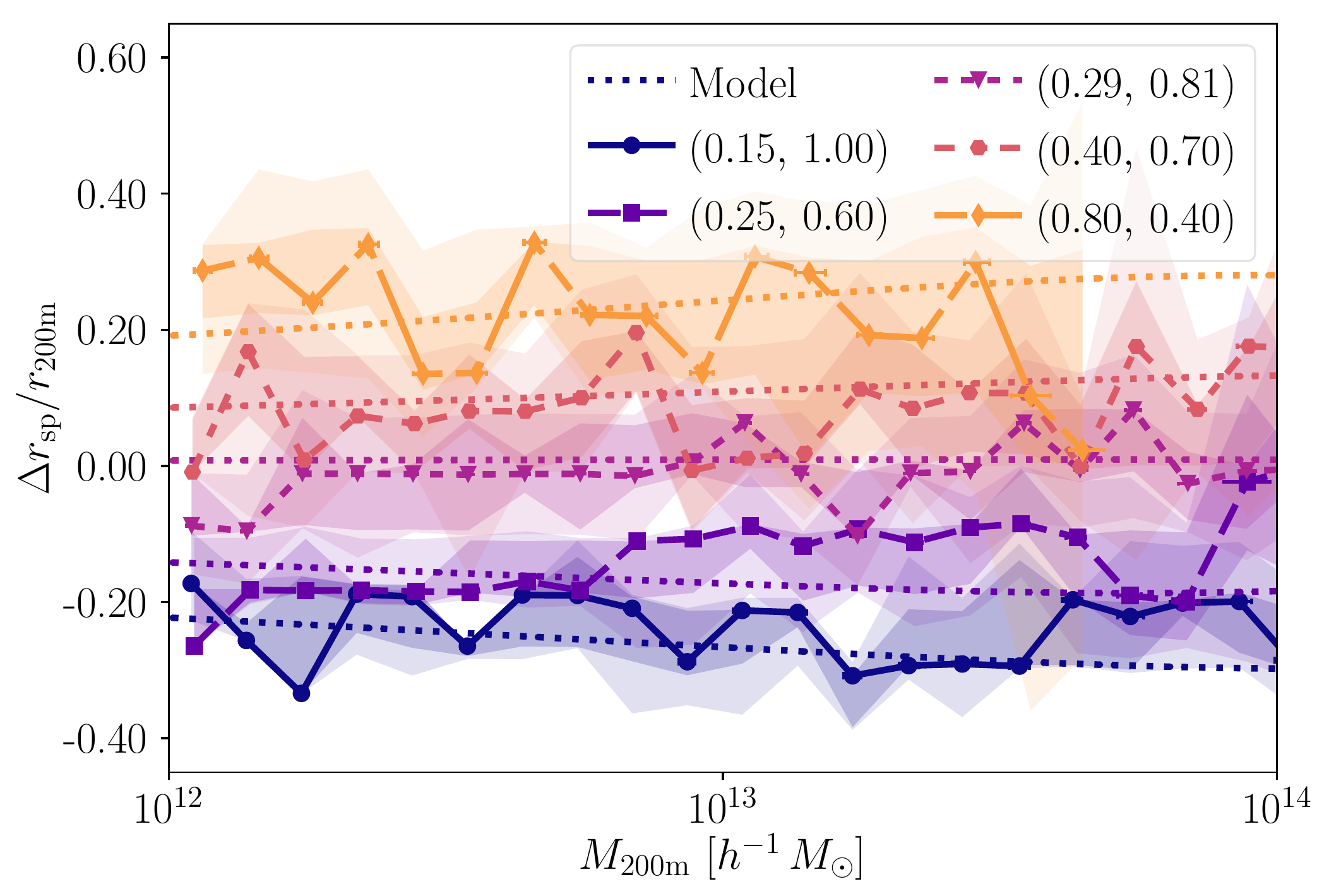}
    \caption{Measured differences in the location of the steepest slope of the density field for matched haloes w.r.t. to the \citet{2017ApJ...843..140D} model, for the 75th percentile. Error regions for $95\, \%$ and $68 \, \%$ are computed from resampled medians from stacks of matched haloes in the direct and rescaled simulation snapshots.}
    \label{fig:rsp75thModelVsDensityFieldDerivatives}
\end{figure}

In Fig.~\ref{fig:rsp75thModelVsDensityFieldDerivatives}, we plot the measured differences in the location of the steepest slope of the density field for matched haloes. We adjust the position of the rescaled profile's steepest slope with $r_{200\text{m}}^\text{(d)}/r_{200\text{m}}^\text{(r)}$ to account for the mismatch in halo mass between the matched samples, which has a minor impact on the result. We compare these differences to the expected offset between the splashback radii $r_\text{sp}$, the apocentre of the first orbit of accreted material \citep[e.g.][]{2014ApJ...789....1D, 2014JCAP...11..019A, 2015ApJ...810...36M, 2016MNRAS.459.3711S, 2017ApJ...841...34M, 2017ApJ...843..140D}, between the direct and rescaled profiles $\Delta r_\text{sp} = r_\text{sp}^{\mathrm{(d)}} - r_\text{sp}^{\mathrm{(r)}}$. We apply the recent fit provided in \citet{2017ApJ...843..140D} to simulation results in \citet{2017ApJS..231....5D} to predict the median splashback radius as a function of halo mass and cosmology. This model has been fitted by tracing billions of particle orbits in haloes spanning from typical cluster to dwarf galaxy host masses in different cosmological simulations up to $z = 8$. Percentiles correspond to the fraction of the first apocenters of the particle orbits contained inside a given radius. Particles which were contained in a subhalo with mass exceeding 1\,\% of the host halo mass at infall are excluded to minimise bias from dynamical friction. In accordance with previous studies \citep[e.g.][]{2014ApJ...789....1D, 2015ApJ...810...36M}, the relation between the halo accretion rate $\Gamma$ and the ratios $r_\text{sp}/r_{200\text{m}}$ and $M_\text{sp}/M_{200\text{m}}$ is found to be well described by a functional form $X_\text{sp} = A + B e^{-\Gamma/C}$ where $X_\text{sp}$ is either ratio and $A, \, B$ and $C$ are free parameters where $B$ and $C$ depend on the matter fraction $\Omega_\text{m}$ and halo peak height $\nu = \delta_\text{c}/(D(z)\sigma(M_\text{h}))$ with $M_\text{h}$ as the halo mass. In addition, the median accretion rate $\Gamma(\nu, \, z)$ can be well captured by a parameterisation $\Gamma = A^\prime \nu + B^\prime \nu^{3/2}$, where $A^\prime$ and $B^\prime$ are polynomials in $z$. We use this expression for the median accretion rate to compute the radii.\footnote{As we are probing the median 3D density profiles, we opt for the 75th percentile of the model which was found to best match the median profiles in \citet{2015ApJ...810...36M}, especially at the high mass end. The splashback radius rescales as $r_\text{sp} \mapsto \alpha r_\text{sp}$ and the predicted position $r_\text{sp}^{\mathrm{(r)}}$ is hence given as the fitted solution in the fiducial simulation at the fiducial redshift with $\beta_\text{m}^{-1}M_{200\text{m}}$ determining the peak height and $r_{200\text{m}}$.}
The measurements trace the model prediction, except for $(0.80, \, 0.40)$ where the scatter is driven by poor statistics due to the small box size.

\begin{figure}
	\includegraphics[width=1.04\columnwidth]{\figrelpath 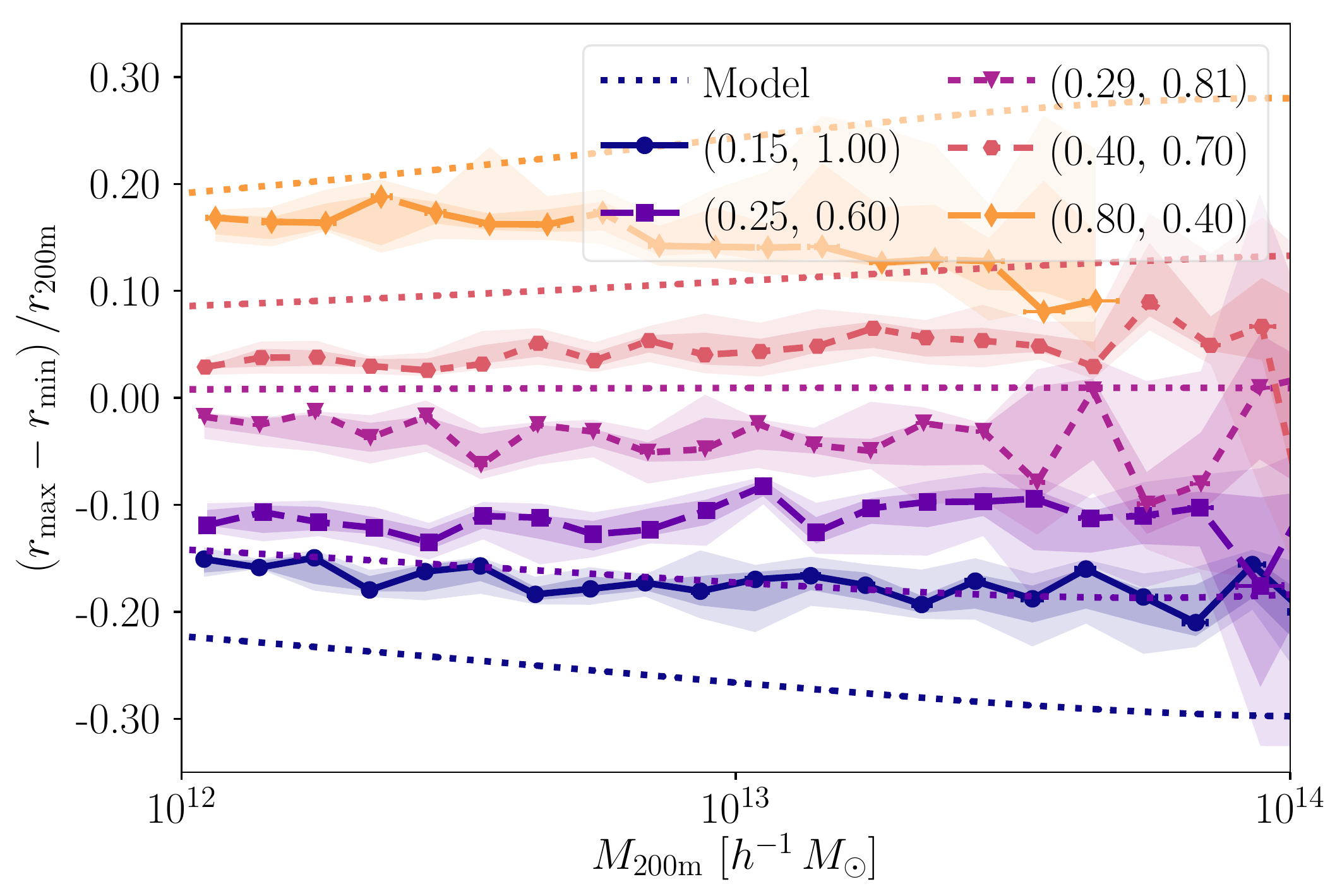}
    \caption{
    \label{fig:interpolatedRspBias} 
    Measured density field outer profile bias for matched haloes vs. the predicted $\Delta r_{\text{sp}}/r_{200\text{m}}$ bias using the model in \citet{2017ApJ...843..140D}.}
\end{figure}

We also compute the radial shifts that minimise the largest relative difference between the direct and rescaled outer density profiles. Between $0.4 < r/r_{200\text{m}} < 2.0$, we locate the maximum of the $1 - \rho_\text{r}(r)/\rho_\text{d}(r)$ residual defining $r = r_\text{max}$ and then shift the interpolated rescaled profile radially to find the radius $r_\text{min}$ that minimises $1 - \rho_\text{r}(r_\text{min})/\rho_\text{d}(r_\text{max})$. The resulting shifts $r_\text{max} - r_\text{min}$ are shown in Fig.~\ref{fig:interpolatedRspBias} for matched haloes with the $r_{200\text{m}}^\text{(d)}/r_{200\text{m}}^\text{(r)}$ correction. This shift is almost constant for haloes, all and matched, with $M_{200\text{m}}$ between $10^{12} - 10^{14}\, h^{-1} \, M_{\sun}$. For higher masses the result is obscured by scatter. The predicted splashback bias do not exactly match the required shifts to remove the radial bias\footnote{Moreover, typically $r_\text{max} \approx 1.3 \, r_{200\text{m}}$, which does \emph{not} coincide with the predicted position of the splashback radius for all masses and cosmologies.}, but they show similar relative amplitudes, signs and weak mass dependence. A splashback radius model may thus provide a good starting point for further improvements of the rescaled profiles and halo masses (an initial attempt to correct the masses is presented in Appendix~\ref{sec:splashbackMassCorrection}).

\begin{figure}
	\includegraphics[width=1.04\columnwidth]{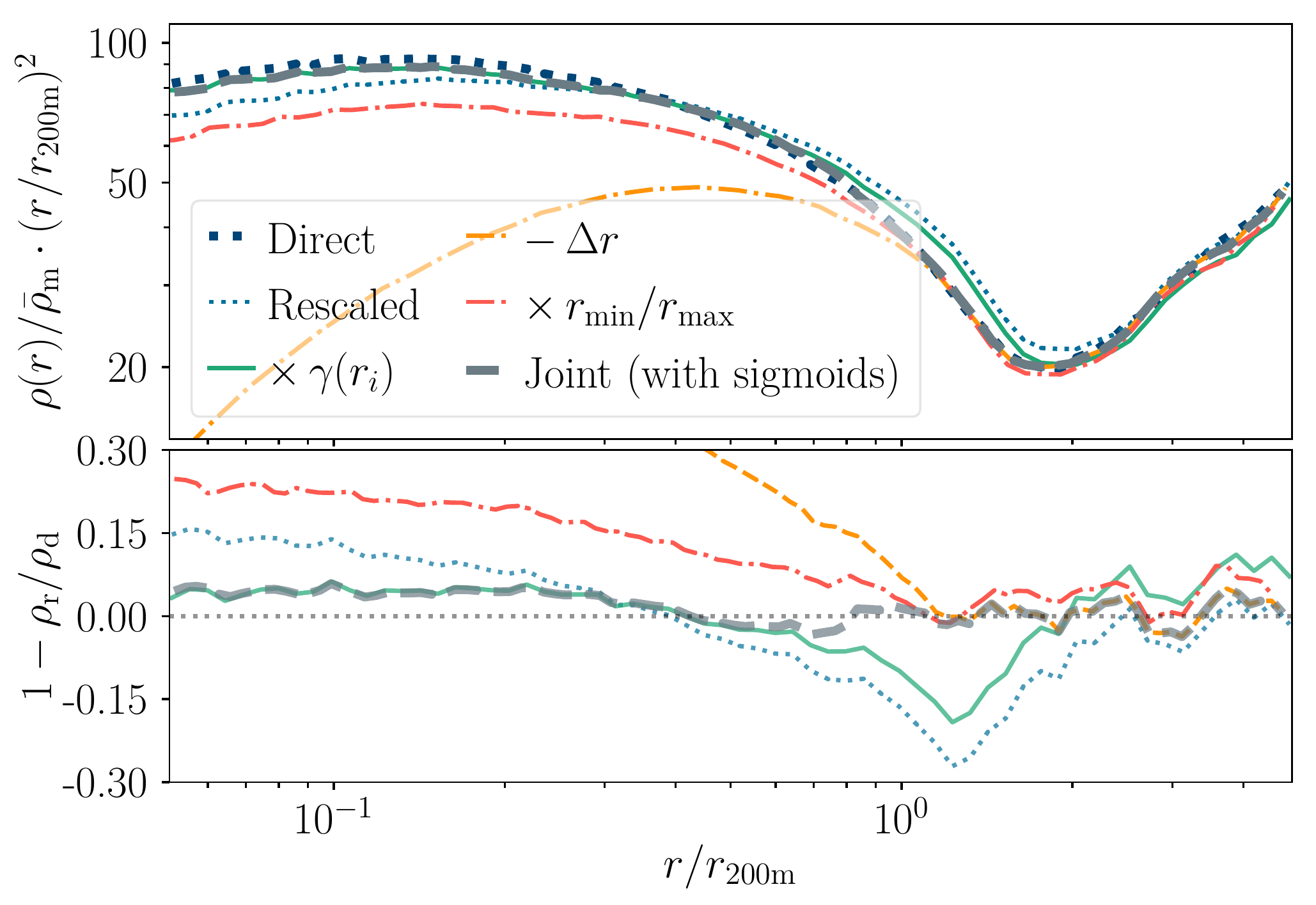}
    \caption{
    \label{fig:effectOfDensityFieldCorrections} 
    Profiles for matched haloes for $(0.25, \, 0.60)$ for $M_{200\text{m}} \in [ 10^{13} , \,10^{13.1} ) \, h^{-1} \, M_{\sun}$ in the direct simulation with different corrections applied (see the text for more detailed descriptions). Although not perfect, the concentration correction '$\times \, \gamma(r_i)$' mitigates the residual in the centre and the shifts remove the outer profile bias. These two corrections can be combined with sigmoids.}
\end{figure}

As Fig.~\ref{fig:effectOfDensityFieldCorrections} illustrates the outer profile bias vanishes, if we shift the rescaled density field values radially by $r \mapsto r_\text{min}/r_\text{max} \times r$ or $r \mapsto r - \Delta r$ with $\Delta r = r_\text{max} - r_\text{min}$. Whereas the multiplicative correction performs better in the halo centre, the additive correction has a better large scale behaviour. To combine the radial shift correction with the concentration correction, we modulate each by a sigmoid function to restrict their actions to their intended radial range:
\begin{equation}
\begin{aligned}
\rho \mapsto \rho^\prime & = \rho(r - \zeta(r)) + \xi(r), \, \zeta(r) = \frac{1}{1 + e^{-k_0 (r - r_{0})}} \Delta r,\\
 \xi(r) & = \frac{1}{1 + e^{-k_1 (r_{0} - r)}} \cdot \left(\rho_\text{NFW}^\prime - \rho_\text{NFW}\right),
\end{aligned}
\end{equation}
where $r_0$ marks the transition scale, $k_0$ and $k_1$ control the sharpness of the onsets of the corrections, and the concentration correction is evaluated at the unshifted radius. Fitting these parameters, $r_0$ in the vicinity of $r_{200\text{m}}$ seems preferred, but all parameters vary with mass and cosmology when fitting the rescaled simulation to the direct simulation. In Fig.~\ref{fig:effectOfDensityFieldCorrections} we plot one possible solution with $(r_0, \, k_0, \, k_1)$ as $(r_{200\text{m}}, \, 9.2, \, 16.4)$, where $\Delta c$ is obtained from the \citetalias{2016MNRAS.460.1214L} model and $\Delta r$ is measured. Future investigations are required to find the best set of parameters.

\section{Discussion}
\label{sec:discussion}

The rescaling predictions for the halo matter and lensing profiles are reasonably accurate even before applying the concentration correction. Partly, this is due to the matched initial conditions. This ensures similar peak heights, proto-halo regions, environments, and tidal fields, which leads to similar growth histories, as the growing density perturbations subsequently cross the collapse threshold. 

After our additional correction, the predictions become accurate at the $5\, \%$ level. In this section, we discuss the expected cosmology dependence of the corrections (Section~\ref{sec:predictingCBias}), the method's accuracy in light of the expected impact of baryons (Section~\ref{sec:baryons}) and large-scale corrections (Section~\ref{sec:2halo}), as well its application for lensing mass estimations (Section~\ref{sec:observationalMassEstimatesForecasts}).

\subsection{Comparison to other approaches and further improvements}
\label{sec:comparisons}

Our approach differs from the setup in~\citet{Mead2014a} since it is a nonlocal operation on the density profiles built from the full 3D and 2D rescaled particle distributions whereas their method involve shifting the halo particle positions. They work with a subset of particles randomly sampled from the fiducial distribution to fill up the predicted density profile where information from the tidal tensor helps to account for the asphericity (this produces better agreement in halo morphology but does not take substructure into account which is problematic for satellite galaxies). It is not evident how much this sampling scheme differs from a refined method working on the actual 3D distribution of particles within the halo. A possible way to implement our algorithm as a localised, discrete mapping is to perform a local measurement of the spherically binned density field around each halo, use the correction to find the closest NFW/Einasto profile and shift the particles between the shells accordingly till some convergence criteria has been met. Preferably, this should prioritise displacements between adjacent shells. One could also account for the shape of the tidal tensor, compute Penna-Dines surfaces for accretion responses \citep[cf.][]{2017ApJ...841...34M} and extract additional phase-space information to preserve the halo shape, composition, stream structure and extension.

\subsection{Predicting the concentration bias as a function of cosmology}
\label{sec:predictingCBias}

\begin{figure}
	\includegraphics[width=1.04\columnwidth]{\figrelpath 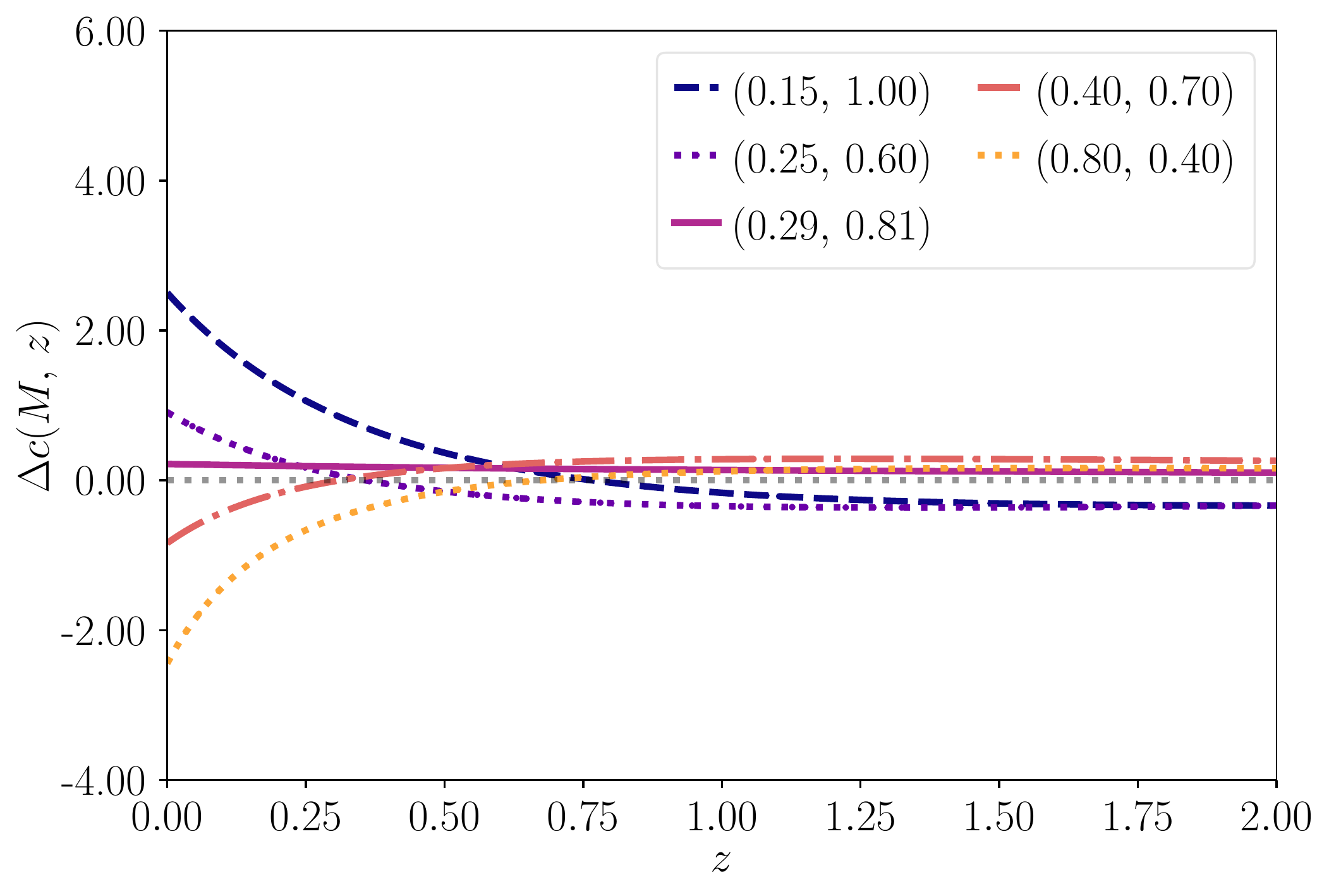}
    \caption{
    \label{fig:redshiftEvolutionDeltaCM}
		Expected bias in the concentration of rescaled haloes based on the \citetalias{2016MNRAS.460.1214L} model, evaluated as the median bias for haloes with $10^{12} < M_{200\text{m}}/(M_{\sun}\,h^{-1}) < 10^{14}$, as a function of redshift.}
\end{figure}

Due to the few simulations in our study, we cannot put strong constraints on a model-independent fitting function for the concentration bias. All cosmologies, with the exception of $(0.25, \, 0.60)$, trace the $\Omega_{\text{m}} - \sigma_8$ degeneracy favoured by weak lensing, which means that we have few constraints perpendicular to this line. We thus use the \citetalias{2016MNRAS.460.1214L} model to predict the rescaled concentration bias for cosmologies and redshifts where we do not have access to a corresponding direct simulation. 

Firstly, we investigate the redshift evolution in the cosmologies already covered. We use the linear growth factor relation in Eq.~\eqref{eq:linGrowthRelation} to calculate the redshifts in the fiducial simulation which correspond to the higher redshifts in the direct simulation. We plot the median concentration bias for haloes with $M_{200\text{m}}$ in $10^{12} - 10^{14}\, h^{-1} \, M_{\sun}$ as a function of redshift from $z = 0$ to $z = 2$ in the direct simulation in Fig.~\ref{fig:redshiftEvolutionDeltaCM}. Overall the difference in concentration decreases with redshift and there is a turnover point for all cosmologies expect $(0.29, \, 0.81)$ where the bias changes sign. This is a consequence of the rescaling parameters being determined by the locally matched growth history. Yet, caution must taken as we have already seen that the model prediction works less well at higher redshifts in Fig.~\ref{fig:concentrationDifference}. To bring about a better agreement with the measurements, the model could be modified to feature a slight redshift dependence which either decreases $C$ and/or raises $f$ since these changes lower the amplitude of the $c(M)-$relation.

\begin{figure}
	\includegraphics[width=1.04\columnwidth]{\figrelpath 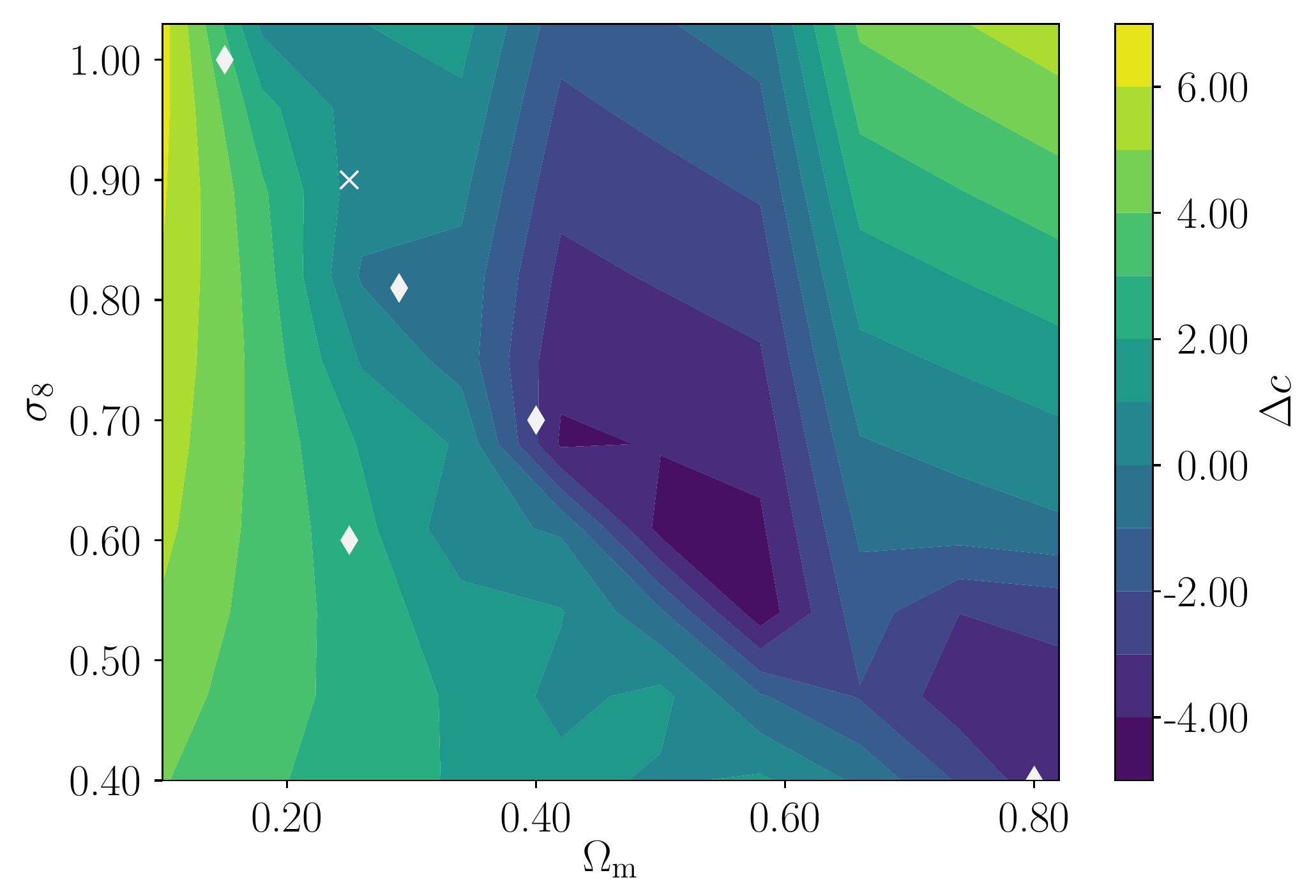}
    \caption{
    \label{fig:contourPlotCM}
		Expected bias in the concentration of rescaled haloes at $z=0$ as a function of the value of $\Omega_\text{m}$ and $\sigma_8$. Our assumed fiducial cosmology is $\Omega_\text{m}=0.25$ and $\sigma_8 = 0.90$ (marked by the white cross). The white diamonds mark the test simulations cosmologies employed in this paper.
		}
\end{figure}

In Fig.~\ref{fig:contourPlotCM}, we plot the expected median $\Delta c$ bias for haloes with masses $M_{200\text{m}}$ in $10^{12} - 10^{14}\, h^{-1} \, M_{\sun}$ when rescaling the Millennium simulation \citep[][]{Springel:2005nw} to match target cosmologies with different $\Omega_\text{m}$ and $\sigma_8$ at $z = 0$, with the target matter power spectra generated by \textsc{CAMB} \citep{2000ApJ...538..473L} combined with linear growth factors \citep[e.g.][]{2001MNRAS.322..419H} assuming a constant baryon fraction $\Omega_\text{baryons}/\Omega_\text{m}$. The corresponding contours for the rescaling parameters $(\alpha, \, z^\ast)$ are shown in Appendix~\ref{sec:alphaZastContours}. Rescaling to a lower $\sigma_8$ at fixed $\Omega_\text{m}$ or a lower $\Omega_\text{m}$ with a higher $\sigma_8$ induces a positive $\Delta c$, whereas raising $\Omega_\text{m}$ and lowering $\sigma_8$ will produce negative $\Delta c$.

If one relaxes the growth history constraint to permit matches in the future, negative redshifts\footnote{An existing $N$-body simulation can cheaply be evolved into the future \citep[see e.g.][]{Angulo:2014gza}.} represent the preferred solutions for the $\Delta \Omega_\text{m} > 0, \, \Delta \sigma_8 > 0$ quadrant. Such solutions yield $\Delta c < 0$. If we instead restrict our redshift range to $z^\ast \gtrsim -0.8$, the concentration bias becomes positive again as we move further away from the degeneracy plane. The contours for the predicted $\Delta r_\text{sp}$-bias (see Appendix~\ref{sec:alphaZastContours}) partly trace the $\Delta c$ contours with the opposite sign over most of the plane except in the $\Delta \Omega_\text{m} > 0, \, \Delta \sigma_8 > 0$ quadrant.

The concentration bias is a smooth function of cosmology, i.e. small changes in the cosmological parameters produce small concentration offsets. A set of well-placed simulations could thus be used together with rescaling to efficiently cover a large region of parameter space accurately.

Lastly, we discuss rescaling to emulate a WMAP7 cosmology \citep{2011ApJS..192...18K} and Planck (2014) cosmology \citep{2014A&A...571A..16P} at $z = 0$ using the Millennium simulation with SAMs in \citet{Guo:2012fy} with the \citetalias{Angulo:2009rc} weighting scheme and in \citet{2015MNRAS.451.2663H} with the \citetalias{Angulo:2014gza} scheme, respectively. The corresponding $(z_\ast, \, \alpha)$ are $(0.28, \, 1.04)$ and $(0.12, \, 0.96)$, respectively, which produce $\Delta c (M)$ relations with shallow slopes with median biases $\Delta c = 0.88$ $(\Delta c_\text{min} = 0.77, \, \Delta c_\text{max} = 0.97)$ and $\Delta c = 0.06$ $(\Delta c_\text{min} = 0.03, \, \Delta c_\text{max} = 0.06)$ for $M_{200\text{m}}$ between $10^{12} - 10^{14}\, h^{-1} \, M_{\sun}$. This means that the concentration bias for haloes in \citet{2015MNRAS.451.2663H} is almost negligible. We plot these relations in Appendix~\ref{sec:wmapPlanckMillennium} with the predicted redshift evolutions, where the biases also are reduced at earlier times. Hence, we can predict the bias of the measured lensing signal around central SAM galaxies in rescaled simulation snapshots.

\subsection{Baryonic effects} 
\label{sec:baryons}

Our method currently does not account for effects baryonic processes have on halo profiles. The impact of baryonic processes on the matter distribution has been investigated in simulations \citep[e.g. by][]{vanDaalen:2013ita, 2014MNRAS.442.2641V, 2015MNRAS.451.1247S, 2017MNRAS.467.3024L, 2017MNRAS.471..227M}. Baryon physics affects the matter clustering by $\sim$ 10\,\% on scales $\lesssim 1\, \text{Mpc}$. The impact on $\Delta \Sigma$ is similar. By matching the haloes in Illustris with their counterparts in a dark matter-only run, the baryonic physics has been found to suppress $\Delta \Sigma$ by $\sim$ 20\,\% from $r \gtrsim 0.4 \, h^{-1} \, \text{Mpc}$ to  $r  \leqslant 4 \, h^{-1} \, \text{Mpc}$ \citep{2017MNRAS.467.3024L}.

Even for cosmologies far from the fiducial cosmology, the rescaling predictions without the concentration corrections are at most off by 40\,\% in the innermost radial bins, and the disagreement decreases to $\sim 10\,\%$  at $r \approx 1 \, h^{-1} \, \text{Mpc}$. The concentration correction substantially improves agreement in the inner region. Moreover, the discrepancies are much smaller for cosmologies closer to the fiducial cosmology. This means that the bias induced by rescaling is subdominant to the baryonic feedback effects below $1 \, h^{-1} \, \text{Mpc}$, except for extreme cosmologies.

\subsection{Large scales}
\label{sec:2halo}

Here, we do not attempt any corrections at very large scales. We have computed the difference between the matter power spectrum in the weakly nonlinear to the nonlinear regime for $(0.15, \, 1.00)$ with and without the large-scale displacement field correction from \citetalias{Angulo:2009rc} and it was found to be negligible. The large-scale halo-matter correlations do not differ significantly between the rescaled and direct simulations for the halo masses we are investigating in 3D. There appears at most a small offset with surrounding scatter. The connection and coupling between this offset and the detected mass bias, as well as the proper response of the large-scale correlations to the rescaling transform are topics for future studies.
In halo models of GGL \citep[e.g.][]{2011PhRvD..83b3008O}, the large-scale lensing signal (2-halo term) is directly related to the projected linear power spectrum. It should thus be straightforward to compute its response to rescaling. Moreover, the proposed recipe in \citetalias{Angulo:2009rc} to correct the displacement field using the Zel'dovich approximation \citep{1970A&A.....5...84Z} should improve the agreement. 

For the linear regime, there already exist fast, accurate large-scale structure solvers, e.g. \textsc{COLA} \citep{2013JCAP...06..036T, 2015arXiv150207751T} and \textsc{FastPM} \citep{2016MNRAS.463.2273F}. Thus, corrections for exclusive large-scale analyses using the rescaling approach are of limited practical importance. However, the benefits of rescaling the small scales become manifest when successfully coupled to such a large-scale solver, as a wide range of cosmologies can be explored on multiple refinement levels.

\subsection{Mass estimation forecasts}
\label{sec:observationalMassEstimatesForecasts}

One application for galaxy-galaxy lensing is halo mass estimation for a selected foreground galaxy sample. We thus examine how the residual statistical and systematic differences in the profiles translate to errors in the measured masses. For simplicity, we focus on the $(0.29, \, 0.81)$ cosmology, and we choose a series of mass-selected samples in the direct simulation: haloes in mass bins of 0.05 dex or 0.1 dex centred on slightly different masses with bin borders shifted with 0.005 dex w.r.t. one another around $10^{12.5} \, h^{-1} M_{\sun}$ (i.e. massive galaxy haloes) or $10^{13.5} \, h^{-1} M_{\sun}$ (galaxy group haloes). The mean $\Delta\Sigma$ profiles for these bins constitute our mock weak lensing observations.

If we fit NFW profiles to these mock lensing observations, we obtain mass estimates that are approx. $5$ to $10\,\%$ below the true mean halo masses as recorded by the halo finder (see Fig~\ref{fig:massBiasFitVsSimMass}). 
We should be able to bypass this bias if we employ the rescaled simulation's stacked profiles (which should be \lq{}biased\rq{} in the same way) as model predictions (instead of analytic NFW profiles) to estimate the mean mass of our mock halo sample. This however requires that the rescaled halo profiles are close enough to the true halo profiles (i.e. the direct simulation's profiles in this exercise), since a mismatch, e.g., in concentration of $\Delta c = 1$ causes an error $\sim 5\,\%$ in the inferred masses \citep[e.g.][]{2016MNRAS.457.1522A, 2018MNRAS.474.2635S}. For the considered example, the concentration mismatches are already small before the correction ($\Delta c \sim 0.30$ and $\Delta c \sim 0.15$), and vanish after the correction. Thus, mass errors due to concentration mismatches are well below $1\,\%$ here (this is not necessarily the case for rescaling to the other, more extreme cosmologies).

\begin{figure}
	\includegraphics[width=1.04\columnwidth]{\figrelpath 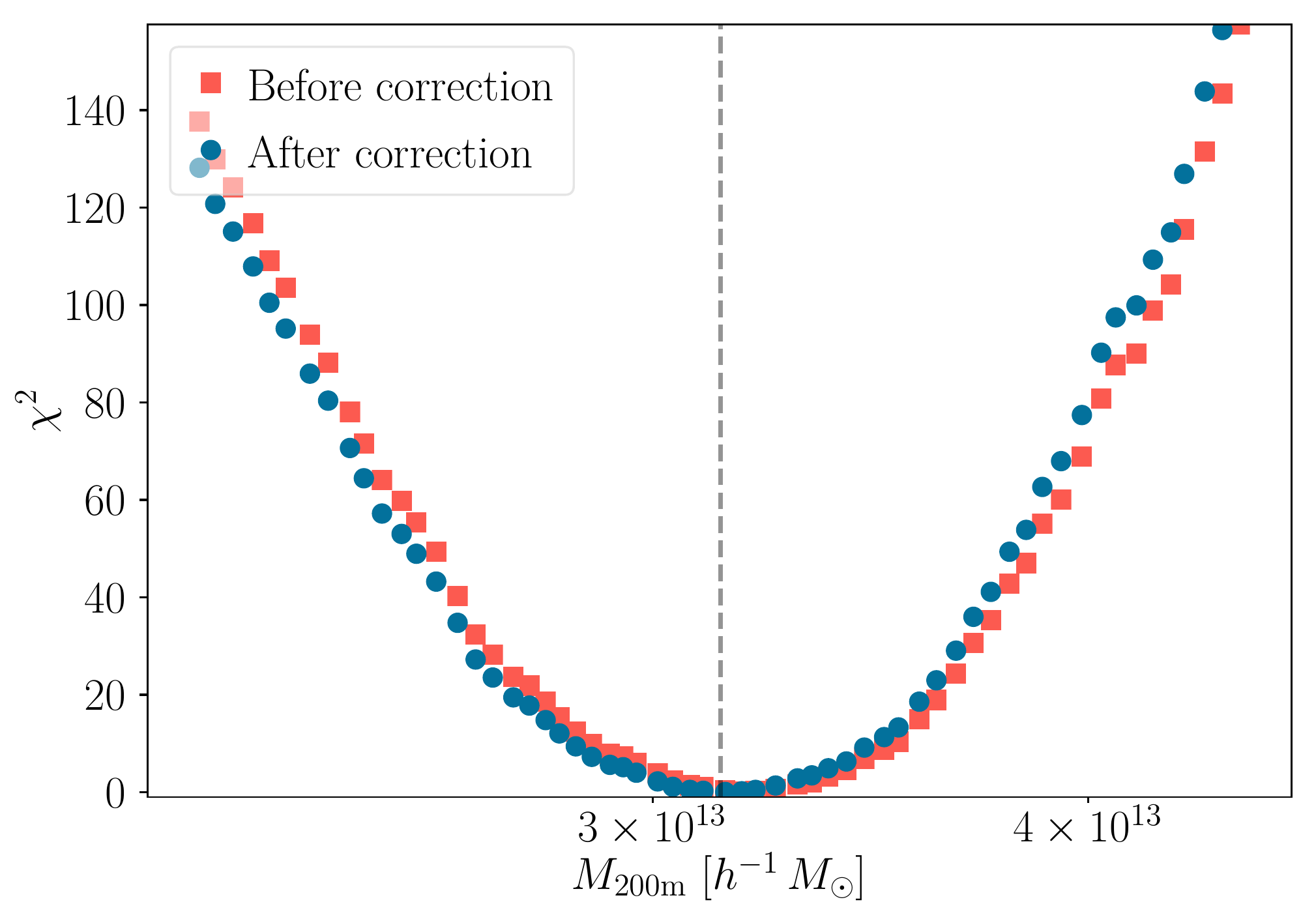}
    \caption{
    \label{fig:chisquareRescaledProfilesFittedToDirectProfiles}
		$\chi^2$-parabolae for rescaled $\Delta \Sigma$ profiles fitted to a direct $\Delta \Sigma$ profile for a stack of galaxy group-size haloes with mean $M_{200\text{m}}$ marked by the vertical dashed line according to Eq.~\eqref{eq:fom_direct_resc}. The minimum determines the best fit rescaled profile, and the corresponding simulation mass the best fit mass. The concentration correction shifts the parabola to be more symmetric around the direct simulation's mean mass, reducing the difference to the best-fit rescaled mass.}
\end{figure}

To fit the rescaled mean profiles (our predictions) to the direct profiles (our mock data), we minimise the figure-of-merit
\begin{equation}
\label{eq:fom_direct_resc}
\chi^2 = \sum^{N_r}_{i} r_i^2 \left[ \Delta\Sigma_\text{direct}(r_i) - \Delta\Sigma_\text{rescaled}(r_i) \right]^2,
\end{equation}
for radial bins $0.05 < r_i/r_{200\text{m}} < 0.8$. Fig.~\ref{fig:chisquareRescaledProfilesFittedToDirectProfiles} illustrates how the figure of merit changes when the mean profile of haloes in the direct simulation in a bin with width 0.1 dex centred on $10^{13.5} \, h^{-1} M_{\sun}$ is fit with rescaled mean profiles of mass bins with the same width but varying mean mass. The concentration correction shifts the $\chi^2$-parabola to be more symmetric around the direct simulation's mean mass.

\begin{table*}
	\centering
\caption{
\label{tab:massEstimationForecastTable}
Maximal disagreements in mass between the direct and rescaled profiles for the different sweep ranges and bin widths. `Group' refers to the $10^{13.45} - 10^{13.55} \, h^{-1} M_{\sun}$ mass range and `Galactic' to the $10^{12.45} - 10^{12.55} \, h^{-1} M_{\sun}$ mass range, respectively. The errors listed are the maximal discrepancies $1 - M_{200\text{m}}^{\text{(r)}}/M_{200\text{m}}^{\text{(d)}}$ for each of the approximately twenty direct profiles in each sweep range and with `Corrected' the concentration correction has been applied to the rescaled profiles in Eq.~\eqref{eq:fom_direct_resc}. `Axial' refers to the largest error for halo profiles compared along a sole projection axis and `Multi-axial' the largest errors for the average profiles along the three spatial axes.
}
\begin{tabular}{c c c c c c}
\hline
		Mass range & Bin size & Max error (multi-axial) & Max error (axial) & Corrected max error (multi-axial) & Corrected max error (axial) \\
\hline		
		Group    &	0.10 dex &	-3.2\,\% & -4.0\,\%  &  -1.1\,\%  &  2.5\,\%  \\  
		Group    &	0.05 dex &	-6.0\,\% & -9.4\,\%  &  -4.8\,\%  & -7.2\,\%  \\  
		Galactic &	0.10 dex &	-2.4\,\% & -3.6\,\%  &   0.2\,\%  & -1.3\,\%  \\  
		Galactic &	0.05 dex &	-2.4\,\% & -4.8\,\%  &   1.2\,\%  & -3.5\,\%  \\  
\hline
\end{tabular}
\end{table*}

The results from the different sweeps are listed in Table~\ref{tab:massEstimationForecastTable}. For smaller halo samples, the $\chi^2$-parabolae feature considerable scatter which cause larger errors for the best-fit mass.  As the number of haloes grow, the $\chi^2$-parabolae become smoother and the errors on the best-fit masses decrease. This behaviour is in line with previous work \citep{2011ApJ...740...25B, 2011MNRAS.412.2095H} where the relative error on the mass was found to be $\sim30\,\%$ per system for group haloes (and around 20\,\% for more massive systems). For example, this yields a relative mass error of $\sim 0.01$ for stacks of $\sim$\,1000 haloes, and $\sim 0.001$ for $\sim$\,10\,000 haloes.

For future dark energy task force stage IV surveys, such as Euclid, statistical errors on mass estimations from $\Delta \Sigma$ profiles are expected to shrink substantially compared to current surveys. We can acquire a rough estimation by scaling corresponding values from CFHTLenS \citep{2014MNRAS.437.2111V}, which has a similar depth but a smaller survey area of $150\,\text{deg}^2$, to an area of $15\,000\,\text{deg}^2$ for Euclid \citep{2011arXiv1110.3193L, 2013LRR....16....6A}. A hundred times larger survey area roughly translates to a reduction of the statistical errors by a factor of ten.
As example, we consider the sample L7 of 344 lenses in \citet{2014MNRAS.437.2111V} with absolute $r$-band magnitudes in the range $\left [ -24.0, \, -23.5 \right ]$, average redshift $\bar{z} = 0.3$, fraction of blue galaxies $f_\text{blue} = 0.03$. The mean halo mass of these lenses estimated from CFHTLenS is $10^{13.51} \, h^{-1} M_{\sun}$ with a quoted $20\,\%$ error. The statistical error for Euclid would shrink to $2\,\%$. This suggests that our proposed method is accurate enough for current halo weak lensing data, and moreover may be viable for much larger future surveys, once baryonic effects on halo profiles have been properly accounted for.

\section{Conclusions}
\label{sec:conclusions}

We have demonstrated the prowess of a refined rescaling algorithm with growth history constraints in predicting halo 3D and GGL profiles. Residual differences in the inner profiles have been parametrised as concentration biases that can be predicted using linear theory combined with excursion sets. Differences in the profile outskirts can be expressed in terms of a shift in the splashback radius. This enables us to correct the profiles and improve the method's accuracy. This represents an important step towards the reusability of $N$-body simulations for cosmic structure analyses.

The algorithm's accuracy is satisfactory for current GGL data. However, small remaining discrepancies in the halo profile outskirts and for the lens mass estimates may require further treatment depending on the application. Further studies could clarify, which of these discrepancies are due to systematic biases, and which are due to scatter in, e.g., halo shapes and line-of-sight structure. With possibly improved corrections capturing biases not addressed so far and large $N$-body simulations to minimise statistical errors, the method may be made suitable for analysing future large (dark energy task force stage IV) surveys.

\section*{Acknowledgements}

We would like to thank the anonymous referee for a comprehensive report which has improved the structure and presentation of our results in this paper. M.R. and S.H. acknowledge support by the DFG cluster of excellence \lq{}Origin and Structure of the Universe\rq{} (\href{http://www.universe-cluster.de}{\texttt{www.universe-cluster.de}}). M.R. and S.H. thank the Max Planck Institute for Astrophysics and the Max Planck Computing and Data Facility for computational resources. R.E.A. acknowledges support from AYA2015-66211-C2-2 and support from the European Research Council through grant number ERC-StG/716151.

%%%%%%%%%%%%%%%%%%%%%%%%%%%%%%%%%%%%%%%%%%%%%%%%%%


\bibliographystyle{mnras}
\bibliography{references}

\begin{thebibliography}{}
\makeatletter
\relax
\def\mn@urlcharsother{\let\do\@makeother \do\$\do\&\do\#\do\^\do\_\do\%\do\~}
\def\mn@doi{\begingroup\mn@urlcharsother \@ifnextchar [ {\mn@doi@}
  {\mn@doi@[]}}
\def\mn@doi@[#1]#2{\def\@tempa{#1}\ifx\@tempa\@empty \href
  {http://dx.doi.org/#2} {doi:#2}\else \href {http://dx.doi.org/#2} {#1}\fi
  \endgroup}
\def\mn@eprint#1#2{\mn@eprint@#1:#2::\@nil}
\def\mn@eprint@arXiv#1{\href {http://arxiv.org/abs/#1} {{\tt arXiv:#1}}}
\def\mn@eprint@dblp#1{\href {http://dblp.uni-trier.de/rec/bibtex/#1.xml}
  {dblp:#1}}
\def\mn@eprint@#1:#2:#3:#4\@nil{\def\@tempa {#1}\def\@tempb {#2}\def\@tempc
  {#3}\ifx \@tempc \@empty \let \@tempc \@tempb \let \@tempb \@tempa \fi \ifx
  \@tempb \@empty \def\@tempb {arXiv}\fi \@ifundefined
  {mn@eprint@\@tempb}{\@tempb:\@tempc}{\expandafter \expandafter \csname
  mn@eprint@\@tempb\endcsname \expandafter{\@tempc}}}

\bibitem[\protect\citeauthoryear{{Adhikari}, {Dalal}  \&
  {Chamberlain}}{{Adhikari} et~al.}{2014}]{2014JCAP...11..019A}
{Adhikari} S.,  {Dalal} N.,   {Chamberlain} R.~T.,  2014, \mn@doi [\jcap]
  {10.1088/1475-7516/2014/11/019}, \href
  {http://adsabs.harvard.edu/abs/2014JCAP...11..019A} {11, 019}

\bibitem[\protect\citeauthoryear{{Amendola} et~al.,}{{Amendola}
  et~al.}{2013}]{2013LRR....16....6A}
{Amendola} L.,  et~al., 2013, \mn@doi [Living Reviews in Relativity]
  {10.12942/lrr-2013-6}, \href
  {http://adsabs.harvard.edu/abs/2013LRR....16....6A} {16}

\bibitem[\protect\citeauthoryear{Angulo \& Hilbert}{Angulo \&
  Hilbert}{2015}]{Angulo:2014gza}
Angulo R.~E.,  Hilbert S.,  2015, \mn@doi [MNRAS] {10.1093/mnras/stv050}, 448,
  364

\bibitem[\protect\citeauthoryear{Angulo \& White}{Angulo \&
  White}{2010}]{Angulo:2009rc}
Angulo R.~E.,  White S.~D.,  2010, \mn@doi [MNRAS]
  {10.1111/j.1365-2966.2010.16459.x}, 405, 143

\bibitem[\protect\citeauthoryear{{Angulo}, {Springel}, {White}, {Jenkins},
  {Baugh}  \& {Frenk}}{{Angulo} et~al.}{2012}]{2012MNRAS.426.2046A}
{Angulo} R.~E.,  {Springel} V.,  {White} S.~D.~M.,  {Jenkins} A.,  {Baugh}
  C.~M.,   {Frenk} C.~S.,  2012, \mn@doi [MNRAS]
  {10.1111/j.1365-2966.2012.21830.x}, \href
  {http://adsabs.harvard.edu/abs/2012MNRAS.426.2046A} {426, 2046}

\bibitem[\protect\citeauthoryear{{Applegate} et~al.,}{{Applegate}
  et~al.}{2016}]{2016MNRAS.457.1522A}
{Applegate} D.~E.,  et~al., 2016, \mn@doi [\mnras] {10.1093/mnras/stw005},
  \href {http://adsabs.harvard.edu/abs/2016MNRAS.457.1522A} {457, 1522}

\bibitem[\protect\citeauthoryear{{Baltz}, {Marshall}  \& {Oguri}}{{Baltz}
  et~al.}{2009}]{2009JCAP...01..015B}
{Baltz} E.~A.,  {Marshall} P.,   {Oguri} M.,  2009, \mn@doi [JCAP]
  {10.1088/1475-7516/2009/01/015}, \href
  {http://adsabs.harvard.edu/abs/2009JCAP...01..015B} {1, 15}

\bibitem[\protect\citeauthoryear{Bartelmann \& Schneider}{Bartelmann \&
  Schneider}{2001}]{Bartelmann:1999yn}
Bartelmann M.,  Schneider P.,  2001, \mn@doi [Phys. Rept.]
  {10.1016/S0370-1573(00)00082-X}, 340, 291

\bibitem[\protect\citeauthoryear{{Becker} \& {Kravtsov}}{{Becker} \&
  {Kravtsov}}{2011}]{2011ApJ...740...25B}
{Becker} M.~R.,  {Kravtsov} A.~V.,  2011, \mn@doi [\apj]
  {10.1088/0004-637X/740/1/25}, \href
  {http://adsabs.harvard.edu/abs/2011ApJ...740...25B} {740, 25}

\bibitem[\protect\citeauthoryear{{Behroozi}, {Conroy}  \&
  {Wechsler}}{{Behroozi} et~al.}{2010}]{2010ApJ...717..379B}
{Behroozi} P.~S.,  {Conroy} C.,   {Wechsler} R.~H.,  2010, \mn@doi [ApJ]
  {10.1088/0004-637X/717/1/379}, \href
  {http://adsabs.harvard.edu/abs/2010ApJ...717..379B} {717, 379}

\bibitem[\protect\citeauthoryear{{Berlind} \& {Weinberg}}{{Berlind} \&
  {Weinberg}}{2002}]{2002ApJ...575..587B}
{Berlind} A.~A.,  {Weinberg} D.~H.,  2002, \mn@doi [\apj] {10.1086/341469},
  \href {http://adsabs.harvard.edu/abs/2002ApJ...575..587B} {575, 587}

\bibitem[\protect\citeauthoryear{{Bett}, {Eke}, {Frenk}, {Jenkins}, {Helly}  \&
  {Navarro}}{{Bett} et~al.}{2007}]{2007MNRAS.376..215B}
{Bett} P.,  {Eke} V.,  {Frenk} C.~S.,  {Jenkins} A.,  {Helly} J.,   {Navarro}
  J.,  2007, \mn@doi [\mnras] {10.1111/j.1365-2966.2007.11432.x}, \href
  {http://adsabs.harvard.edu/abs/2007MNRAS.376..215B} {376, 215}

\bibitem[\protect\citeauthoryear{{Bond}, {Cole}, {Efstathiou}  \&
  {Kaiser}}{{Bond} et~al.}{1991}]{1991ApJ...379..440B}
{Bond} J.~R.,  {Cole} S.,  {Efstathiou} G.,   {Kaiser} N.,  1991, \mn@doi [ApJ]
  {10.1086/170520}, \href {http://adsabs.harvard.edu/abs/1991ApJ...379..440B}
  {379, 440}

\bibitem[\protect\citeauthoryear{{Bower}, {Benson}, {Malbon}, {Helly}, {Frenk},
  {Baugh}, {Cole}  \& {Lacey}}{{Bower} et~al.}{2006}]{2006MNRAS.370..645B}
{Bower} R.~G.,  {Benson} A.~J.,  {Malbon} R.,  {Helly} J.~C.,  {Frenk} C.~S.,
  {Baugh} C.~M.,  {Cole} S.,   {Lacey} C.~G.,  2006, \mn@doi [\mnras]
  {10.1111/j.1365-2966.2006.10519.x}, \href
  {http://adsabs.harvard.edu/abs/2006MNRAS.370..645B} {370, 645}

\bibitem[\protect\citeauthoryear{{Brainerd}, {Blandford}  \&
  {Smail}}{{Brainerd} et~al.}{1996}]{1996ApJ...466..623B}
{Brainerd} T.~G.,  {Blandford} R.~D.,   {Smail} I.,  1996, \mn@doi [\apj]
  {10.1086/177537}, \href {http://adsabs.harvard.edu/abs/1996ApJ...466..623B}
  {466, 623}

\bibitem[\protect\citeauthoryear{{Cole} \& {Lacey}}{{Cole} \&
  {Lacey}}{1996}]{1996MNRAS.281..716C}
{Cole} S.,  {Lacey} C.,  1996, \mn@doi [\mnras] {10.1093/mnras/281.2.716},
  \href {http://adsabs.harvard.edu/abs/1996MNRAS.281..716C} {281, 716}

\bibitem[\protect\citeauthoryear{{Conroy} \& {Wechsler}}{{Conroy} \&
  {Wechsler}}{2009}]{2009ApJ...696..620C}
{Conroy} C.,  {Wechsler} R.~H.,  2009, \mn@doi [ApJ]
  {10.1088/0004-637X/696/1/620}, \href
  {http://adsabs.harvard.edu/abs/2009ApJ...696..620C} {696, 620}

\bibitem[\protect\citeauthoryear{{Conroy}, {Wechsler}  \& {Kravtsov}}{{Conroy}
  et~al.}{2006}]{2006ApJ...647..201C}
{Conroy} C.,  {Wechsler} R.~H.,   {Kravtsov} A.~V.,  2006, \mn@doi [ApJ]
  {10.1086/503602}, \href {http://adsabs.harvard.edu/abs/2006ApJ...647..201C}
  {647, 201}

\bibitem[\protect\citeauthoryear{{Cooray} \& {Sheth}}{{Cooray} \&
  {Sheth}}{2002}]{2002PhR...372....1C}
{Cooray} A.,  {Sheth} R.,  2002, \mn@doi [\physrep]
  {10.1016/S0370-1573(02)00276-4}, \href
  {http://adsabs.harvard.edu/abs/2002PhR...372....1C} {372, 1}

\bibitem[\protect\citeauthoryear{{Correa}, {Wyithe}, {Schaye}  \&
  {Duffy}}{{Correa} et~al.}{2015}]{2015MNRAS.452.1217C}
{Correa} C.~A.,  {Wyithe} J.~S.~B.,  {Schaye} J.,   {Duffy} A.~R.,  2015,
  \mn@doi [MNRAS] {10.1093/mnras/stv1363}, \href
  {http://adsabs.harvard.edu/abs/2015MNRAS.452.1217C} {452, 1217}

\bibitem[\protect\citeauthoryear{{Crain} et~al.,}{{Crain}
  et~al.}{2015}]{2015MNRAS.450.1937C}
{Crain} R.~A.,  et~al., 2015, \mn@doi [MNRAS] {10.1093/mnras/stv725}, \href
  {http://adsabs.harvard.edu/abs/2015MNRAS.450.1937C} {450, 1937}

\bibitem[\protect\citeauthoryear{{Davis}, {Efstathiou}, {Frenk}  \&
  {White}}{{Davis} et~al.}{1985}]{1985ApJ...292..371D}
{Davis} M.,  {Efstathiou} G.,  {Frenk} C.~S.,   {White} S.~D.~M.,  1985,
  \mn@doi [\apj] {10.1086/163168}, \href
  {http://adsabs.harvard.edu/abs/1985ApJ...292..371D} {292, 371}

\bibitem[\protect\citeauthoryear{{De Lucia} \& {Blaizot}}{{De Lucia} \&
  {Blaizot}}{2007}]{2007MNRAS.375....2D}
{De Lucia} G.,  {Blaizot} J.,  2007, \mn@doi [\mnras]
  {10.1111/j.1365-2966.2006.11287.x}, \href
  {http://adsabs.harvard.edu/abs/2007MNRAS.375....2D} {375, 2}

\bibitem[\protect\citeauthoryear{{Diemer}}{{Diemer}}{2017}]{2017ApJS..231....5D}
{Diemer} B.,  2017, \mn@doi [\apjs] {10.3847/1538-4365/aa799c}, \href
  {http://adsabs.harvard.edu/abs/2017ApJS..231....5D} {231, 5}

\bibitem[\protect\citeauthoryear{{Diemer} \& {Kravtsov}}{{Diemer} \&
  {Kravtsov}}{2014}]{2014ApJ...789....1D}
{Diemer} B.,  {Kravtsov} A.~V.,  2014, \mn@doi [\apj]
  {10.1088/0004-637X/789/1/1}, \href
  {http://adsabs.harvard.edu/abs/2014ApJ...789....1D} {789, 1}

\bibitem[\protect\citeauthoryear{{Diemer}, {Mansfield}, {Kravtsov}  \&
  {More}}{{Diemer} et~al.}{2017}]{2017ApJ...843..140D}
{Diemer} B.,  {Mansfield} P.,  {Kravtsov} A.~V.,   {More} S.,  2017, \mn@doi
  [\apj] {10.3847/1538-4357/aa79ab}, \href
  {http://adsabs.harvard.edu/abs/2017ApJ...843..140D} {843, 140}

\bibitem[\protect\citeauthoryear{Efron}{Efron}{1979}]{Efron:1979}
Efron B.,  1979, \mn@doi [Annals of Statistics] {10.1214/aos/1176344552}, 7, 1

\bibitem[\protect\citeauthoryear{{Einasto}}{{Einasto}}{1965}]{1965TrAlm...5...87E}
{Einasto} J.,  1965, Trudy Astrofizicheskogo Instituta Alma-Ata, \href
  {http://adsabs.harvard.edu/abs/1965TrAlm...5...87E} {5, 87}

\bibitem[\protect\citeauthoryear{{Feng}, {Chu}, {Seljak}  \& {McDonald}}{{Feng}
  et~al.}{2016}]{2016MNRAS.463.2273F}
{Feng} Y.,  {Chu} M.-Y.,  {Seljak} U.,   {McDonald} P.,  2016, \mn@doi [\mnras]
  {10.1093/mnras/stw2123}, \href
  {http://adsabs.harvard.edu/abs/2016MNRAS.463.2273F} {463, 2273}

\bibitem[\protect\citeauthoryear{{Gao}, {Navarro}, {Cole}, {Frenk}, {White},
  {Springel}, {Jenkins}  \& {Neto}}{{Gao} et~al.}{2008}]{2008MNRAS.387..536G}
{Gao} L.,  {Navarro} J.~F.,  {Cole} S.,  {Frenk} C.~S.,  {White} S.~D.~M.,
  {Springel} V.,  {Jenkins} A.,   {Neto} A.~F.,  2008, \mn@doi [\mnras]
  {10.1111/j.1365-2966.2008.13277.x}, \href
  {http://adsabs.harvard.edu/abs/2008MNRAS.387..536G} {387, 536}

\bibitem[\protect\citeauthoryear{{Genel} et~al.,}{{Genel}
  et~al.}{2014}]{2014MNRAS.445..175G}
{Genel} S.,  et~al., 2014, \mn@doi [MNRAS] {10.1093/mnras/stu1654}, \href
  {http://adsabs.harvard.edu/abs/2014MNRAS.445..175G} {445, 175}

\bibitem[\protect\citeauthoryear{{Gillis} et~al.,}{{Gillis}
  et~al.}{2013}]{2013MNRAS.431.1439G}
{Gillis} B.~R.,  et~al., 2013, \mn@doi [\mnras] {10.1093/mnras/stt274}, \href
  {http://adsabs.harvard.edu/abs/2013MNRAS.431.1439G} {431, 1439}

\bibitem[\protect\citeauthoryear{Guo, White, Boylan-Kolchin, De~Lucia,
  Kauffmann  et~al.}{Guo et~al.}{2011}]{Guo:2010ap}
Guo Q.,  White S.,  Boylan-Kolchin M.,  De~Lucia G.,  Kauffmann G.,   et~al.,
  2011, \mn@doi [MNRAS] {10.1111/j.1365-2966.2010.18114.x}, 413, 101

\bibitem[\protect\citeauthoryear{Guo, White, Angulo, Henriques, Lemson,
  Boylan-Kolchin, Thomas  \& Short}{Guo et~al.}{2013}]{Guo:2012fy}
Guo Q.,  White S.,  Angulo R.~E.,  Henriques B.,  Lemson G.,  Boylan-Kolchin
  M.,  Thomas P.,   Short C.,  2013, \mn@doi [MNRAS] {10.1093/mnras/sts115},
  428, 1351

\bibitem[\protect\citeauthoryear{{Hamilton}}{{Hamilton}}{2001}]{2001MNRAS.322..419H}
{Hamilton} A.~J.~S.,  2001, \mn@doi [\mnras]
  {10.1046/j.1365-8711.2001.04137.x}, \href
  {http://adsabs.harvard.edu/abs/2001MNRAS.322..419H} {322, 419}

\bibitem[\protect\citeauthoryear{{Henriques}, {White}, {Thomas}, {Angulo},
  {Guo}, {Lemson}  \& {Springel}}{{Henriques}
  et~al.}{2013}]{2013MNRAS.431.3373H}
{Henriques} B.~M.~B.,  {White} S.~D.~M.,  {Thomas} P.~A.,  {Angulo} R.~E.,
  {Guo} Q.,  {Lemson} G.,   {Springel} V.,  2013, \mn@doi [\mnras]
  {10.1093/mnras/stt415}, \href
  {http://adsabs.harvard.edu/abs/2013MNRAS.431.3373H} {431, 3373}

\bibitem[\protect\citeauthoryear{{Henriques}, {White}, {Thomas}, {Angulo},
  {Guo}, {Lemson}, {Springel}  \& {Overzier}}{{Henriques}
  et~al.}{2015}]{2015MNRAS.451.2663H}
{Henriques} B.~M.~B.,  {White} S.~D.~M.,  {Thomas} P.~A.,  {Angulo} R.,  {Guo}
  Q.,  {Lemson} G.,  {Springel} V.,   {Overzier} R.,  2015, \mn@doi [\mnras]
  {10.1093/mnras/stv705}, \href
  {http://adsabs.harvard.edu/abs/2015MNRAS.451.2663H} {451, 2663}

\bibitem[\protect\citeauthoryear{{Hilbert} \& {White}}{{Hilbert} \&
  {White}}{2010}]{2010MNRAS.404..486H}
{Hilbert} S.,  {White} S.~D.~M.,  2010, \mn@doi [\mnras]
  {10.1111/j.1365-2966.2010.16310.x}, \href
  {http://adsabs.harvard.edu/abs/2010MNRAS.404..486H} {404, 486}

\bibitem[\protect\citeauthoryear{{Hilbert}, {Hartlap}, {White}  \&
  {Schneider}}{{Hilbert} et~al.}{2009}]{2009A&A...499...31H}
{Hilbert} S.,  {Hartlap} J.,  {White} S.~D.~M.,   {Schneider} P.,  2009,
  \mn@doi [\aap] {10.1051/0004-6361/200811054}, \href
  {http://adsabs.harvard.edu/abs/2009A%26A...499...31H} {499, 31}

\bibitem[\protect\citeauthoryear{{Hinshaw} et~al.,}{{Hinshaw}
  et~al.}{2013}]{2013ApJS..208...19H}
{Hinshaw} G.,  et~al., 2013, \mn@doi [\apjs] {10.1088/0067-0049/208/2/19},
  \href {http://adsabs.harvard.edu/abs/2013ApJS..208...19H} {208, 19}

\bibitem[\protect\citeauthoryear{{Hoekstra}, {Hartlap}, {Hilbert}  \& {van
  Uitert}}{{Hoekstra} et~al.}{2011}]{2011MNRAS.412.2095H}
{Hoekstra} H.,  {Hartlap} J.,  {Hilbert} S.,   {van Uitert} E.,  2011, \mn@doi
  [\mnras] {10.1111/j.1365-2966.2010.18053.x}, \href
  {http://adsabs.harvard.edu/abs/2011MNRAS.412.2095H} {412, 2095}

\bibitem[\protect\citeauthoryear{{Jiang} \& {van den Bosch}}{{Jiang} \& {van
  den Bosch}}{2016}]{2016MNRAS.458.2848J}
{Jiang} F.,  {van den Bosch} F.~C.,  2016, \mn@doi [\mnras]
  {10.1093/mnras/stw439}, \href
  {http://adsabs.harvard.edu/abs/2016MNRAS.458.2848J} {458, 2848}

\bibitem[\protect\citeauthoryear{{Jing} \& {Suto}}{{Jing} \&
  {Suto}}{2002}]{2002ApJ...574..538J}
{Jing} Y.~P.,  {Suto} Y.,  2002, \mn@doi [\apj] {10.1086/341065}, \href
  {http://adsabs.harvard.edu/abs/2002ApJ...574..538J} {574, 538}

\bibitem[\protect\citeauthoryear{{Kauffmann}, {Colberg}, {Diaferio}  \&
  {White}}{{Kauffmann} et~al.}{1999}]{1999MNRAS.303..188K}
{Kauffmann} G.,  {Colberg} J.~M.,  {Diaferio} A.,   {White} S.~D.~M.,  1999,
  \mn@doi [\mnras] {10.1046/j.1365-8711.1999.02202.x}, \href
  {http://adsabs.harvard.edu/abs/1999MNRAS.303..188K} {303, 188}

\bibitem[\protect\citeauthoryear{{Komatsu} et~al.,}{{Komatsu}
  et~al.}{2009}]{2009ApJS..180..330K}
{Komatsu} E.,  et~al., 2009, \mn@doi [\apjs] {10.1088/0067-0049/180/2/330},
  \href {http://adsabs.harvard.edu/abs/2009ApJS..180..330K} {180, 330}

\bibitem[\protect\citeauthoryear{{Komatsu} et~al.,}{{Komatsu}
  et~al.}{2011}]{2011ApJS..192...18K}
{Komatsu} E.,  et~al., 2011, \mn@doi [\apjs] {10.1088/0067-0049/192/2/18},
  \href {http://adsabs.harvard.edu/abs/2011ApJS..192...18K} {192, 18}

\bibitem[\protect\citeauthoryear{{Kravtsov}, {Berlind}, {Wechsler}, {Klypin},
  {Gottl{\"o}ber}, {Allgood}  \& {Primack}}{{Kravtsov}
  et~al.}{2004}]{2004ApJ...609...35K}
{Kravtsov} A.~V.,  {Berlind} A.~A.,  {Wechsler} R.~H.,  {Klypin} A.~A.,
  {Gottl{\"o}ber} S.,  {Allgood} B.,   {Primack} J.~R.,  2004, \mn@doi [ApJ]
  {10.1086/420959}, \href {http://adsabs.harvard.edu/abs/2004ApJ...609...35K}
  {609, 35}

\bibitem[\protect\citeauthoryear{{Lacey} \& {Cole}}{{Lacey} \&
  {Cole}}{1993}]{1993MNRAS.262..627L}
{Lacey} C.,  {Cole} S.,  1993, \mn@doi [MNRAS] {10.1093/mnras/262.3.627}, \href
  {http://adsabs.harvard.edu/abs/1993MNRAS.262..627L} {262, 627}

\bibitem[\protect\citeauthoryear{{Laureijs} et~al.,}{{Laureijs}
  et~al.}{2011}]{2011arXiv1110.3193L}
{Laureijs} R.,  et~al., 2011, preprint, \href
  {http://adsabs.harvard.edu/abs/2011arXiv1110.3193L} {} (\mn@eprint {arXiv}
  {1110.3193})

\bibitem[\protect\citeauthoryear{{Leauthaud}, {Tinker}, {Behroozi}, {Busha}  \&
  {Wechsler}}{{Leauthaud} et~al.}{2011}]{2011ApJ...738...45L}
{Leauthaud} A.,  {Tinker} J.,  {Behroozi} P.~S.,  {Busha} M.~T.,   {Wechsler}
  R.~H.,  2011, \mn@doi [ApJ] {10.1088/0004-637X/738/1/45}, \href
  {http://adsabs.harvard.edu/abs/2011ApJ...738...45L} {738, 45}

\bibitem[\protect\citeauthoryear{{Leauthaud} et~al.}{{Leauthaud}
  et~al.}{2012}]{2012ApJ...744..159L}
{Leauthaud} A.,  et~al., 2012, \mn@doi [ApJ] {10.1088/0004-637X/744/2/159},
  \href {http://adsabs.harvard.edu/abs/2012ApJ...744..159L} {744, 159}

\bibitem[\protect\citeauthoryear{{Leauthaud} et~al.,}{{Leauthaud}
  et~al.}{2017}]{2017MNRAS.467.3024L}
{Leauthaud} A.,  et~al., 2017, \mn@doi [\mnras] {10.1093/mnras/stx258}, \href
  {http://adsabs.harvard.edu/abs/2017MNRAS.467.3024L} {467, 3024}

\bibitem[\protect\citeauthoryear{{Lewis}, {Challinor}  \& {Lasenby}}{{Lewis}
  et~al.}{2000}]{2000ApJ...538..473L}
{Lewis} A.,  {Challinor} A.,   {Lasenby} A.,  2000, \mn@doi [\apj]
  {10.1086/309179}, \href {http://adsabs.harvard.edu/abs/2000ApJ...538..473L}
  {538, 473}

\bibitem[\protect\citeauthoryear{{Ludlow} \& {Angulo}}{{Ludlow} \&
  {Angulo}}{2017}]{2017MNRAS.465L..84L}
{Ludlow} A.~D.,  {Angulo} R.~E.,  2017, \mn@doi [\mnras]
  {10.1093/mnrasl/slw216}, \href
  {http://adsabs.harvard.edu/abs/2017MNRAS.465L..84L} {465, L84}

\bibitem[\protect\citeauthoryear{{Ludlow}, {Navarro}, {Li}, {Angulo},
  {Boylan-Kolchin}  \& {Bett}}{{Ludlow} et~al.}{2012}]{2012MNRAS.427.1322L}
{Ludlow} A.~D.,  {Navarro} J.~F.,  {Li} M.,  {Angulo} R.~E.,  {Boylan-Kolchin}
  M.,   {Bett} P.~E.,  2012, \mn@doi [MNRAS]
  {10.1111/j.1365-2966.2012.21892.x}, \href
  {http://adsabs.harvard.edu/abs/2012MNRAS.427.1322L} {427, 1322}

\bibitem[\protect\citeauthoryear{{Ludlow}, {Navarro}, {Angulo},
  {Boylan-Kolchin}, {Springel}, {Frenk}  \& {White}}{{Ludlow}
  et~al.}{2014}]{2014MNRAS.441..378L}
{Ludlow} A.~D.,  {Navarro} J.~F.,  {Angulo} R.~E.,  {Boylan-Kolchin} M.,
  {Springel} V.,  {Frenk} C.,   {White} S.~D.~M.,  2014, \mn@doi [MNRAS]
  {10.1093/mnras/stu483}, \href
  {http://adsabs.harvard.edu/abs/2014MNRAS.441..378L} {441, 378}

\bibitem[\protect\citeauthoryear{{Ludlow}, {Bose}, {Angulo}, {Wang},
  {Hellwing}, {Navarro}, {Cole}  \& {Frenk}}{{Ludlow}
  et~al.}{2016}]{2016MNRAS.460.1214L}
{Ludlow} A.~D.,  {Bose} S.,  {Angulo} R.~E.,  {Wang} L.,  {Hellwing} W.~A.,
  {Navarro} J.~F.,  {Cole} S.,   {Frenk} C.~S.,  2016, \mn@doi [MNRAS]
  {10.1093/mnras/stw1046}, \href
  {http://adsabs.harvard.edu/abs/2016MNRAS.460.1214L} {460, 1214}

\bibitem[\protect\citeauthoryear{{Macci{\`o}}, {Dutton}, {van den Bosch},
  {Moore}, {Potter}  \& {Stadel}}{{Macci{\`o}}
  et~al.}{2007}]{2007MNRAS.378...55M}
{Macci{\`o}} A.~V.,  {Dutton} A.~A.,  {van den Bosch} F.~C.,  {Moore} B.,
  {Potter} D.,   {Stadel} J.,  2007, \mn@doi [MNRAS]
  {10.1111/j.1365-2966.2007.11720.x}, \href
  {http://adsabs.harvard.edu/abs/2007MNRAS.378...55M} {378, 55}

\bibitem[\protect\citeauthoryear{{Mansfield}, {Kravtsov}  \&
  {Diemer}}{{Mansfield} et~al.}{2017}]{2017ApJ...841...34M}
{Mansfield} P.,  {Kravtsov} A.~V.,   {Diemer} B.,  2017, \mn@doi [\apj]
  {10.3847/1538-4357/aa7047}, \href
  {http://adsabs.harvard.edu/abs/2017ApJ...841...34M} {841, 34}

\bibitem[\protect\citeauthoryear{{Mead} \& {Peacock}}{{Mead} \&
  {Peacock}}{2014a}]{Mead2014a}
{Mead} A.~J.,  {Peacock} J.~A.,  2014a, \mn@doi [\mnras]
  {10.1093/mnras/stu345}, \href
  {http://adsabs.harvard.edu/abs/2014MNRAS.440.1233M} {440, 1233}

\bibitem[\protect\citeauthoryear{{Mead} \& {Peacock}}{{Mead} \&
  {Peacock}}{2014b}]{Mead2014b}
{Mead} A.~J.,  {Peacock} J.~A.,  2014b, \mn@doi [\mnras]
  {10.1093/mnras/stu1964}, \href
  {http://adsabs.harvard.edu/abs/2014MNRAS.445.3453M} {445, 3453}

\bibitem[\protect\citeauthoryear{{Miralda-Escud\'e}}{{Miralda-Escud\'e}}{1991}]{1991ApJ...370....1M}
{Miralda-Escud\'e} J.,  1991, \mn@doi [ApJ] {10.1086/169789}, \href
  {http://adsabs.harvard.edu/abs/1991ApJ...370....1M} {370, 1}

\bibitem[\protect\citeauthoryear{{More}, {Diemer}  \& {Kravtsov}}{{More}
  et~al.}{2015}]{2015ApJ...810...36M}
{More} S.,  {Diemer} B.,   {Kravtsov} A.~V.,  2015, \mn@doi [\apj]
  {10.1088/0004-637X/810/1/36}, \href
  {http://adsabs.harvard.edu/abs/2015ApJ...810...36M} {810, 36}

\bibitem[\protect\citeauthoryear{{Moster}, {Somerville}, {Maulbetsch}, {van den
  Bosch}, {Macci{\`o}}, {Naab}  \& {Oser}}{{Moster}
  et~al.}{2010}]{2010ApJ...710..903M}
{Moster} B.~P.,  {Somerville} R.~S.,  {Maulbetsch} C.,  {van den Bosch} F.~C.,
  {Macci{\`o}} A.~V.,  {Naab} T.,   {Oser} L.,  2010, \mn@doi [\apj]
  {10.1088/0004-637X/710/2/903}, \href
  {http://adsabs.harvard.edu/abs/2010ApJ...710..903M} {710, 903}

\bibitem[\protect\citeauthoryear{{Mummery}, {McCarthy}, {Bird}  \&
  {Schaye}}{{Mummery} et~al.}{2017}]{2017MNRAS.471..227M}
{Mummery} B.~O.,  {McCarthy} I.~G.,  {Bird} S.,   {Schaye} J.,  2017, \mn@doi
  [\mnras] {10.1093/mnras/stx1469}, \href
  {http://adsabs.harvard.edu/abs/2017MNRAS.471..227M} {471, 227}

\bibitem[\protect\citeauthoryear{{Navarro}, {Frenk}  \& {White}}{{Navarro}
  et~al.}{1996}]{Navarro:1995iw}
{Navarro} J.~F.,  {Frenk} C.~S.,   {White} S.~D.~M.,  1996, \mn@doi [ApJ]
  {10.1086/177173}, \href {http://adsabs.harvard.edu/abs/1996ApJ...462..563N}
  {462, 563}

\bibitem[\protect\citeauthoryear{{Navarro}, {Frenk}  \& {White}}{{Navarro}
  et~al.}{1997}]{Navarro:1996gj}
{Navarro} J.~F.,  {Frenk} C.~S.,   {White} S.~D.~M.,  1997, \mn@doi [ApJ]
  {10.1086/304888}, \href {http://adsabs.harvard.edu/abs/1997ApJ...490..493N}
  {490, 493}

\bibitem[\protect\citeauthoryear{{Neto} et~al.,}{{Neto}
  et~al.}{2007}]{2007MNRAS.381.1450N}
{Neto} A.~F.,  et~al., 2007, \mn@doi [MNRAS]
  {10.1111/j.1365-2966.2007.12381.x}, \href
  {http://adsabs.harvard.edu/abs/2007MNRAS.381.1450N} {381, 1450}

\bibitem[\protect\citeauthoryear{{Oguri} \& {Takada}}{{Oguri} \&
  {Takada}}{2011}]{2011PhRvD..83b3008O}
{Oguri} M.,  {Takada} M.,  2011, \mn@doi [Phys. Rev.]
  {10.1103/PhysRevD.83.023008}, \href
  {http://adsabs.harvard.edu/abs/2011PhRvD..83b3008O} {D83, 023008}

\bibitem[\protect\citeauthoryear{{Pastor Mira}, {Hilbert}, {Hartlap}  \&
  {Schneider}}{{Pastor Mira} et~al.}{2011}]{2011A&A...531A.169P}
{Pastor Mira} E.,  {Hilbert} S.,  {Hartlap} J.,   {Schneider} P.,  2011,
  \mn@doi [\aap] {10.1051/0004-6361/201116851}, \href
  {http://adsabs.harvard.edu/abs/2011A%26A...531A.169P} {531, A169}

\bibitem[\protect\citeauthoryear{{Peacock} \& {Smith}}{{Peacock} \&
  {Smith}}{2000}]{2000MNRAS.318.1144P}
{Peacock} J.~A.,  {Smith} R.~E.,  2000, \mn@doi [\mnras]
  {10.1046/j.1365-8711.2000.03779.x}, \href
  {http://adsabs.harvard.edu/abs/2000MNRAS.318.1144P} {318, 1144}

\bibitem[\protect\citeauthoryear{{Planck Collaboration}}{{Planck
  Collaboration}}{2014}]{2014A&A...571A..16P}
{Planck Collaboration} 2014, \mn@doi [\aap] {10.1051/0004-6361/201321591},
  \href {http://adsabs.harvard.edu/abs/2014A%26A...571A..16P} {571, A16}

\bibitem[\protect\citeauthoryear{{Press} \& {Schechter}}{{Press} \&
  {Schechter}}{1974}]{1974ApJ...187..425P}
{Press} W.~H.,  {Schechter} P.,  1974, \mn@doi [ApJ] {10.1086/152650}, \href
  {http://adsabs.harvard.edu/abs/1974ApJ...187..425P} {187, 425}

\bibitem[\protect\citeauthoryear{{Retana-Montenegro}, {van Hese}, {Gentile},
  {Baes}  \& {Frutos-Alfaro}}{{Retana-Montenegro}
  et~al.}{2012}]{2012A&A...540A..70R}
{Retana-Montenegro} E.,  {van Hese} E.,  {Gentile} G.,  {Baes} M.,
  {Frutos-Alfaro} F.,  2012, \mn@doi [A\&A] {10.1051/0004-6361/201118543},
  \href {http://adsabs.harvard.edu/abs/2012A%26A...540A..70R} {540, A70}

\bibitem[\protect\citeauthoryear{{Ruiz}, {Padilla}, {Dom{\'{\i}}nguez}  \&
  {Cora}}{{Ruiz} et~al.}{2011}]{Ruiz2011}
{Ruiz} A.~N.,  {Padilla} N.~D.,  {Dom{\'{\i}}nguez} M.~J.,   {Cora} S.~A.,
  2011, \mn@doi [\mnras] {10.1111/j.1365-2966.2011.19635.x}, \href
  {http://adsabs.harvard.edu/abs/2011MNRAS.418.2422R} {418, 2422}

\bibitem[\protect\citeauthoryear{{Saghiha}, {Hilbert}, {Schneider}  \&
  {Simon}}{{Saghiha} et~al.}{2012}]{2012A&A...547A..77S}
{Saghiha} H.,  {Hilbert} S.,  {Schneider} P.,   {Simon} P.,  2012, \mn@doi
  [\aap] {10.1051/0004-6361/201219358}, \href
  {http://adsabs.harvard.edu/abs/2012A%26A...547A..77S} {547, A77}

\bibitem[\protect\citeauthoryear{{Saghiha}, {Simon}, {Schneider}  \&
  {Hilbert}}{{Saghiha} et~al.}{2017}]{2017A&A...601A..98S}
{Saghiha} H.,  {Simon} P.,  {Schneider} P.,   {Hilbert} S.,  2017, \mn@doi
  [\aap] {10.1051/0004-6361/201629608}, \href
  {http://adsabs.harvard.edu/abs/2017A%26A...601A..98S} {601, A98}

\bibitem[\protect\citeauthoryear{{Schaller} et~al.,}{{Schaller}
  et~al.}{2015}]{2015MNRAS.451.1247S}
{Schaller} M.,  et~al., 2015, \mn@doi [\mnras] {10.1093/mnras/stv1067}, \href
  {http://adsabs.harvard.edu/abs/2015MNRAS.451.1247S} {451, 1247}

\bibitem[\protect\citeauthoryear{{Schaye} et~al.,}{{Schaye}
  et~al.}{2015}]{2015MNRAS.446..521S}
{Schaye} J.,  et~al., 2015, \mn@doi [MNRAS] {10.1093/mnras/stu2058}, \href
  {http://adsabs.harvard.edu/abs/2015MNRAS.446..521S} {446, 521}

\bibitem[\protect\citeauthoryear{{Schrabback} et~al.,}{{Schrabback}
  et~al.}{2015}]{2015MNRAS.454.1432S}
{Schrabback} T.,  et~al., 2015, \mn@doi [\mnras] {10.1093/mnras/stv2053}, \href
  {http://adsabs.harvard.edu/abs/2015MNRAS.454.1432S} {454, 1432}

\bibitem[\protect\citeauthoryear{{Schrabback} et~al.,}{{Schrabback}
  et~al.}{2018}]{2018MNRAS.474.2635S}
{Schrabback} T.,  et~al., 2018, \mn@doi [\mnras] {10.1093/mnras/stx2666}, \href
  {https://ui.adsabs.harvard.edu/#abs/2018MNRAS.474.2635S} {474, 2635}

\bibitem[\protect\citeauthoryear{{Seljak}}{{Seljak}}{2000}]{2000MNRAS.318..203S}
{Seljak} U.,  2000, \mn@doi [\mnras] {10.1046/j.1365-8711.2000.03715.x}, \href
  {http://adsabs.harvard.edu/abs/2000MNRAS.318..203S} {318, 203}

\bibitem[\protect\citeauthoryear{{Sereno}, {Fedeli}  \& {Moscardini}}{{Sereno}
  et~al.}{2016}]{2016JCAP...01..042S}
{Sereno} M.,  {Fedeli} C.,   {Moscardini} L.,  2016, \mn@doi [JCAP]
  {10.1088/1475-7516/2016/01/042}, \href
  {http://adsabs.harvard.edu/abs/2016JCAP...01..042S} {1, 042}

\bibitem[\protect\citeauthoryear{{Shi}}{{Shi}}{2016}]{2016MNRAS.459.3711S}
{Shi} X.,  2016, \mn@doi [\mnras] {10.1093/mnras/stw925}, \href
  {http://adsabs.harvard.edu/abs/2016MNRAS.459.3711S} {459, 3711}

\bibitem[\protect\citeauthoryear{{Spergel} et~al.,}{{Spergel}
  et~al.}{2003}]{2003ApJS..148..175S}
{Spergel} D.~N.,  et~al., 2003, \mn@doi [ApJS] {10.1086/377226}, \href
  {http://adsabs.harvard.edu/abs/2003ApJS..148..175S} {148, 175}

\bibitem[\protect\citeauthoryear{{Spergel} et~al.,}{{Spergel}
  et~al.}{2007}]{2007ApJS..170..377S}
{Spergel} D.~N.,  et~al., 2007, \mn@doi [\apjs] {10.1086/513700}, \href
  {http://adsabs.harvard.edu/abs/2007ApJS..170..377S} {170, 377}

\bibitem[\protect\citeauthoryear{{Springel}}{{Springel}}{2005}]{2005MNRAS.364.1105S}
{Springel} V.,  2005, \mn@doi [\mnras] {10.1111/j.1365-2966.2005.09655.x},
  \href {http://adsabs.harvard.edu/abs/2005MNRAS.364.1105S} {364, 1105}

\bibitem[\protect\citeauthoryear{{Springel}, {White}, {Tormen}  \&
  {Kauffmann}}{{Springel} et~al.}{2001}]{2001MNRAS.328..726S}
{Springel} V.,  {White} S.~D.~M.,  {Tormen} G.,   {Kauffmann} G.,  2001,
  \mn@doi [\mnras] {10.1046/j.1365-8711.2001.04912.x}, \href
  {http://adsabs.harvard.edu/abs/2001MNRAS.328..726S} {328, 726}

\bibitem[\protect\citeauthoryear{Springel et~al.}{Springel
  et~al.}{2005}]{Springel:2005nw}
Springel V.,  et~al., 2005, \mn@doi [Nature] {10.1038/nature03597}, 435, 629

\bibitem[\protect\citeauthoryear{{Squires} \& {Kaiser}}{{Squires} \&
  {Kaiser}}{1996}]{1996ApJ...473...65S}
{Squires} G.,  {Kaiser} N.,  1996, \mn@doi [ApJ] {10.1086/178127}, \href
  {http://adsabs.harvard.edu/abs/1996ApJ...473...65S} {473, 65}

\bibitem[\protect\citeauthoryear{{Tasitsiomi}, {Kravtsov}, {Wechsler}  \&
  {Primack}}{{Tasitsiomi} et~al.}{2004}]{2004ApJ...614..533T}
{Tasitsiomi} A.,  {Kravtsov} A.~V.,  {Wechsler} R.~H.,   {Primack} J.~R.,
  2004, \mn@doi [\apj] {10.1086/423784}, \href
  {http://adsabs.harvard.edu/abs/2004ApJ...614..533T} {614, 533}

\bibitem[\protect\citeauthoryear{{Tassev}, {Zaldarriaga}  \&
  {Eisenstein}}{{Tassev} et~al.}{2013}]{2013JCAP...06..036T}
{Tassev} S.,  {Zaldarriaga} M.,   {Eisenstein} D.~J.,  2013, \mn@doi [\jcap]
  {10.1088/1475-7516/2013/06/036}, \href
  {http://adsabs.harvard.edu/abs/2013JCAP...06..036T} {6, 36}

\bibitem[\protect\citeauthoryear{{Tassev}, {Eisenstein}, {Wandelt}  \&
  {Zaldarriaga}}{{Tassev} et~al.}{2015}]{2015arXiv150207751T}
{Tassev} S.,  {Eisenstein} D.~J.,  {Wandelt} B.~D.,   {Zaldarriaga} M.,  2015,
  preprint, \href {http://adsabs.harvard.edu/abs/2015arXiv150207751T} {}
  (\mn@eprint {arXiv} {1502.07751})

\bibitem[\protect\citeauthoryear{{Thomas}, {Muanwong}, {Pearce}, {Couchman},
  {Edge}, {Jenkins}  \& {Onuora}}{{Thomas} et~al.}{2001}]{2001MNRAS.324..450T}
{Thomas} P.~A.,  {Muanwong} O.,  {Pearce} F.~R.,  {Couchman} H.~M.~P.,  {Edge}
  A.~C.,  {Jenkins} A.,   {Onuora} L.,  2001, \mn@doi [MNRAS]
  {10.1046/j.1365-8711.2001.04330.x}, \href
  {http://adsabs.harvard.edu/abs/2001MNRAS.324..450T} {324, 450}

\bibitem[\protect\citeauthoryear{{Vale} \& {Ostriker}}{{Vale} \&
  {Ostriker}}{2006}]{2006MNRAS.371.1173V}
{Vale} A.,  {Ostriker} J.~P.,  2006, \mn@doi [MNRAS]
  {10.1111/j.1365-2966.2006.10605.x}, \href
  {http://adsabs.harvard.edu/abs/2006MNRAS.371.1173V} {371, 1173}

\bibitem[\protect\citeauthoryear{{Velander} et~al.}{{Velander}
  et~al.}{2014}]{2014MNRAS.437.2111V}
{Velander} M.,  et~al., 2014, \mn@doi [MNRAS] {10.1093/mnras/stt2013}, \href
  {http://adsabs.harvard.edu/abs/2014MNRAS.437.2111V} {437, 2111}

\bibitem[\protect\citeauthoryear{{Velliscig}, {van Daalen}, {Schaye},
  {McCarthy}, {Cacciato}, {Le Brun}  \& {Dalla Vecchia}}{{Velliscig}
  et~al.}{2014}]{2014MNRAS.442.2641V}
{Velliscig} M.,  {van Daalen} M.~P.,  {Schaye} J.,  {McCarthy} I.~G.,
  {Cacciato} M.,  {Le Brun} A.~M.~C.,   {Dalla Vecchia} C.,  2014, \mn@doi
  [\mnras] {10.1093/mnras/stu1044}, \href
  {http://adsabs.harvard.edu/abs/2014MNRAS.442.2641V} {442, 2641}

\bibitem[\protect\citeauthoryear{{Velliscig} et~al.,}{{Velliscig}
  et~al.}{2017}]{2017MNRAS.471.2856V}
{Velliscig} M.,  et~al., 2017, \mn@doi [\mnras] {10.1093/mnras/stx1789}, \href
  {http://adsabs.harvard.edu/abs/2017MNRAS.471.2856V} {471, 2856}

\bibitem[\protect\citeauthoryear{{Viola} et~al.,}{{Viola}
  et~al.}{2015}]{2015MNRAS.452.3529V}
{Viola} M.,  et~al., 2015, \mn@doi [\mnras] {10.1093/mnras/stv1447}, \href
  {http://adsabs.harvard.edu/abs/2015MNRAS.452.3529V} {452, 3529}

\bibitem[\protect\citeauthoryear{{Vogelsberger} et~al.,}{{Vogelsberger}
  et~al.}{2014a}]{2014MNRAS.444.1518V}
{Vogelsberger} M.,  et~al., 2014a, \mn@doi [MNRAS] {10.1093/mnras/stu1536},
  \href {http://adsabs.harvard.edu/abs/2014MNRAS.444.1518V} {444, 1518}

\bibitem[\protect\citeauthoryear{{Vogelsberger} et~al.,}{{Vogelsberger}
  et~al.}{2014b}]{2014Natur.509..177V}
{Vogelsberger} M.,  et~al., 2014b, \mn@doi [Nature] {10.1038/nature13316},
  \href {http://adsabs.harvard.edu/abs/2014Natur.509..177V} {509, 177}

\bibitem[\protect\citeauthoryear{{Wang}, {White}, {Mandelbaum}, {Henriques},
  {Anderson}  \& {Han}}{{Wang} et~al.}{2016}]{2016MNRAS.456.2301W}
{Wang} W.,  {White} S.~D.~M.,  {Mandelbaum} R.,  {Henriques} B.,  {Anderson}
  M.~E.,   {Han} J.,  2016, \mn@doi [MNRAS] {10.1093/mnras/stv2809}, \href
  {http://adsabs.harvard.edu/abs/2016MNRAS.456.2301W} {456, 2301}

\bibitem[\protect\citeauthoryear{{White} \& {Frenk}}{{White} \&
  {Frenk}}{1991}]{1991ApJ...379...52W}
{White} S.~D.~M.,  {Frenk} C.~S.,  1991, \mn@doi [\apj] {10.1086/170483}, \href
  {http://adsabs.harvard.edu/abs/1991ApJ...379...52W} {379, 52}

\bibitem[\protect\citeauthoryear{{Wilson}, {Kaiser}, {Luppino}  \&
  {Cowie}}{{Wilson} et~al.}{2001}]{2001ApJ...555..572W}
{Wilson} G.,  {Kaiser} N.,  {Luppino} G.~A.,   {Cowie} L.~L.,  2001, \mn@doi
  [\apj] {10.1086/321441}, \href
  {http://adsabs.harvard.edu/abs/2001ApJ...555..572W} {555, 572}

\bibitem[\protect\citeauthoryear{{Wright} \& {Brainerd}}{{Wright} \&
  {Brainerd}}{2000}]{2000ApJ...534...34W}
{Wright} C.~O.,  {Brainerd} T.~G.,  2000, \mn@doi [ApJ] {10.1086/308744}, \href
  {http://adsabs.harvard.edu/abs/2000ApJ...534...34W} {534, 34}

\bibitem[\protect\citeauthoryear{{Zel'dovich}}{{Zel'dovich}}{1970}]{1970A&A.....5...84Z}
{Zel'dovich} Y.~B.,  1970, \aap, \href
  {http://adsabs.harvard.edu/abs/1970A%26A.....5...84Z} {5, 84}

\bibitem[\protect\citeauthoryear{{Zu} \& {Mandelbaum}}{{Zu} \&
  {Mandelbaum}}{2015}]{2015MNRAS.454.1161Z}
{Zu} Y.,  {Mandelbaum} R.,  2015, \mn@doi [MNRAS] {10.1093/mnras/stv2062},
  \href {http://adsabs.harvard.edu/abs/2015MNRAS.454.1161Z} {454, 1161}

\bibitem[\protect\citeauthoryear{van Daalen, Schaye, McCarthy, Booth  \&
  Vecchia}{van Daalen et~al.}{2014}]{vanDaalen:2013ita}
van Daalen M.~P.,  Schaye J.,  McCarthy I.~G.,  Booth C.,   Vecchia C.~D.,
  2014, \mn@doi [MNRAS] {10.1093/mnras/stu482}, 440, 2997

\bibitem[\protect\citeauthoryear{{van den Bosch}, {Tormen}  \& {Giocoli}}{{van
  den Bosch} et~al.}{2005}]{2005MNRAS.359.1029V}
{van den Bosch} F.~C.,  {Tormen} G.,   {Giocoli} C.,  2005, \mn@doi [MNRAS]
  {10.1111/j.1365-2966.2005.08964.x}, \href
  {http://adsabs.harvard.edu/abs/2005MNRAS.359.1029V} {359, 1029}

\makeatother
\end{thebibliography}

%%%%%%%%%%%%%%%%%%%%%%%%%%%%%%%%%%%%%%%%%%%%%%%%%%

%%%%%%%%%%%%%%%%% APPENDICES %%%%%%%%%%%%%%%%%%%%%

\appendix

\section{Impact of radial binning and field residual variances for $\Delta \Sigma$ profiles}
\label{sec:radialBinningImpact}

\begin{figure}
	\includegraphics[width=\columnwidth]{\figrelpath 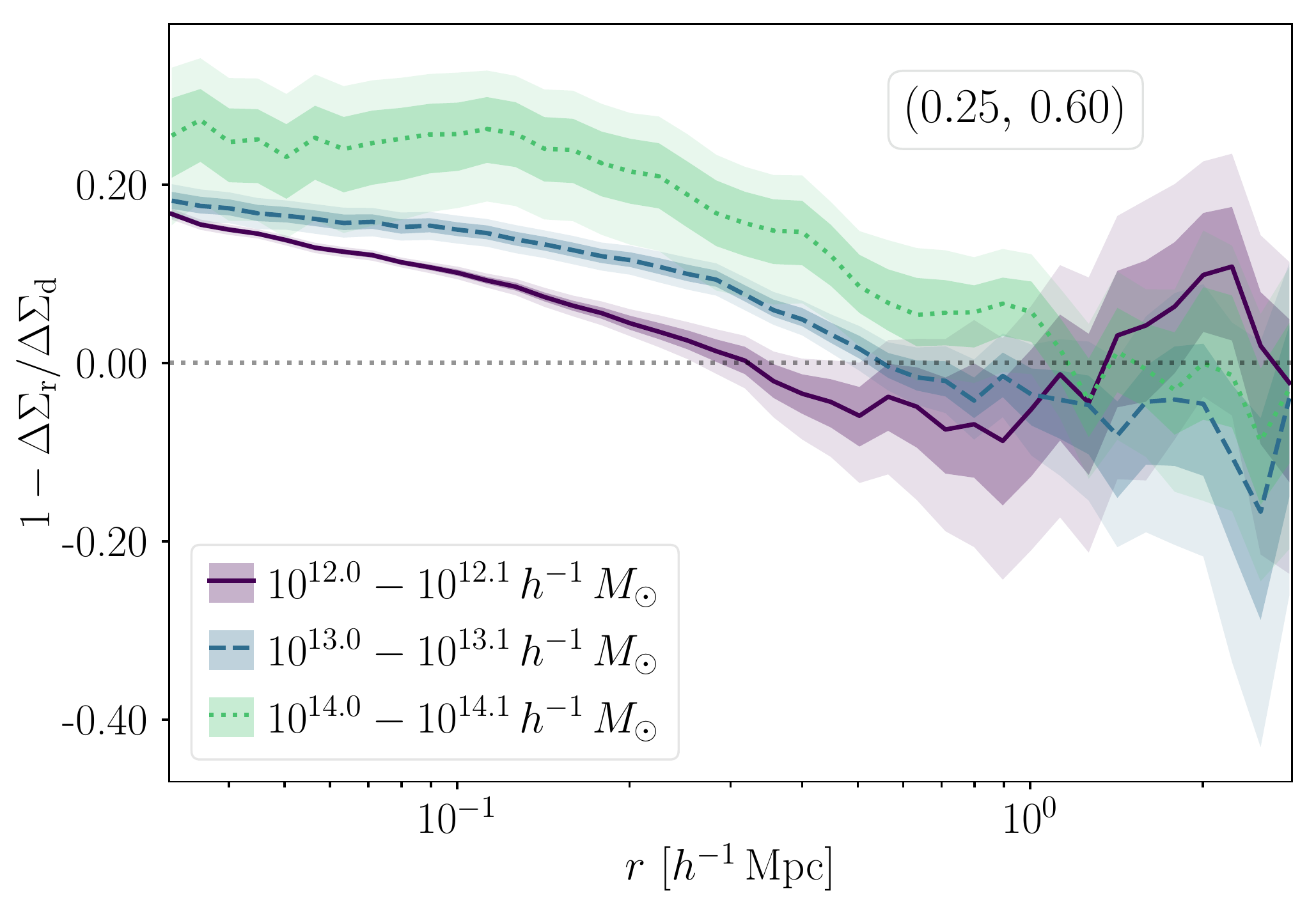}
    \caption{Residuals from three different mass bins' $\Delta \Sigma$ profiles for $(0.25, \, 0.60)$ in the rescaled simulation w.r.t. the direct simulation.}
    \label{fig:deltaSigmaSom025Residuals}
\end{figure}

\begin{figure}
	\includegraphics[width=\columnwidth]{\figrelpath 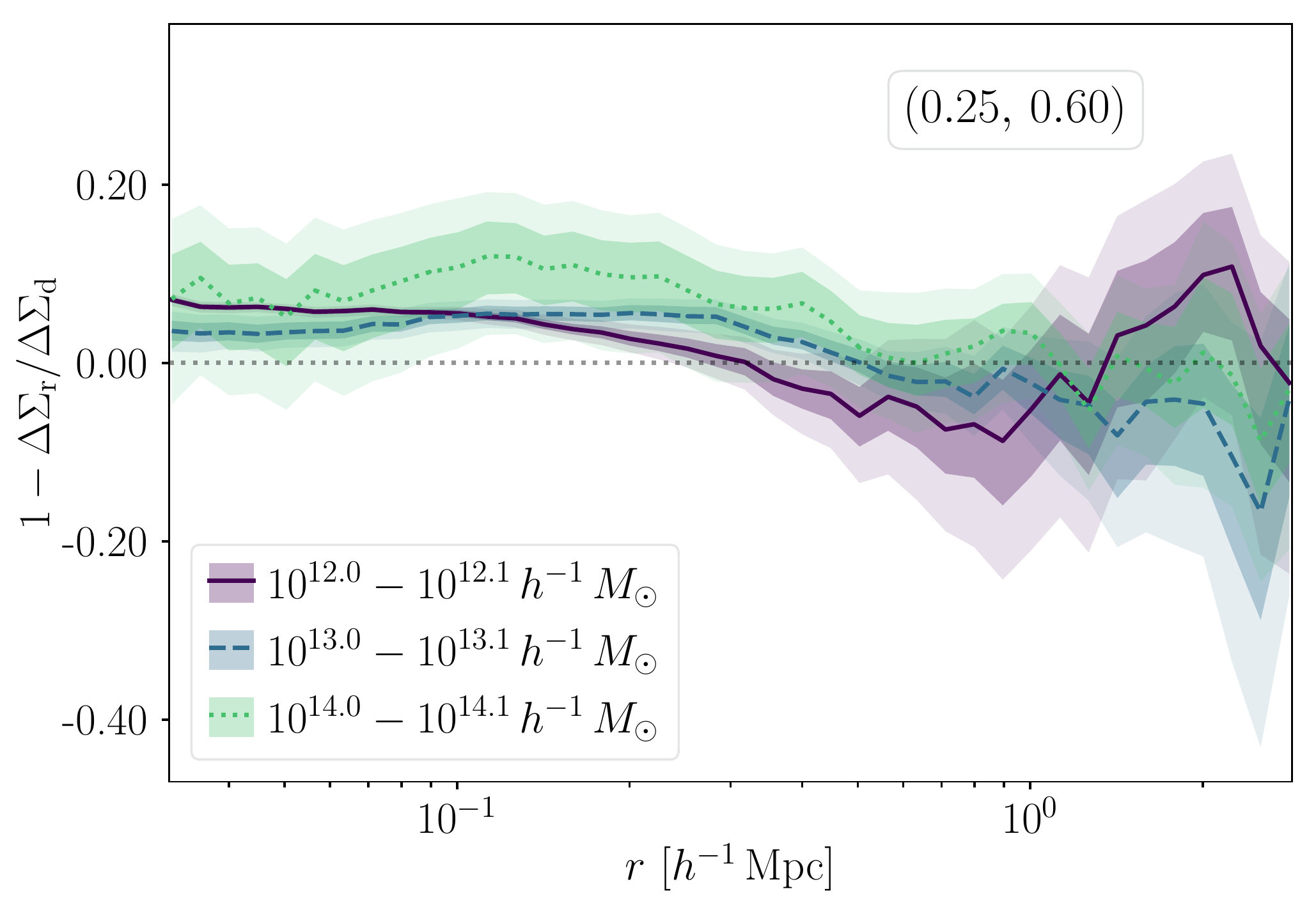}
    \caption{Residuals from concentration corrected $\Delta \Sigma$ profiles for $(0.25, \, 0.60)$.}
    \label{fig:deltaSigmaSom025ResidualsCorrected}
\end{figure}

The measured differences between direct and rescaled halo profiles presented in Section~\ref{sec:results} could depend on the radial binning. To investigate the impact of the bin width, we compute $\Delta \Sigma$ profiles with twice as many bins. For $\Delta \Sigma$, the new values for $( 0.15, \, 1.00)$ are $41\, \%$ and $32\, \%$ (pre-correction) and $15\, \%$ and $7.0\, \%$ (post-correction), which represent the largest differences owing to the lower resolution of this simulation. For $(0.29, \,  0.81)$, the differences increase to $7.0 \, \%$ and $2.6 \%$ (pre-correction) and $6.1 \, \%$ and $-1.2 \, \%$ (post-correction) which implies an increase with $1\,\%$ for the total maxima and less than $1\, \%$ for the median maximum values. For $( 0.25, \, 0.60)$ and $( 0.40, \, 0.70)$, the resulting changes are below or maximally 1\,\%. The same is true for $(0.80, \, 0.40)$, though the median maximum deviation changes signs to $-4.2\,\%$ post-correction.

Concerning the cosmic variances of these residuals, we plot the residuals from the bootstrapped profiles for $(0.25, \, 0.60)$ using all haloes in three mass bins in Figs.~\ref{fig:deltaSigmaSom025Residuals} and \ref{fig:deltaSigmaSom025ResidualsCorrected}, before and after applying concentration correction (the results are qualitatively the same for the other simulations). For galaxy and galaxy group class haloes, the spread in the differences in the inner regions are quite narrow and they widen as one approaches the 1-halo to 2-halo transition regime. For cluster size haloes, there is a larger variance in the inner regions which is both driven by poor statistics and the impact of unrelaxed systems. This is reflected in the spread in concentrations. Overall, the correction preserves the variance with slightly larger error bars for cluster mass haloes as the haloes are not necessarily matched in each bootstrapped stack w.r.t. one another.
\section{Results for $(0.80, \, 0.40)$}
\label{sec:resultsForAlmostEinsteinDeSitter}

\begin{figure}
	\includegraphics[width=\columnwidth]{\figrelpath 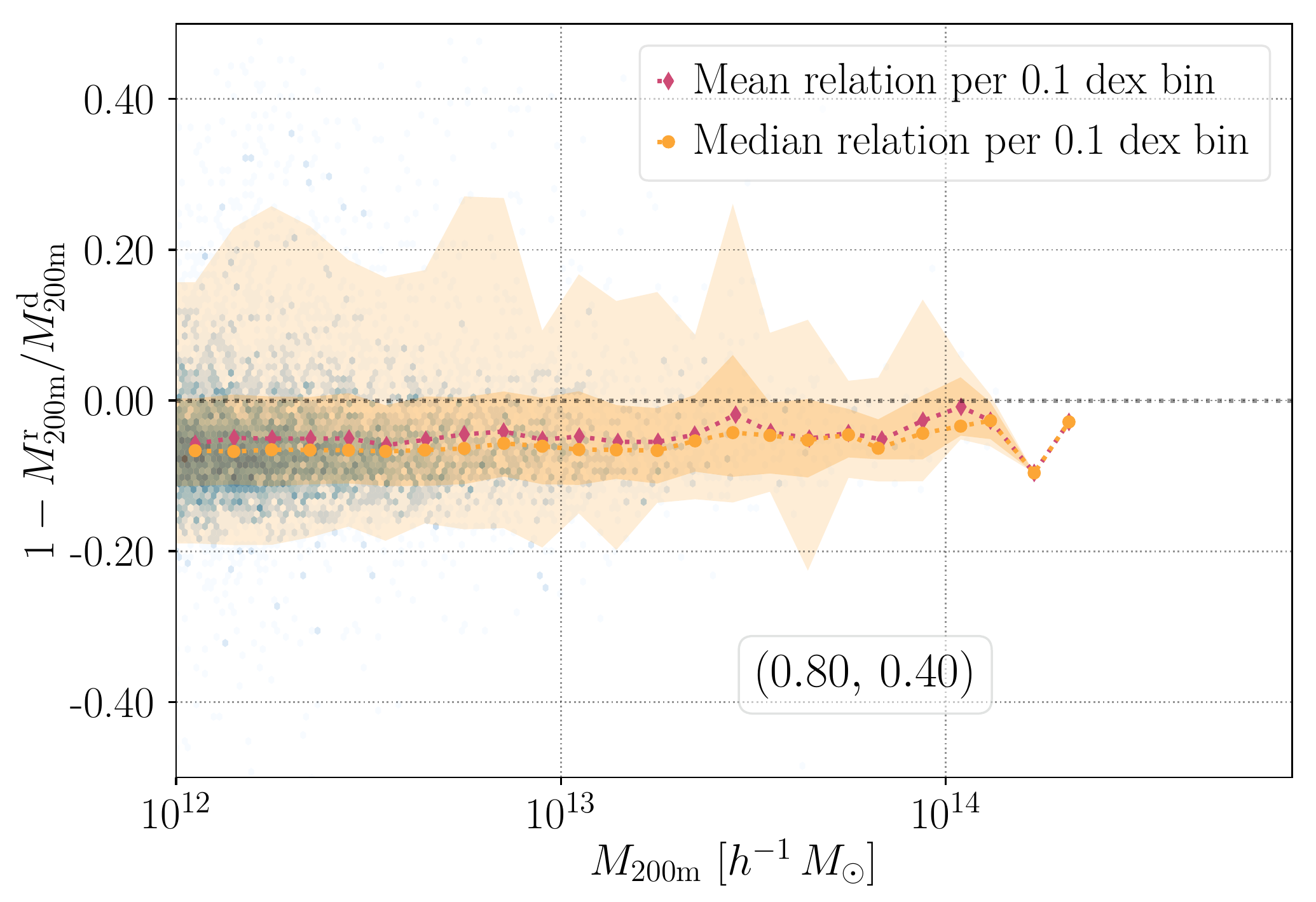}
    \caption{Mass bias for matched haloes in the $(0.80, \, 0.40)$ simulation. The rescaled haloes are consistently more massive than their counterparts in the direct simulation across the whole mass range, with some outliers among galaxy class haloes.}
    \label{fig:matchedHaloMassBiasSom08}
\end{figure}

\begin{figure}
	\includegraphics[width=\columnwidth]{\figrelpath 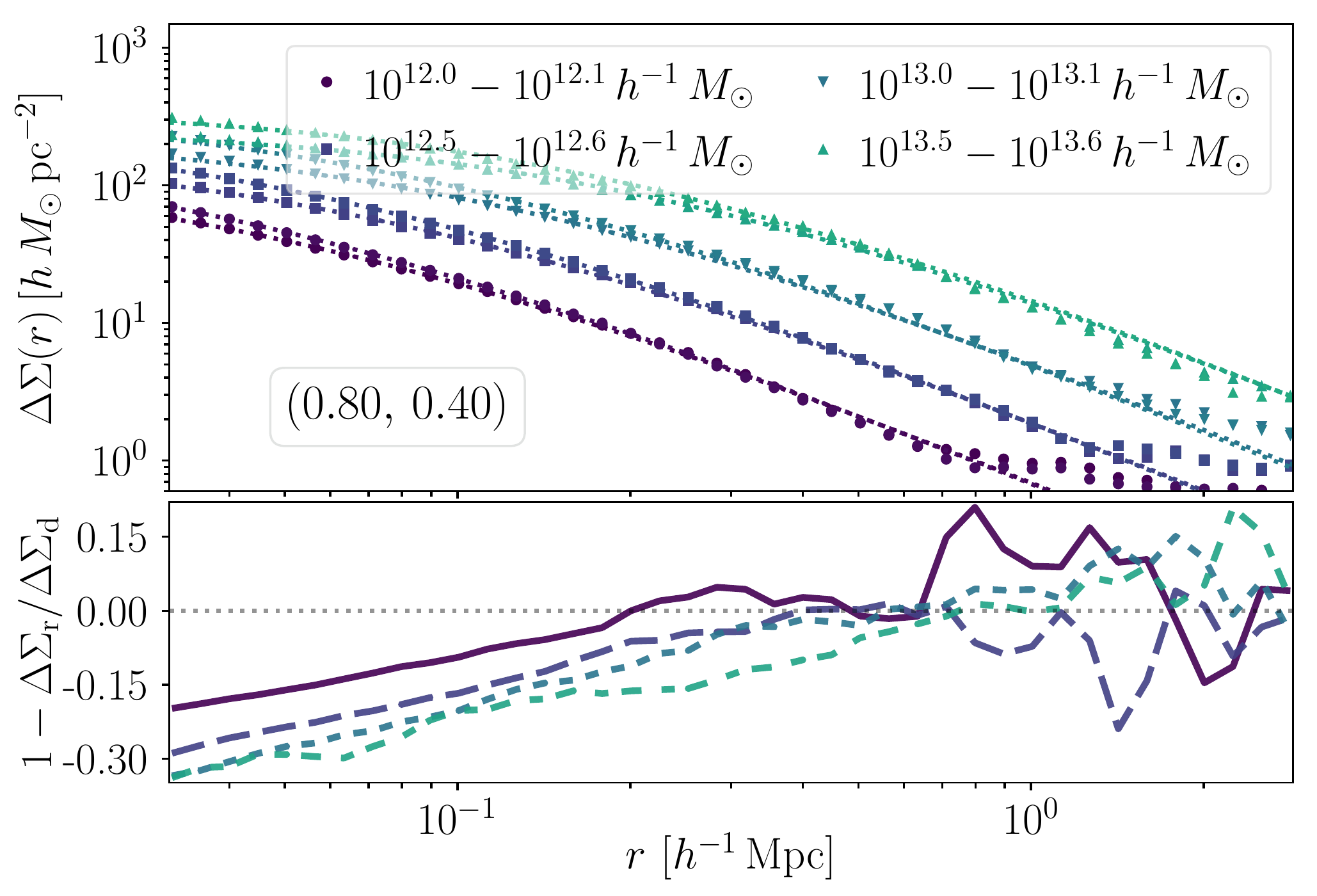}
    \caption{$\Delta \Sigma$ profiles for $(0.80, \, 0.40)$.}
    \label{fig:deltaSigmaSom08}
\end{figure}

\begin{figure}
	\includegraphics[width=\columnwidth]{\figrelpath 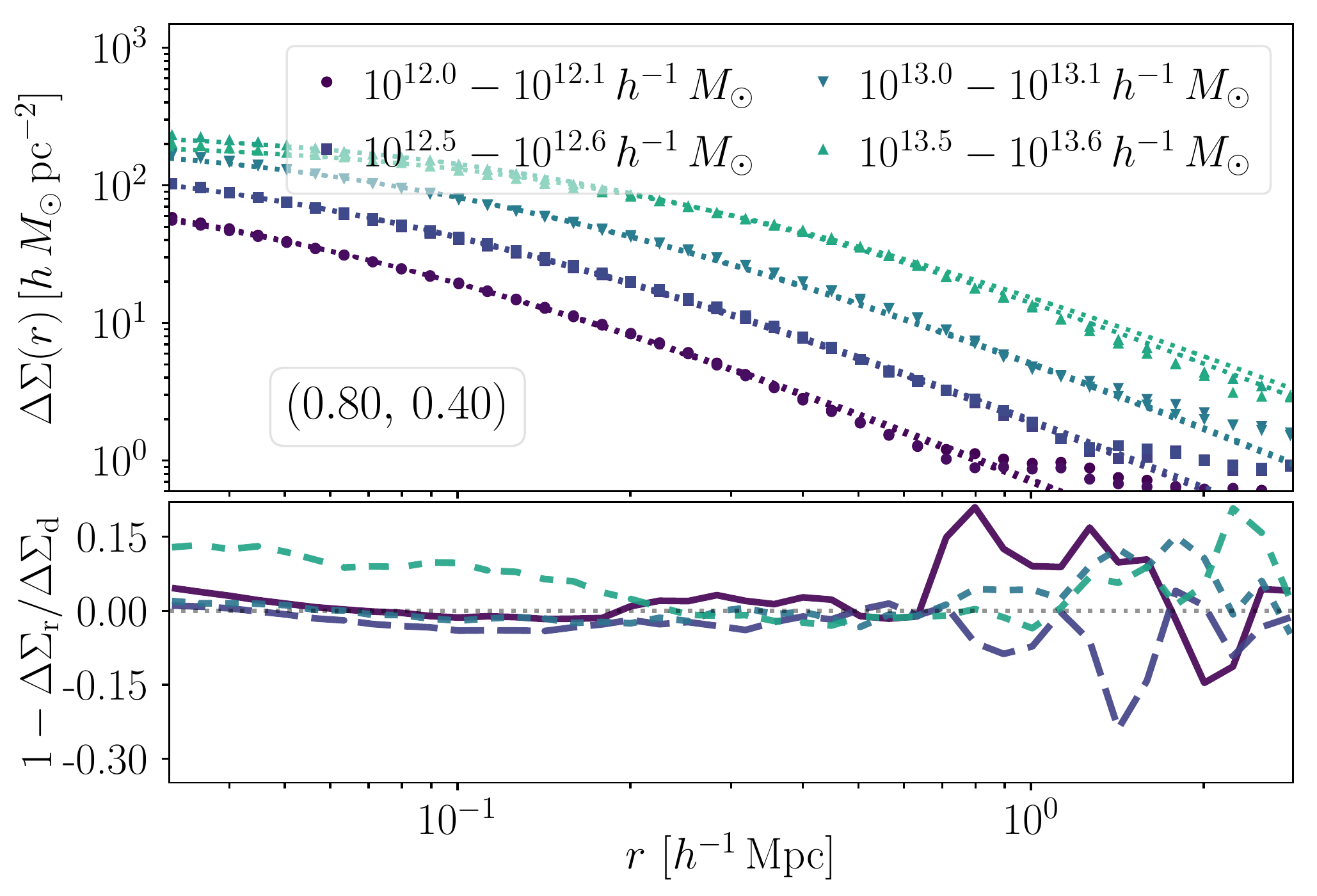}
    \caption{Concentration corrected $\Delta \Sigma$ profiles for $(0.80, \, 0.40)$.}
    \label{fig:correctedDeltaSigmaSom08}
\end{figure}

\begin{figure}
	\includegraphics[width=1.04\columnwidth]{\figrelpath 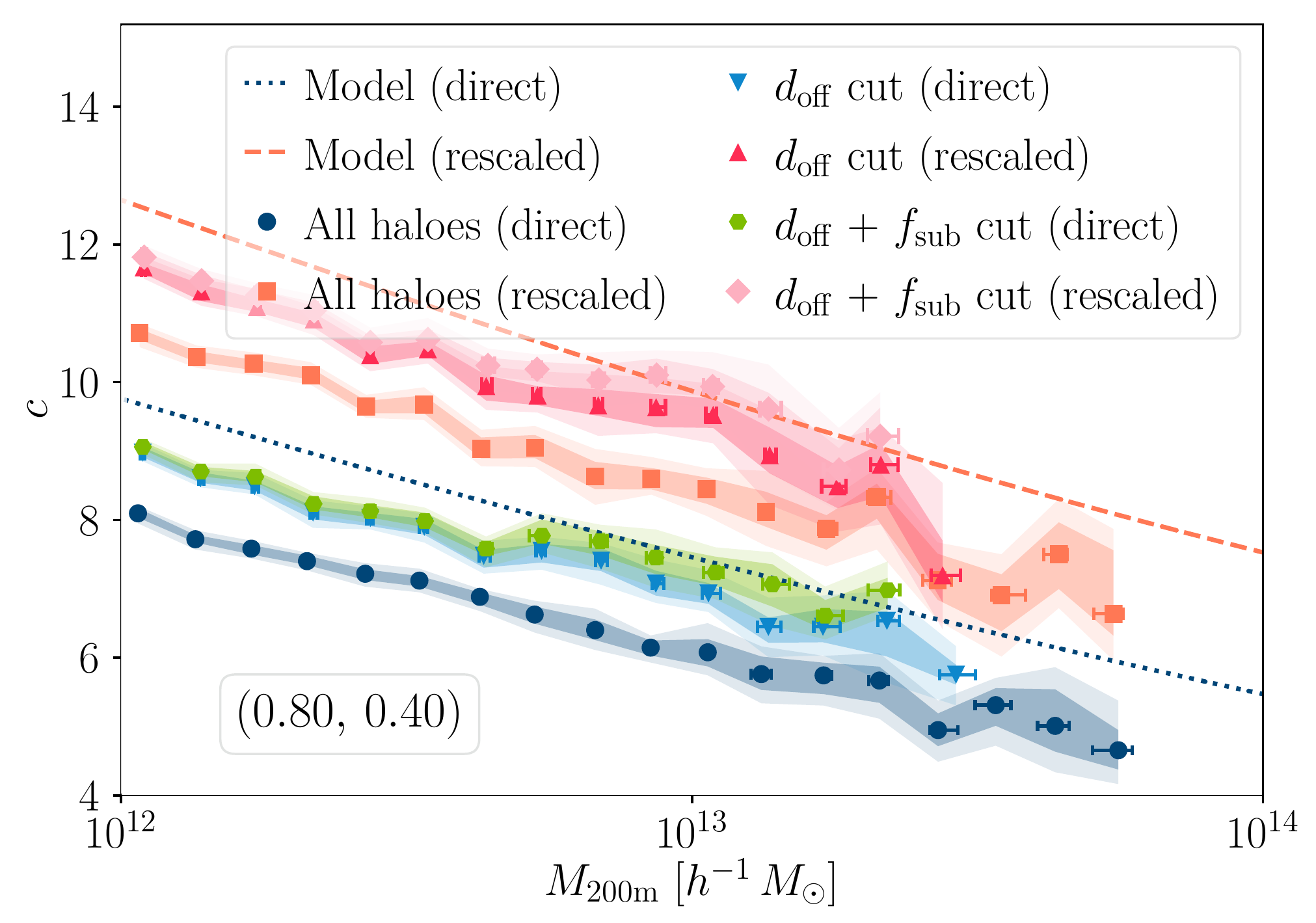}
    \caption{
    \label{fig:som08CMNFWWithCuts}
		NFW $c(M)$-relations for $(0.80, \, 0.40)$ for all haloes and with different relaxation cuts enforced.
		}
\end{figure}

The almost Einstein-de Sitter cosmology represents our most extreme sample, and its cosmological parameters deviate strongly from what is favoured by observations. The masses differ substantially between the matched haloes in the direct and rescaled simulation, see Fig.~\ref{fig:matchedHaloMassBiasSom08}, with haloes in the rescaled simulation on average more massive. In Fig.~\ref{fig:deltaSigmaSom08}, we show the measured $\Delta \Sigma$ profiles together with the fitted NFW lens profiles and in Fig.~\ref{fig:correctedDeltaSigmaSom08} the profiles post-correction. Due to the small volume of the simulation as listed in Table~\ref{tab:simulationTable}, we do not have any mass bins beyond $10^{14} \, h^{-1} \, M_{\sun}$ with more than twenty haloes in both the direct and rescaled snapshot. Since the amplitude of the 2-halo term is directly proportional to the matter fraction of the Universe, its influence kicks in at smaller scales than for the other simulations. The inner profile bias is negative and can be quantified as $\Delta c \approx - 2$ as seen in Figs.~\ref{fig:concentrationDifference} and \ref{fig:som08CMNFWWithCuts} where we plot the 3D density profile NFW $c(M)$-relations. The Einasto $c(M)$-relations, see Fig.~\ref{fig:som08CMEinastoWithCuts}, perform slightly better at the low mass end w.r.t. the \citetalias{2016MNRAS.460.1214L} predictions.

\section{Matched halo results}
\label{sec:matchedHaloes}

\begin{figure*}
\begin{centering}
\includegraphics[width=2.1\columnwidth]{\figrelpath 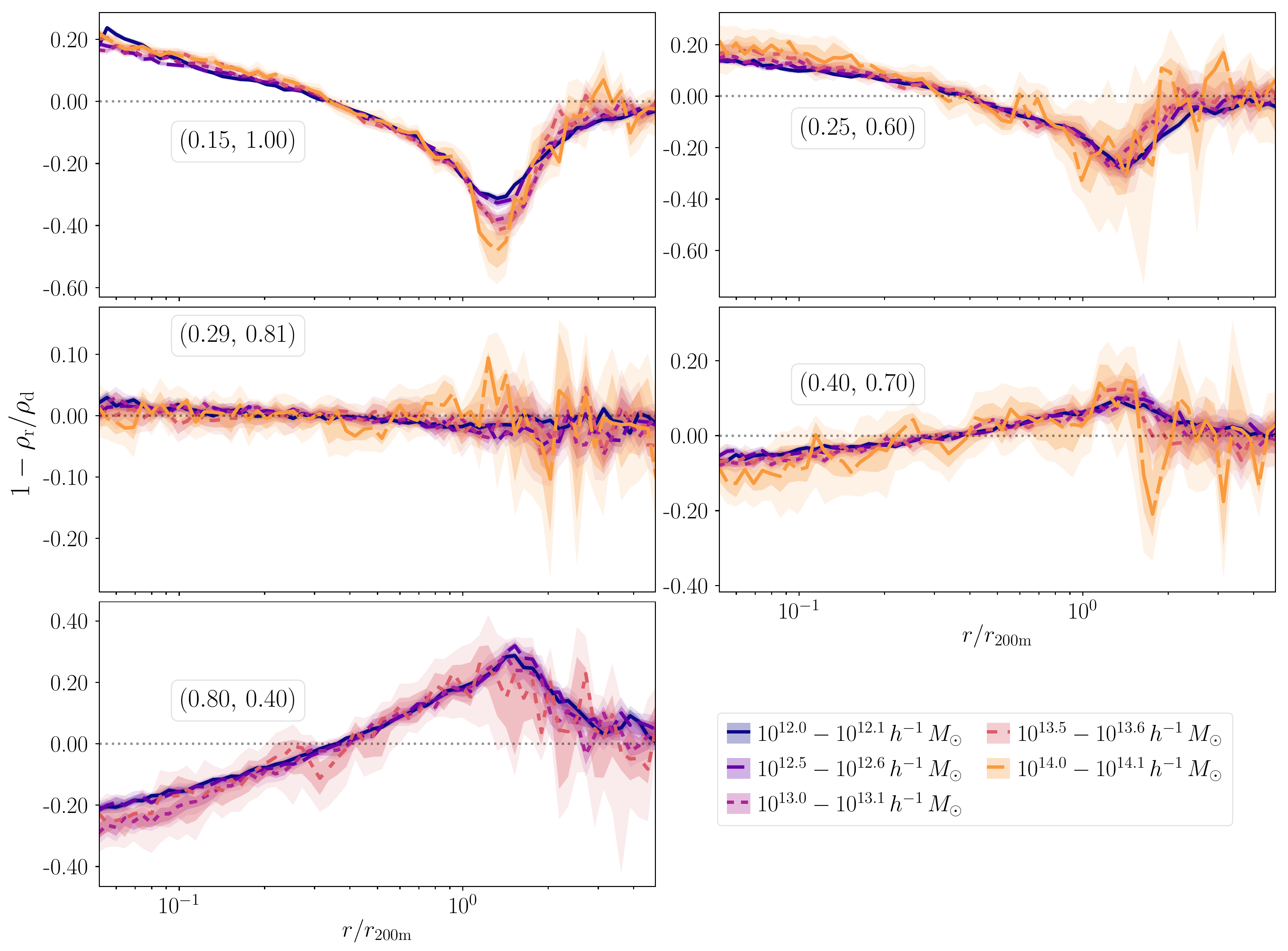}

\caption{Matched halo density field residuals from 64 $\log$-equidistant radial bins with error regions from bootstrapped stacks of matched haloes in each simulation. These error regions trace the median results well.}
\label{fig:som0RhoMatchedHalos}
\end{centering}
\end{figure*}

\begin{figure*}
\begin{centering}
\large
\stackon[5pt]{
\includegraphics[width=1.02\columnwidth]{\figrelpath 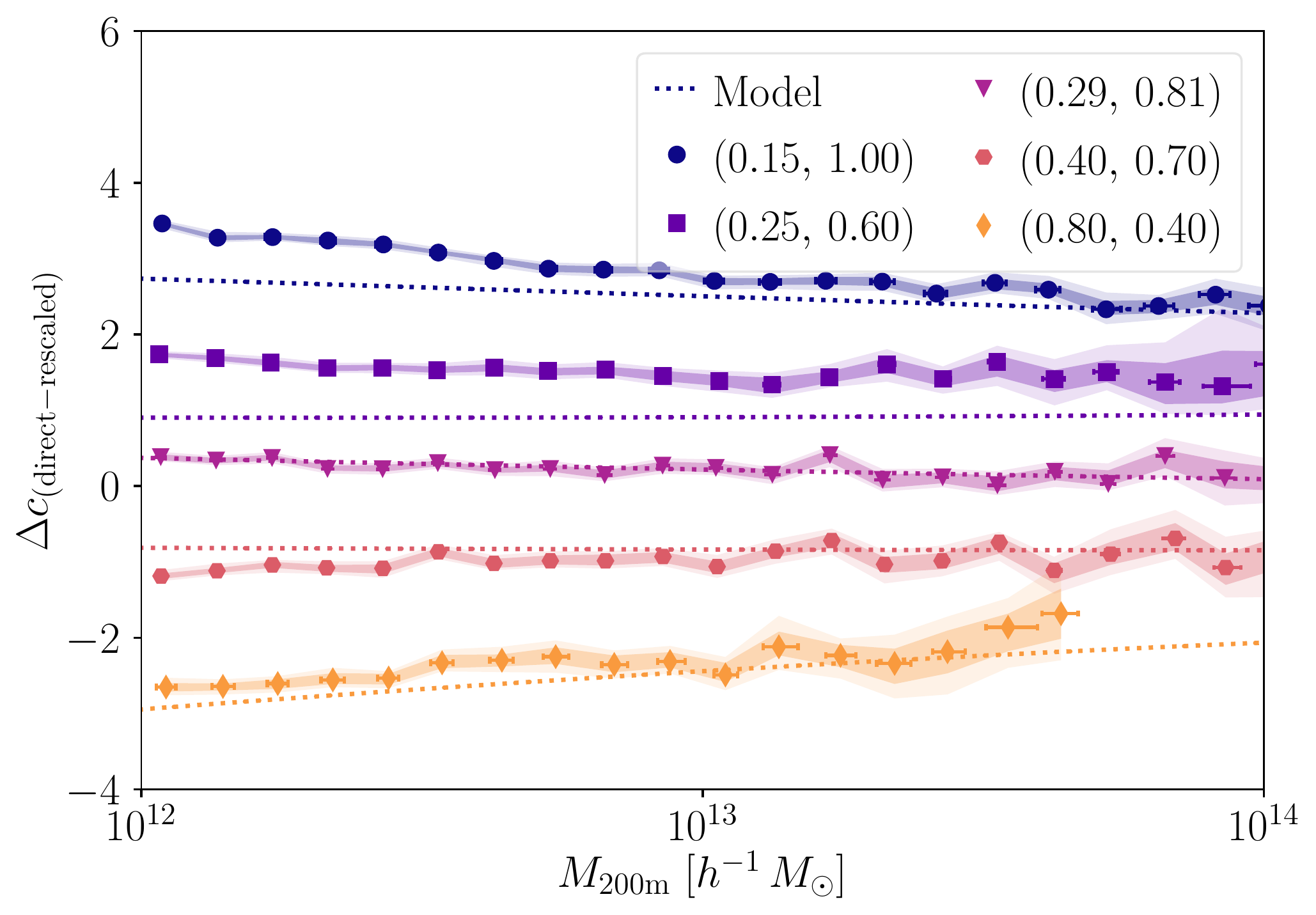}}{{\bf Matched haloes: before correction}}
\stackon[5pt]{
\includegraphics[width=1.02\columnwidth]{\figrelpath 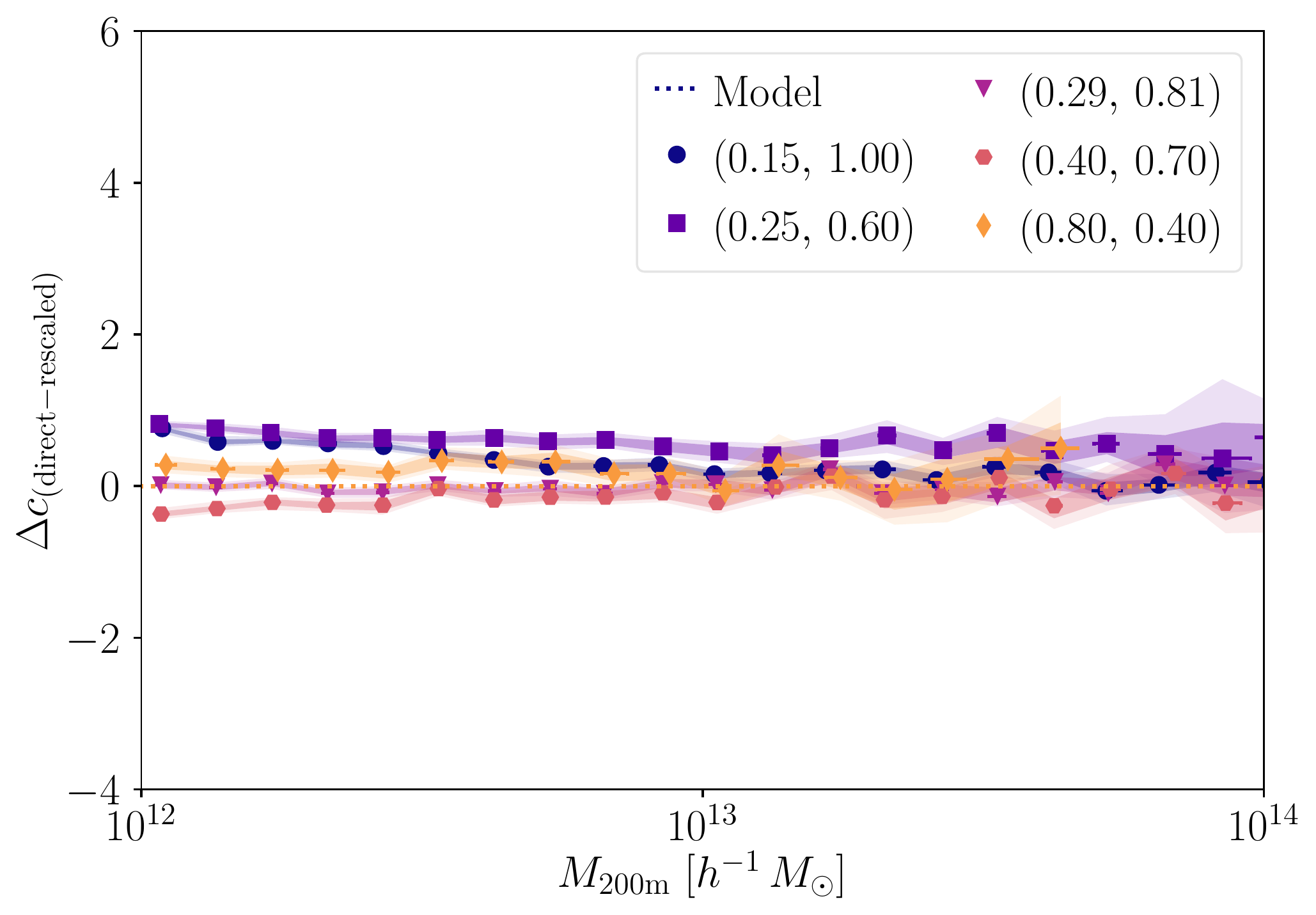}}{{\bf Matched haloes: after correction}}
\caption{
\label{fig:correctedConcentrationDifferenceMatchedHaloes}
Difference in concentration estimated from density profiles (as in the left panel of Fig.~\ref{fig:concentrationDifference}) for matched haloes before and after applying our correction to rescaled haloes.
}
\end{centering}
\end{figure*}

In Fig.~\ref{fig:som0RhoMatchedHalos} we show the fractional differences in the median density profiles between matched haloes in the direct and rescaled simulations binned according to the mass in the direct run for all test cosmologies. The error regions are calculated from comparing the median differences between the same bootstrapped matched haloes in the direct and rescaled simulations. With respect to the differences shown in Fig.~\ref{fig:som0Rho}, the two biases are slightly more discernible, especially the outer profile bias and the (small) concentration bias for $(0.29, \, 0.81)$. Re-sampling the matched population for each mass bin yields similar results. For all cosmologies and mass bins the profile bias changes signs at $\approx 0.3 - 0.4 \, r/r_{200\text{m}}$ which was also observed previously for all haloes. The median $\Delta c$-biases for these matched haloes are illustrated in Fig.~\ref{fig:correctedConcentrationDifferenceMatchedHaloes} where the error regions are computed from bootstrap resamples of the same matched haloes in the direct and rescaled simulations. 

\begin{figure*}
\begin{centering}
\includegraphics[width=1.\columnwidth]{\figrelpath 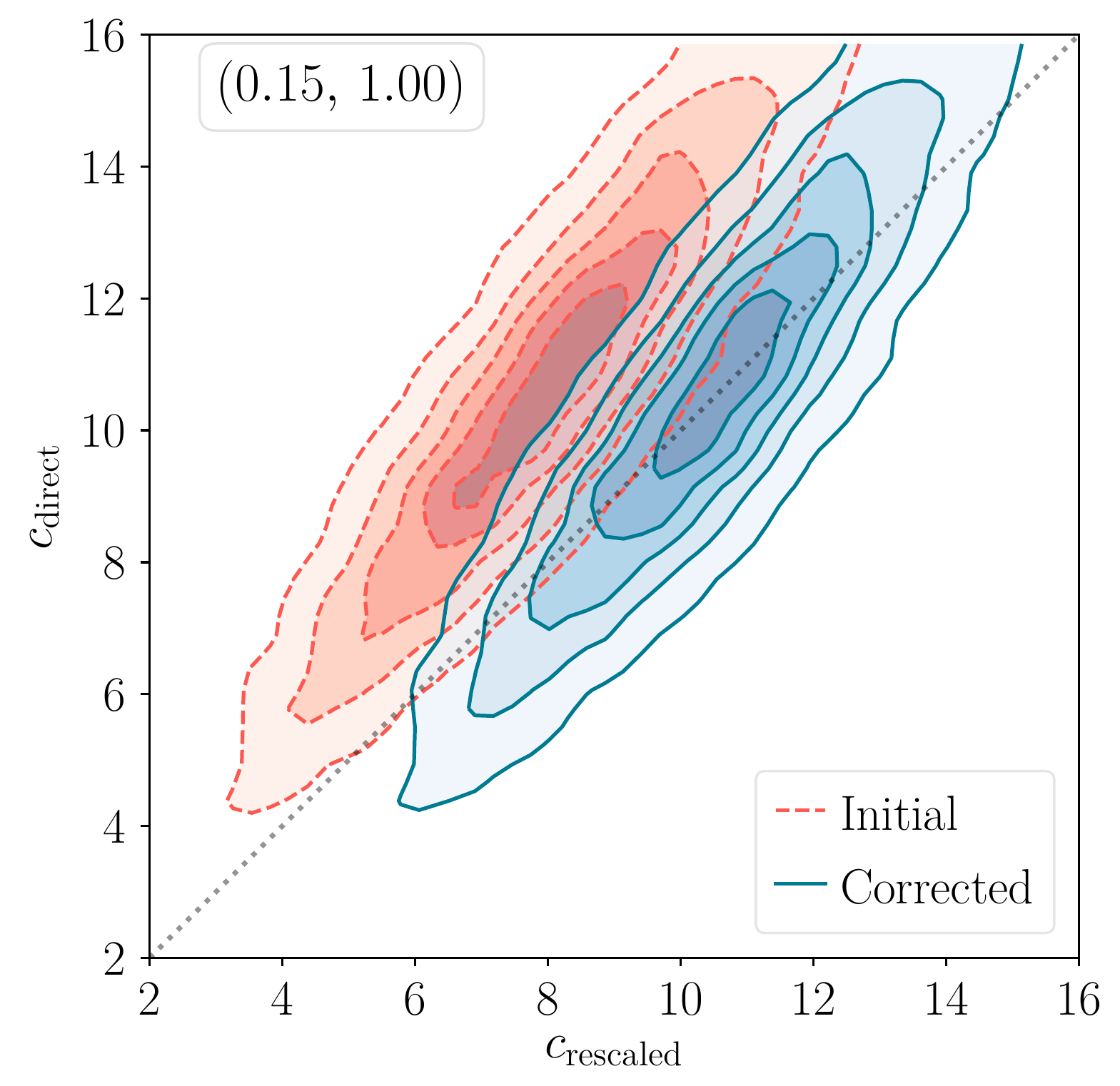}
\includegraphics[width=1.\columnwidth]{\figrelpath 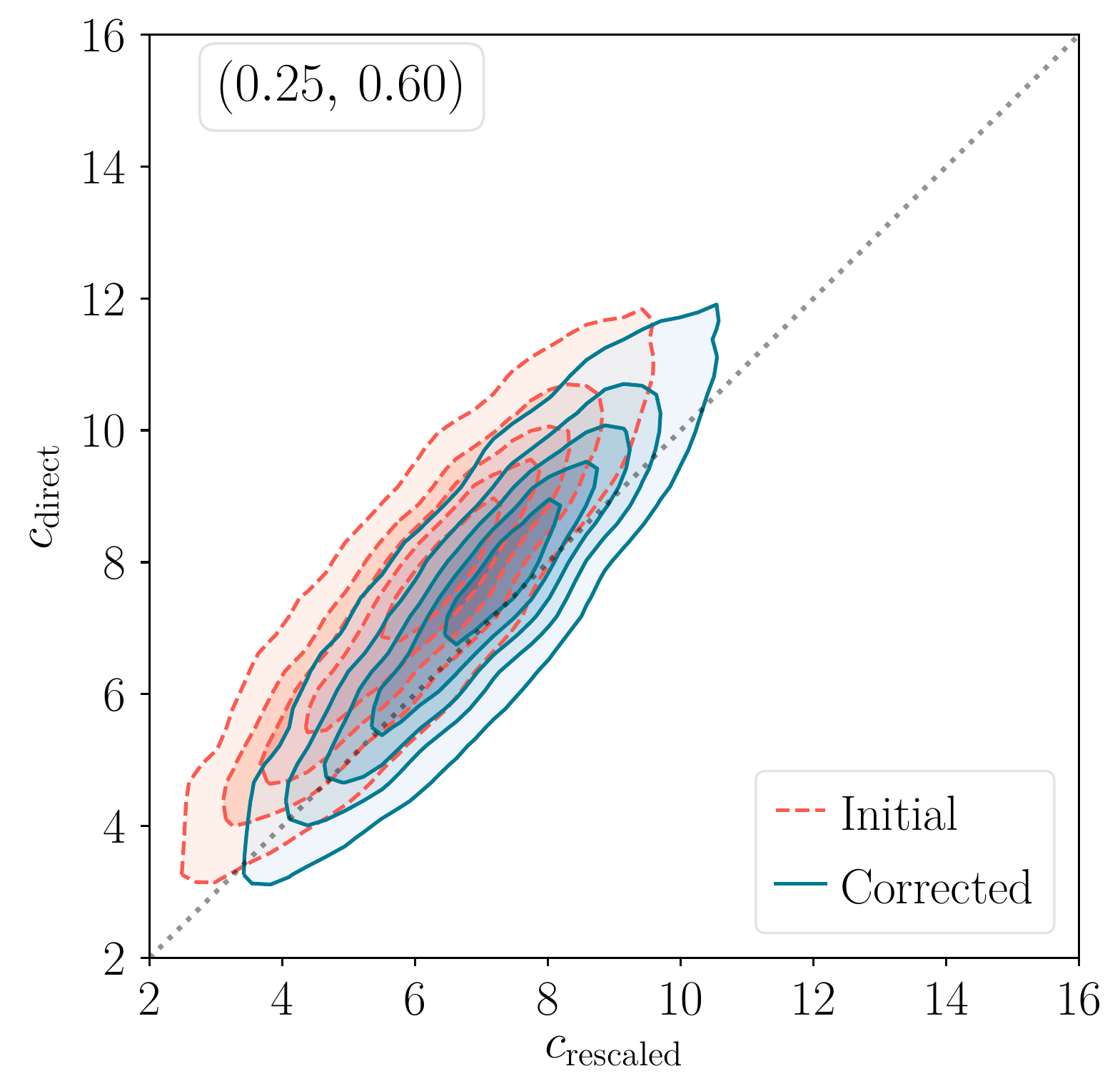}
\includegraphics[width=1.\columnwidth]{\figrelpath 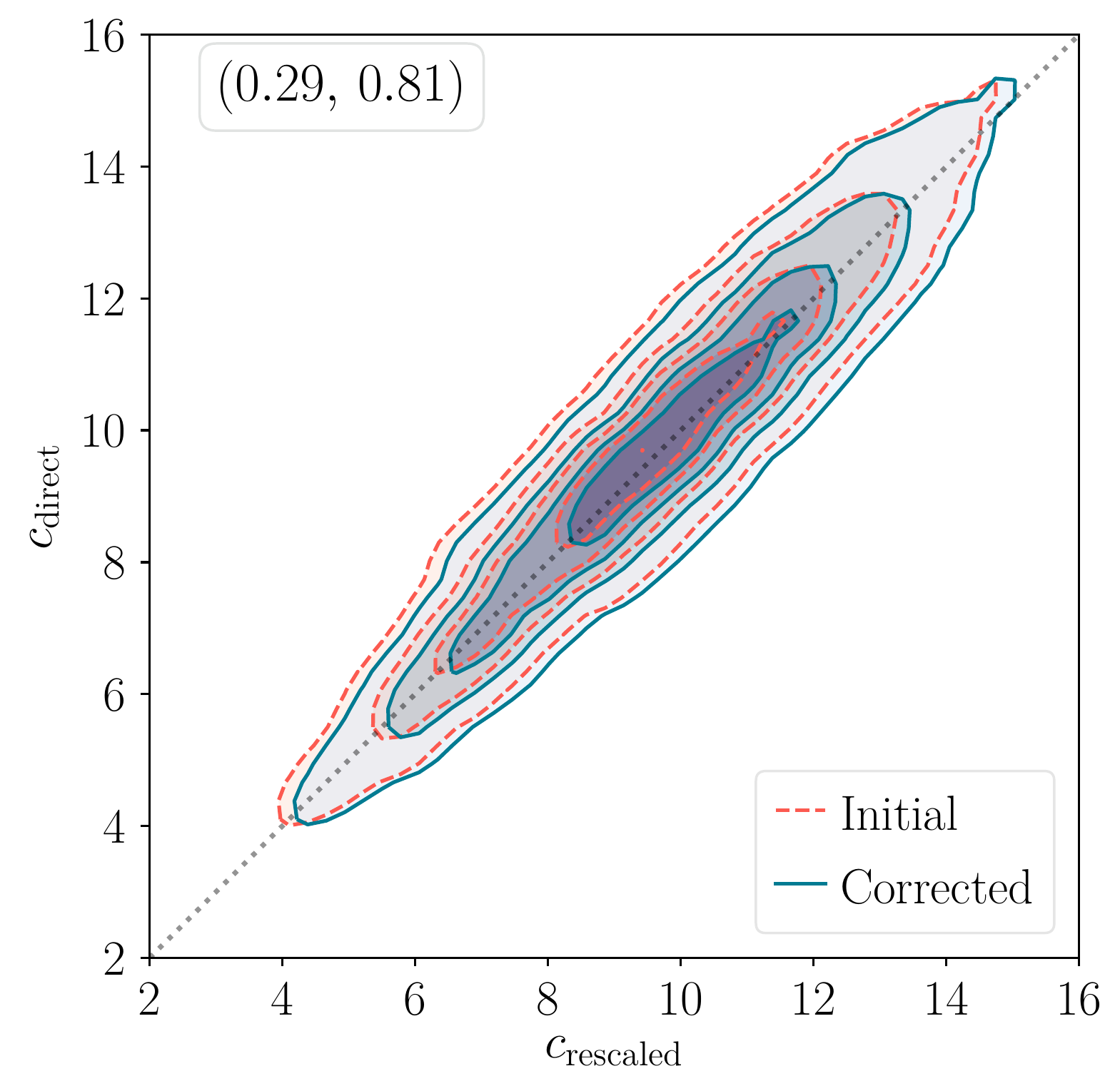}
\includegraphics[width=1.\columnwidth]{\figrelpath 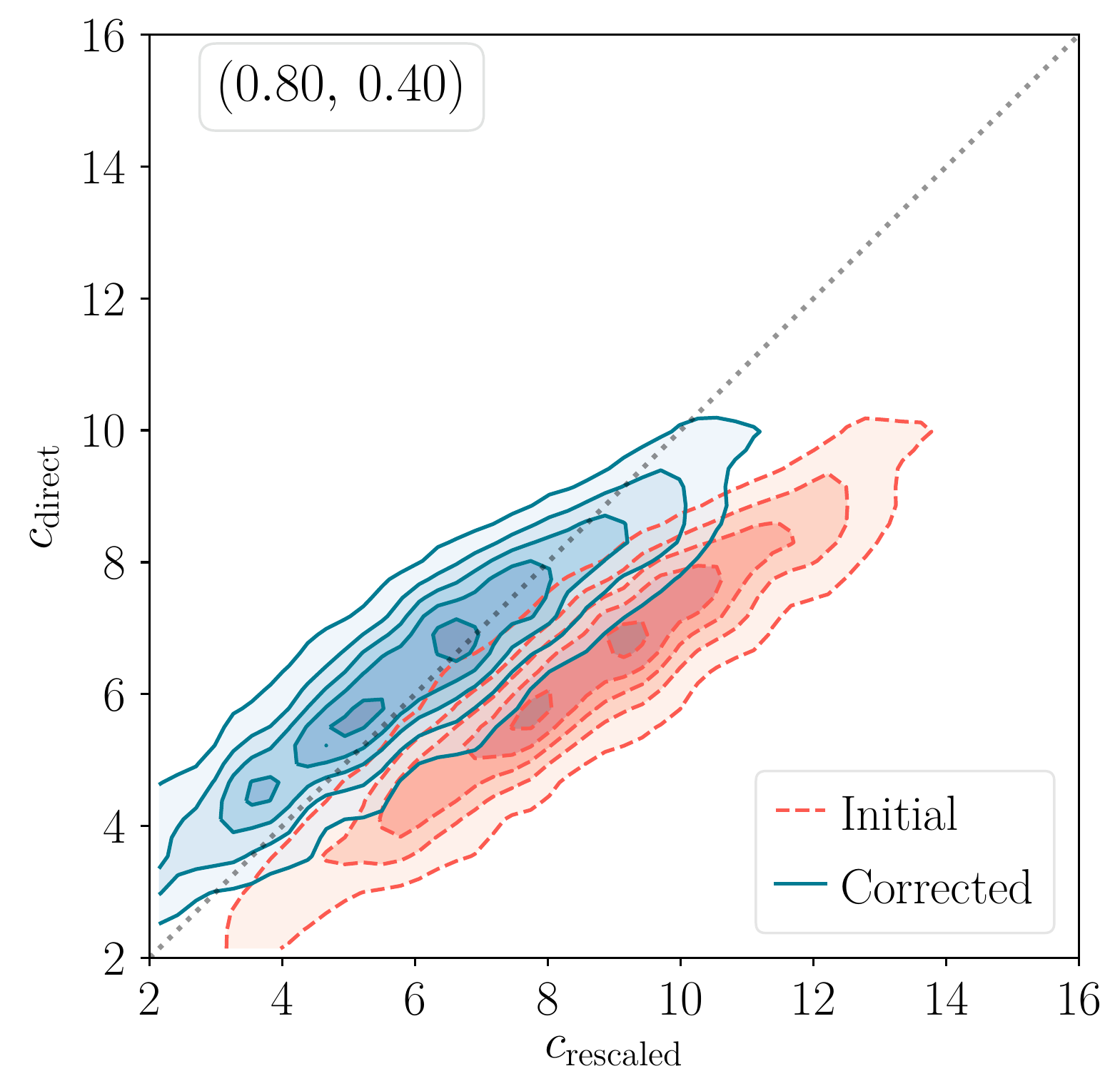}
\caption{Concentration difference for matched haloes quantified with 3D NFW profiles, pre- and post-correction. Note that the lower mass threshold in the direct simulation for $(0.15, \, 1.00)$ is $ M_{200\text{m}} >10^{12.7} \, h^{-1} \, M_{\sun}$ instead of $ >10^{12.5} \, h^{-1} \, M_{\sun}$ for the others to mitigate resolution effects.}
\label{fig:correctedConcentrationDifferenceIndHaloes}
\end{centering}
\end{figure*}

In Fig.~\ref{fig:correctedConcentrationDifferenceIndHaloes} we plot the individual concentration relations in the direct and rescaled simulation for all cosmologies except for $(0.40, \, 0.70)$ which was already shown in Fig.~\ref{fig:som04IndHaloDensityFieldCEstimatesPrePostcorrection} with the same setup. We only correct the profiles if the fitted $c + \Delta c > 0$. This chiefly affects massive haloes in the $(0.80, \, 0.40)$ simulation and it has a negligible impact on the shape of the contours. The concentration correction induces a translation towards the diagonal but rotations are required for $(0.15, \, 1.00)$ and $(0.80, \, 0.40)$ to bring about agreement. Slight rotational adjustments might improve the concordance for $(0.25, \, 0.60)$ and $(0.29, \, 0.81)$. For $(0.25, \, 0.60)$, a larger translation correction is required. Imposing relaxation cuts and demanding that haloes pass them in both simulations does not affect the tilt of the distributions, but removes low concentration haloes as expected.

\section{Einasto concentrations}
\label{sec:einastoConcentrations}

\begin{figure}
	\includegraphics[width=\columnwidth]{\figrelpath 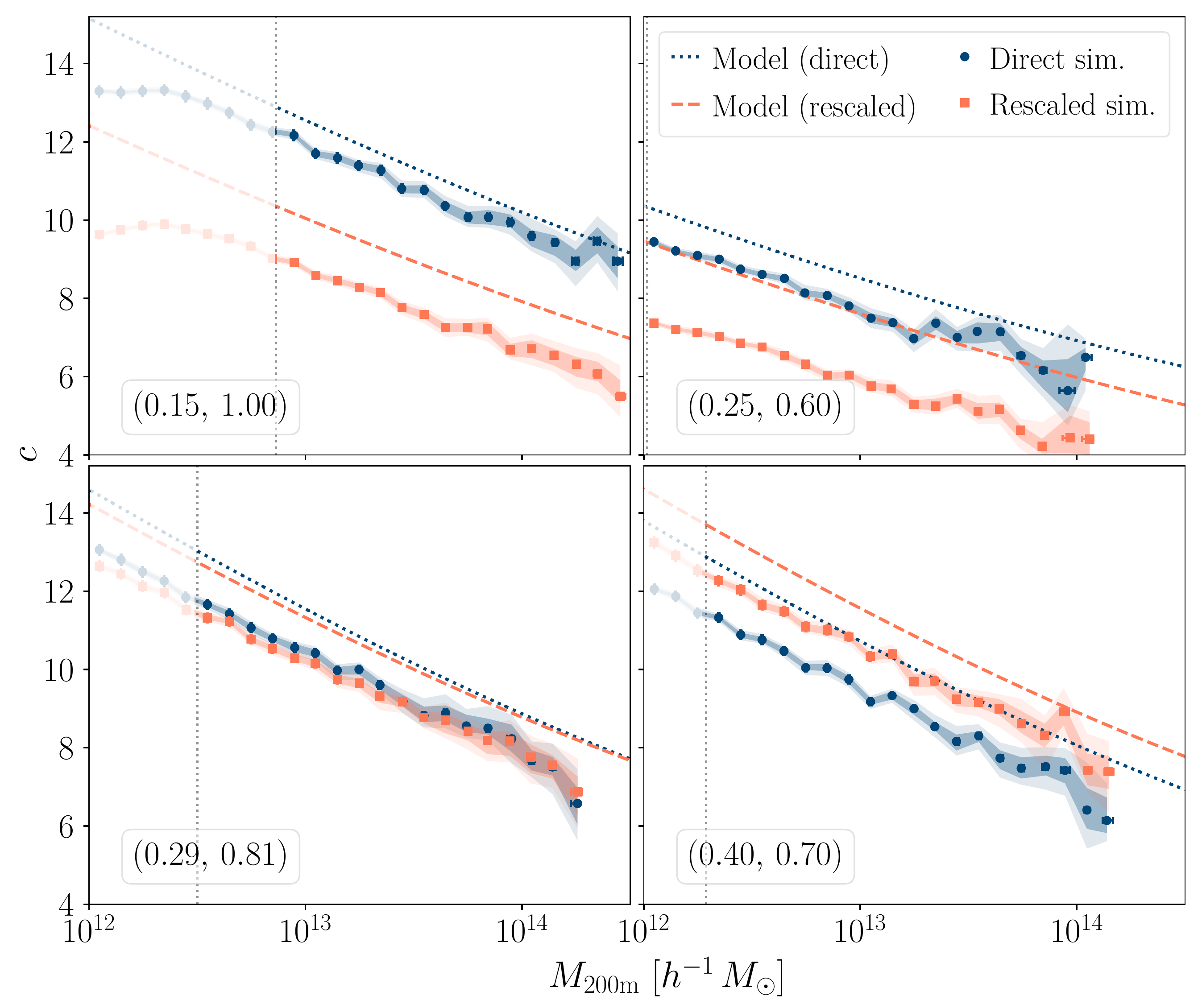}
    \caption{Concentration-mass relations for Einasto fits with $\alpha = 0.18$ for direct and rescaled simulations w.r.t. the \citetalias{2016MNRAS.460.1214L} model predictions.}
    \label{fig:3DMeanMassEinasto}
\end{figure}

\begin{figure}
	\includegraphics[width=1.04\columnwidth]{\figrelpath 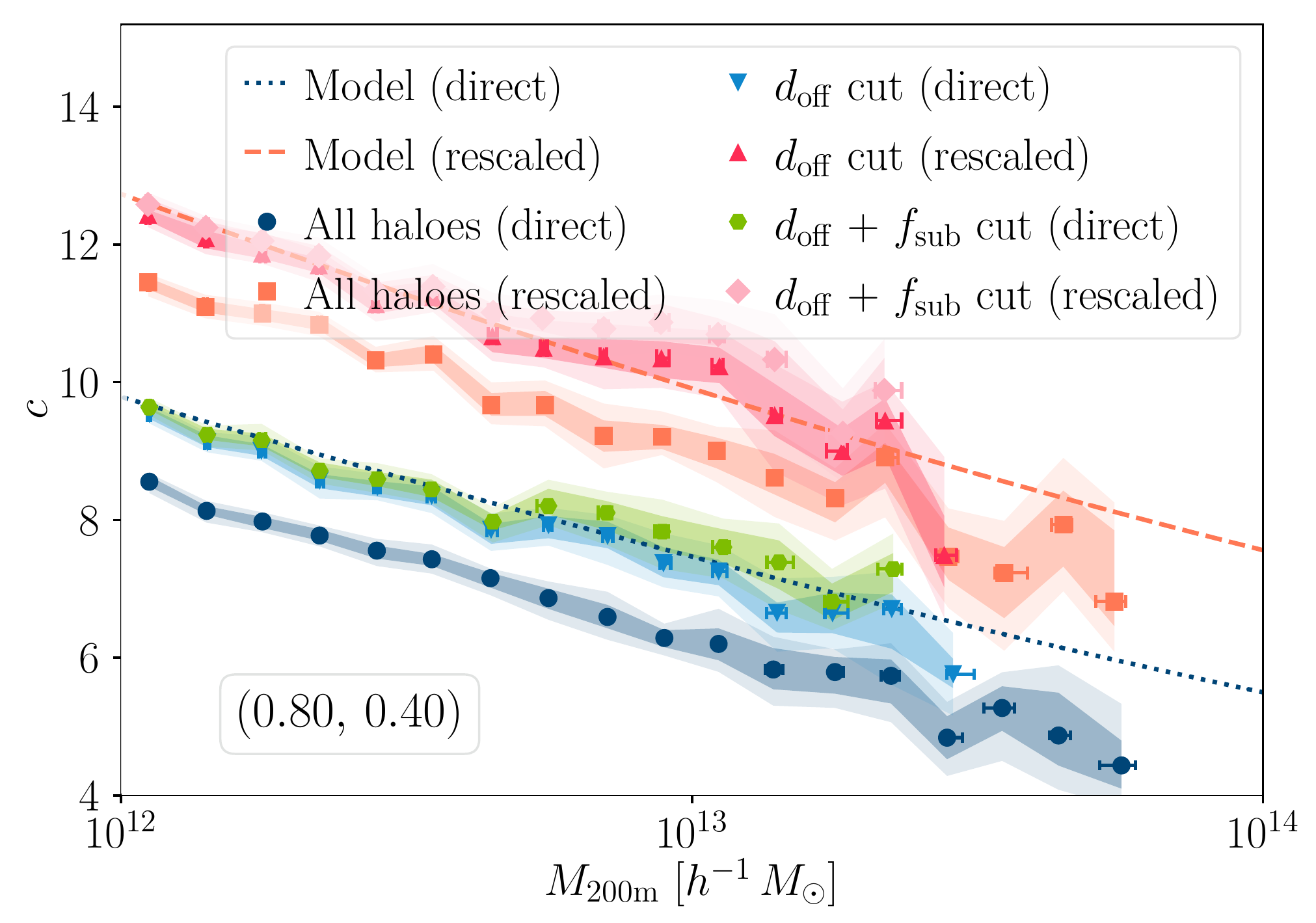}
    \caption{
    \label{fig:som08CMEinastoWithCuts}
		Einasto $c(M)$-relations for $(0.80, \, 0.40)$ for all haloes and with different relaxation cuts enforced. The corresponding NFW relations are similar though the Einasto measurements correspond better to the theory values at the low mass end.
		}
\end{figure}

\begin{figure}
	\includegraphics[width=\columnwidth]{\figrelpath 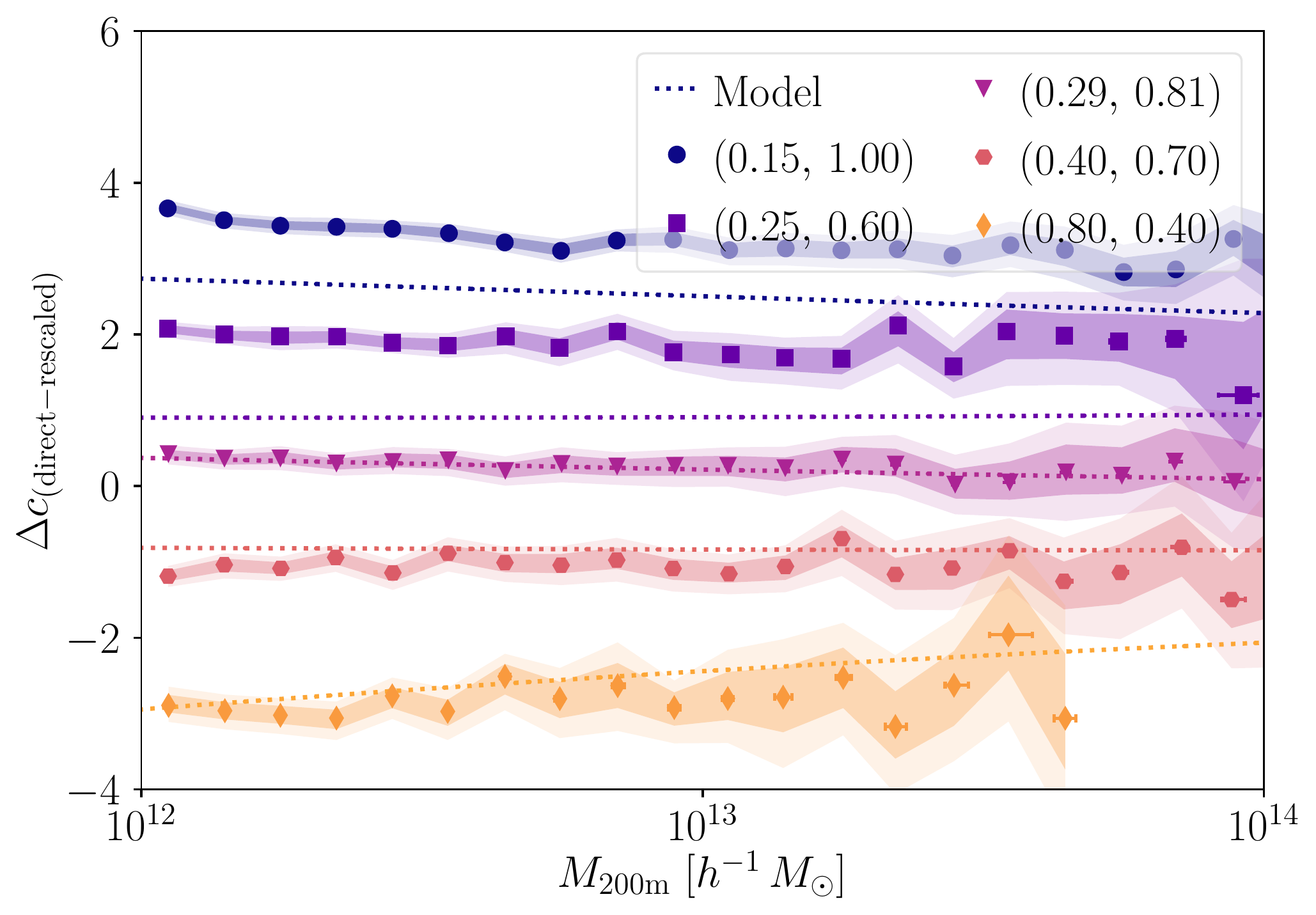}
    \caption{The measured differences for Einasto concentrations with $\alpha = 0.18$ and $r_\text{s}$ and $\rho_\text{s}$ free.}
    \label{fig:deltaCEinasto}
\end{figure}

\begin{figure}
	\includegraphics[width=\columnwidth]{\figrelpath 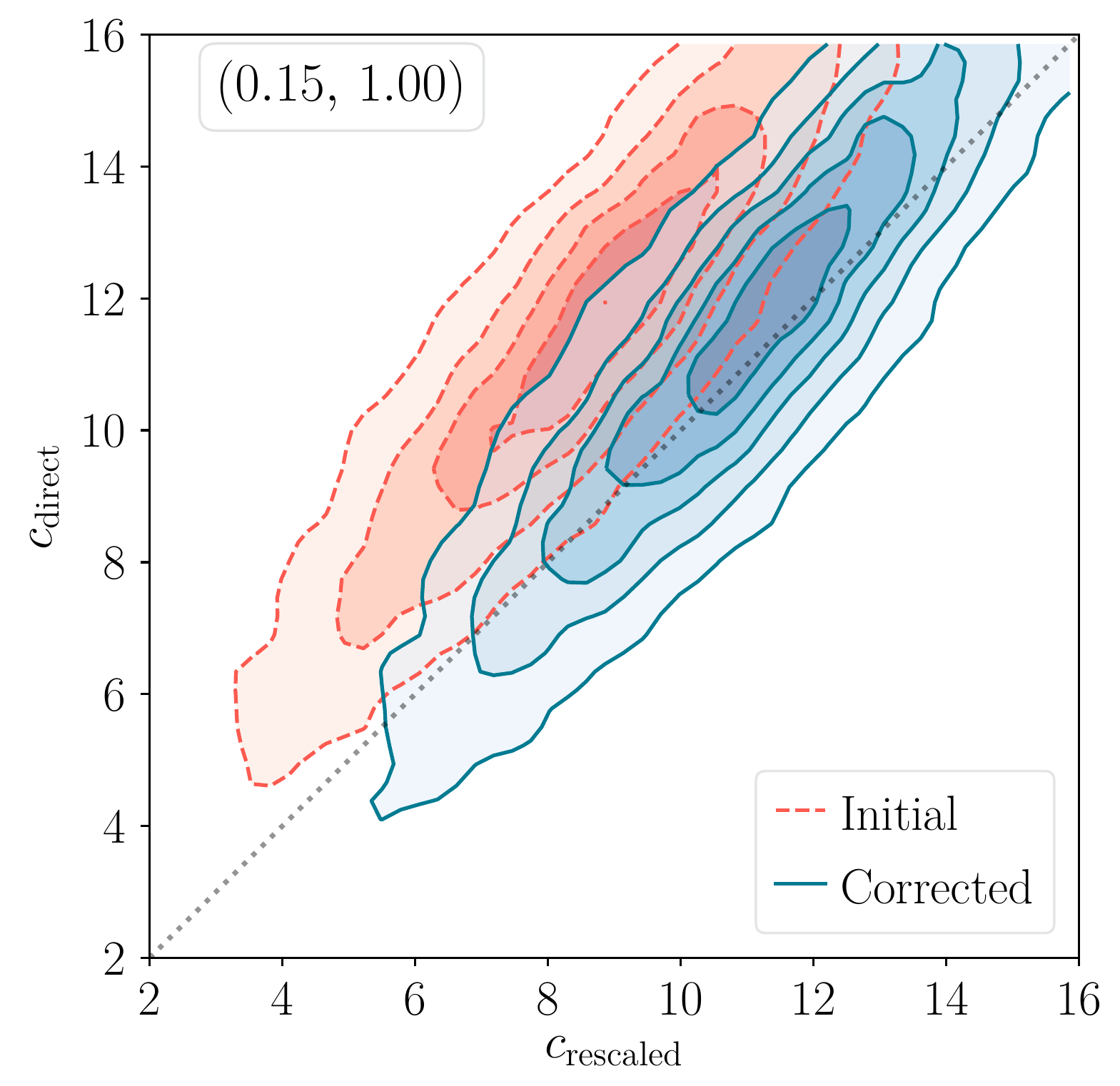}
    \caption{Einasto estimated concentrations for matched haloes in the direct and rescaled simulation with $ M_{200\text{m}} >10^{12.7} \, h^{-1} \, M_{\sun}$ for haloes in the direct simulation.}
    \label{fig:som015CDirectRescaledEinasto}
\end{figure}

\begin{figure}
	\includegraphics[width=\columnwidth]{\figrelpath 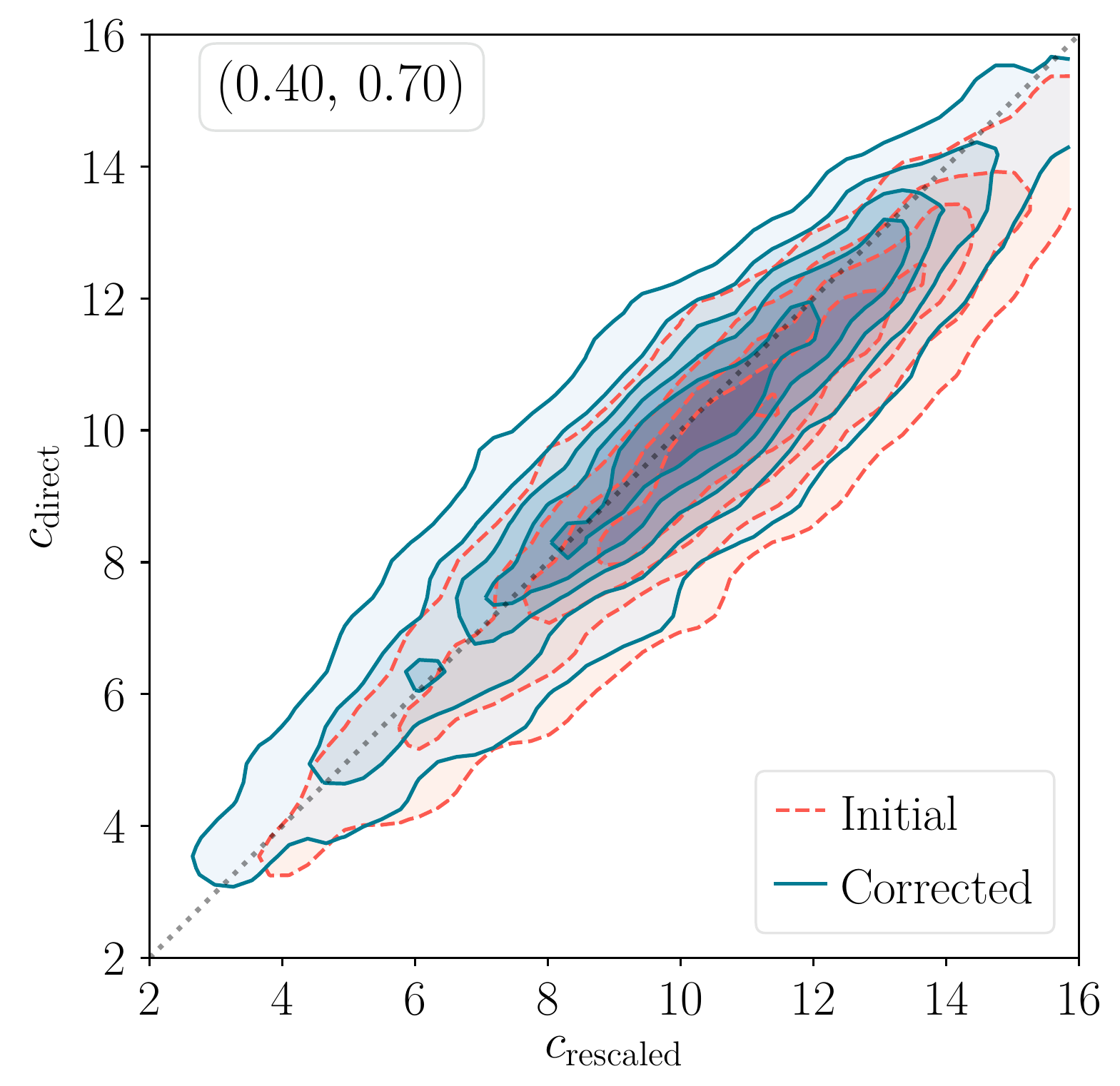}
    \caption{Einasto estimated concentrations for matched haloes in the direct and rescaled simulation with $ M_{200\text{m}} >10^{12.5} \, h^{-1} \, M_{\sun}$ for haloes in the direct simulation.}
    \label{fig:som04CDirectRescaledEinasto}
\end{figure}

In Figs.~\ref{fig:3DMeanMassEinasto} and \ref{fig:som08CMEinastoWithCuts} we plot the measured $c(M)$-relations and relaxation cut impacts for an Einasto parametrisation of the density field, and in Fig.~\ref{fig:deltaCEinasto} the corresponding $\Delta c$ biases. To compute the rescaling mappings we rephrase the density profile in Eq.~\eqref{eq:einastoDensityProfile} in terms of the average density $\expVal{\rho_\text{Einasto}}(r)$ for the enclosed mass $M( < r)$:
\begin{equation}
\label{eq:einastoProfileEnclosedDensity}
\expVal{\rho_\text{Einasto}}(r) = \frac{M( < r)}{4\pi/3 \, r^3} = \frac{\Delta}{y^3} \frac{\gamma(3/\alpha; \, 2/\alpha (y c_\Delta)^\alpha)}{\gamma(3/\alpha; \, 2/\alpha c_\Delta^\alpha)} \rho_\text{crit}(z_0),
\end{equation}
where $y = r/r_\Delta$ and $\gamma(a; \, b)$ is the lower incomplete gamma function, readily replace the NFW density profile in Eq.~\eqref{eq:densityProfileCorrection} and calculate the correction accordingly. Evaluating Eq.~\eqref{eq:einastoProfileEnclosedDensity} at the scale radius, the concentration w.r.t. the mean density $c_{\Delta\text{m}}$ is then the solution to
\begin{equation}
\frac{c_{\Delta\text{m}}^3}{\gamma(3/\alpha; \, 2/\alpha c_{\Delta \text{m}}^\alpha)} = \frac{c_{\Delta\text{c}}^3}{\gamma(3/\alpha; \, 2/\alpha c_{\Delta \text{c}}^\alpha)} \frac{E(z)^2}{\Omega_\text{m} (1 + z)^3}.
\end{equation}
The masses are rescaled in the same manner as in Section~\ref{sec:rescaledCMZ} and the resulting $c(M)$-relations with $\alpha = 0.18$ and $\Delta = 200$ differ negligibly from the NFW curves. 

With relaxation cuts enforced, the measured Einasto $c(M)$-relations are close to the \citetalias{2016MNRAS.460.1214L} model predictions, as Fig.~\ref{fig:som08CMEinastoWithCuts} shows. Overall the the model predictions better match measured relations for Einasto profiles (see Fig.~\ref{fig:3DMeanMassEinasto}) than for NFW profiles (see Fig.~\ref{fig:cMM200MeanMean3D2D}). While the concentration biases are similar to those measured for the NFW relations in Fig.~\ref{fig:concentrationDifference}, we have a slightly larger bias for $(0.15, \, 1.00)$ and $(0.25, \, 0.60)$, and for the low mass bins for $(0.40, \, 0.70)$ in Fig.~\ref{fig:deltaCEinasto}. Since the masses are fixed, the small horizontal scatter stems from the different median $M_{200\text{m}}$ masses of the bootstrap samples. These values do not deviate significantly from one another until the sparsely populated high mass end for some cosmologies.

Fig.~\ref{fig:som015CDirectRescaledEinasto} shows the Einasto estimated concentration distribution for individual haloes pre- and post-correction for $(0.15, \, 1.00)$. Compared to the NFW distributions, the Einasto fits favour higher concentrations for low mass haloes which is seen for the median $c(M)-$relations in Fig.~\ref{fig:3DMeanMassEinasto} and also in the shift of the distributions between Fig.~\ref{fig:som015CDirectRescaledEinasto} and the $(0.15, \, 1.00)$ panel in Fig.~\ref{fig:correctedConcentrationDifferenceIndHaloes}. In addition, the slightly larger mismatch between the \citetalias{2016MNRAS.460.1214L} model prediction and the measured median concentration relations for $(0.15, \, 1.00)$ is visible as an offset between the diagonal and the centre of the densest contour in Fig.~\ref{fig:som015CDirectRescaledEinasto} (cf. Figs.~\ref{fig:correctedConcentrationDifferenceIndHaloes}, \ref{fig:correctedConcentrationDifferenceMatchedHaloes} and \ref{fig:concentrationDifference}). A larger spread of concentrations is also possible, which can be noted by comparing the contours for $(0.40, \, 0.70)$ in Fig.~\ref{fig:som04IndHaloDensityFieldCEstimatesPrePostcorrection} (NFW) to those in Fig.~\ref{fig:som04CDirectRescaledEinasto} (Einasto). Still, the tilt is preserved by the two parameterisations for all cosmologies. The results in general are qualitatively quite similar.

\section{Splashback mass correction}
\label{sec:splashbackMassCorrection}

\begin{figure}
	\includegraphics[width=1.04\columnwidth]{\figrelpath 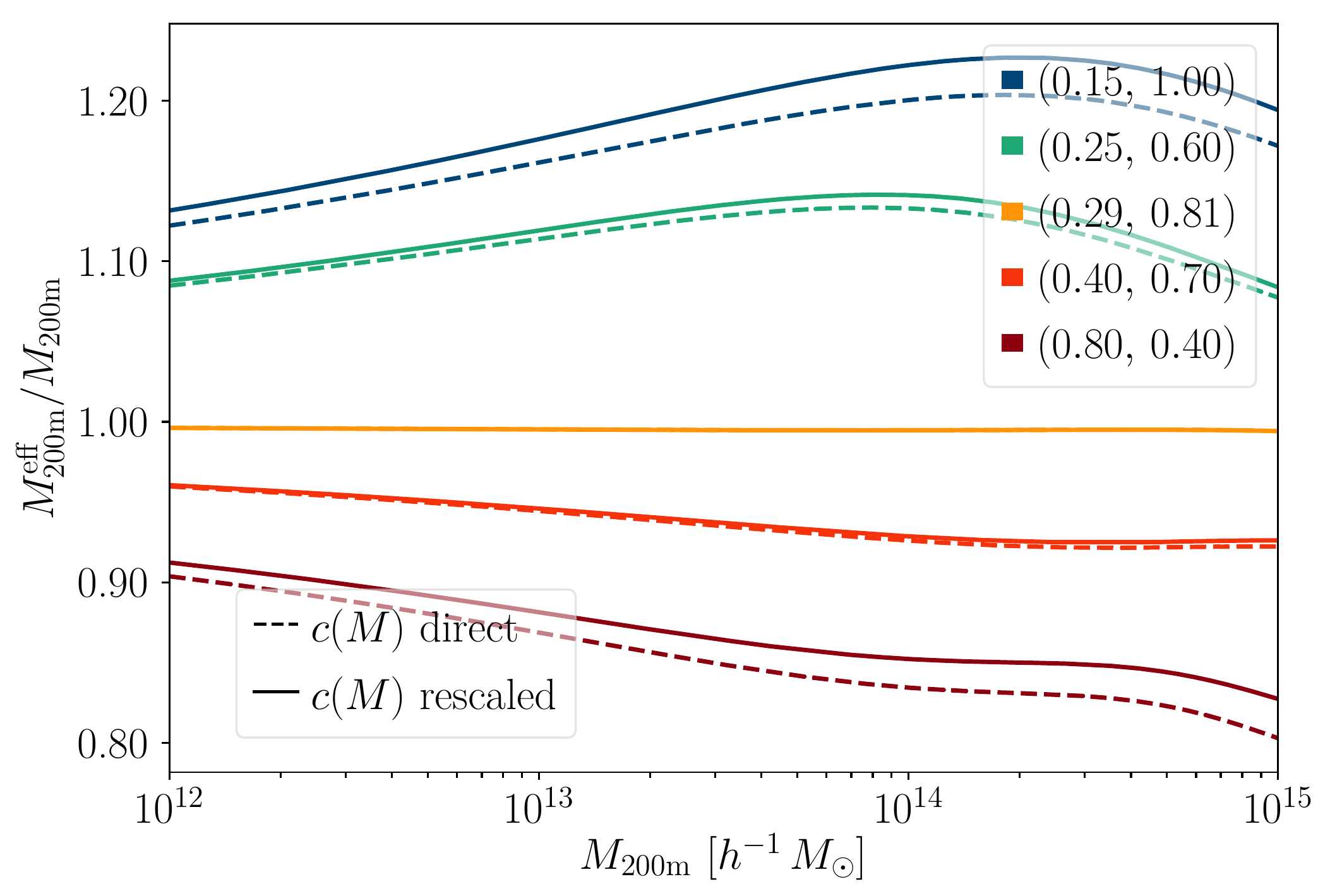}
    \caption{Effective mass correction with the NFW density field correction before and after the concentrations are corrected.}
    \label{fig:effectiveMassCorrection}
\end{figure}

\begin{figure}
	\includegraphics[width=1.04\columnwidth]{\figrelpath 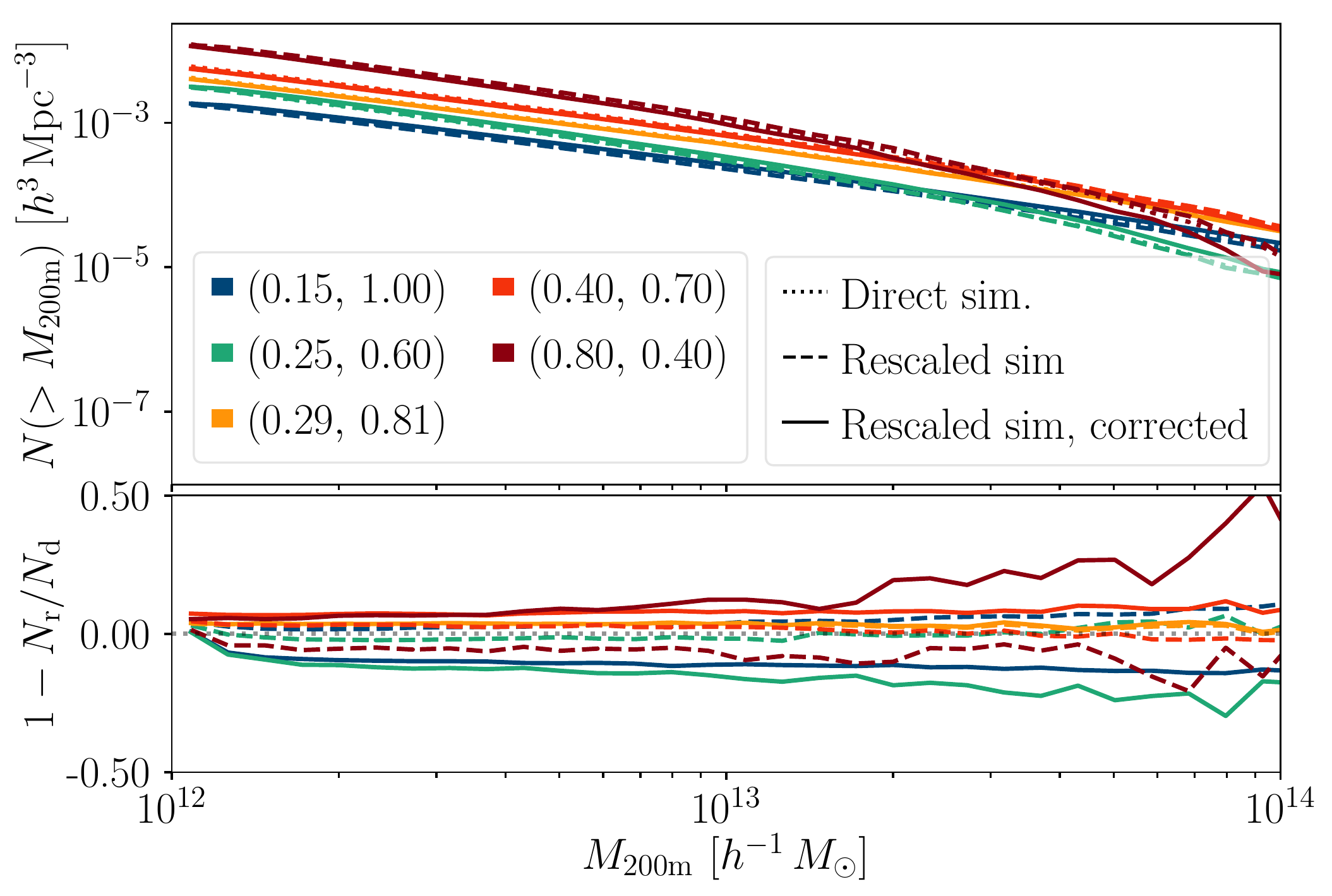}
    \caption{Halo mass function before and after the mass correction.}
    \label{fig:hmfCorrection}
\end{figure}

The outer profile correction can be used to build a na{\"i}ve mass correction, if we redefine the $M_{200\text{m}}$ masses in the rescaled simulation as masses within the perturbed $r_{200\text{m}}^\prime$, which is set such that $r_\text{sp}^\text{(d)}/r_{200\text{m}}^\text{(d)} = r_\text{sp}^\text{(r)}/r_{200\text{m}}^\prime$. Assuming that the density field just beyond $r_{200\text{m}}$ is dominated by the 1-halo term which is well captured by an NFW profile, one could extend the integration to $r_{200\text{m}}^\prime = (1 + \delta) r_{200\text{m}}$ where $1 + \delta = 1/(1 + \Delta r_\text{sp})$. This simplifies to the following expression for the mass correction:
\begin{equation}
\label{eq:jointMassCorrection}
\frac{M_{200\text{m}}^\text{eff.}}{M_{200\text{m}}}= \frac{1 - 1/(1 + c/(1 + \Delta r_\text{sp})) - 
 \ln{(1 + c/(1 + \Delta r_\text{sp}))}}{1 - 1/(1 + c) - \ln{(1 + c)}},
\end{equation}
where $c = c(M)$, and $\Delta r_\text{sp}$ could be predicted with the \citetalias{2016MNRAS.460.1214L} and \citet{2017ApJ...843..140D} model fits, respectively. The weak mass evolution of this correction factor for the different cosmologies is plotted in Fig.~\ref{fig:effectiveMassCorrection} for the uncorrected and corrected rescaled density field. The concentration correction affects the relation marginally. Due to the mismatch between the detected outer profile bias for $(0.15, \, 1.00)$, $(0.25, \, 0.60)$ and $(0.40, \, 0.70)$ and the model prediction in Fig.~\ref{fig:interpolatedRspBias}, as well as the mismatch between the \citetalias{2016MNRAS.460.1214L} model and the measured $c(M)$-relations, the correction is too large. This is reflected in the cumulative halo mass function in Fig.~\ref{fig:hmfCorrection} for matched haloes pre- and post-mass correction, where the agreement is worse. For $(0.15, \, 1.00)$ and $(0.80, \, 0.40)$, however, the bias changes signs at the low mass end, and for $(0.80, \, 0.40)$, the situation improves somewhat at the low mass end.

We can interpret these results in light of the discrepancies in mass between the matched direct and rescaled haloes in Fig.~\ref{fig:hmfMatchedMassDifference}, where the median relations for these two simulations are off (see appendix~\ref{sec:resultsForAlmostEinsteinDeSitter} for $(0.80, \, 0.40)$) and the mass correction shifts these median levels in the right direction. Still, there is a mass evolution of the discrepancy between the direct and rescaled haloes which must be modelled by a more elaborate correction. For the other simulations, this tilt dominates over the wrong offset level, and for $(0.29, \, 0.81)$ there is a very small predicted shift.

\section{Cosmological contour plots for the rescaling parameters}
\label{sec:alphaZastContours}

\begin{figure}
	\includegraphics[width=\columnwidth]{\figrelpath 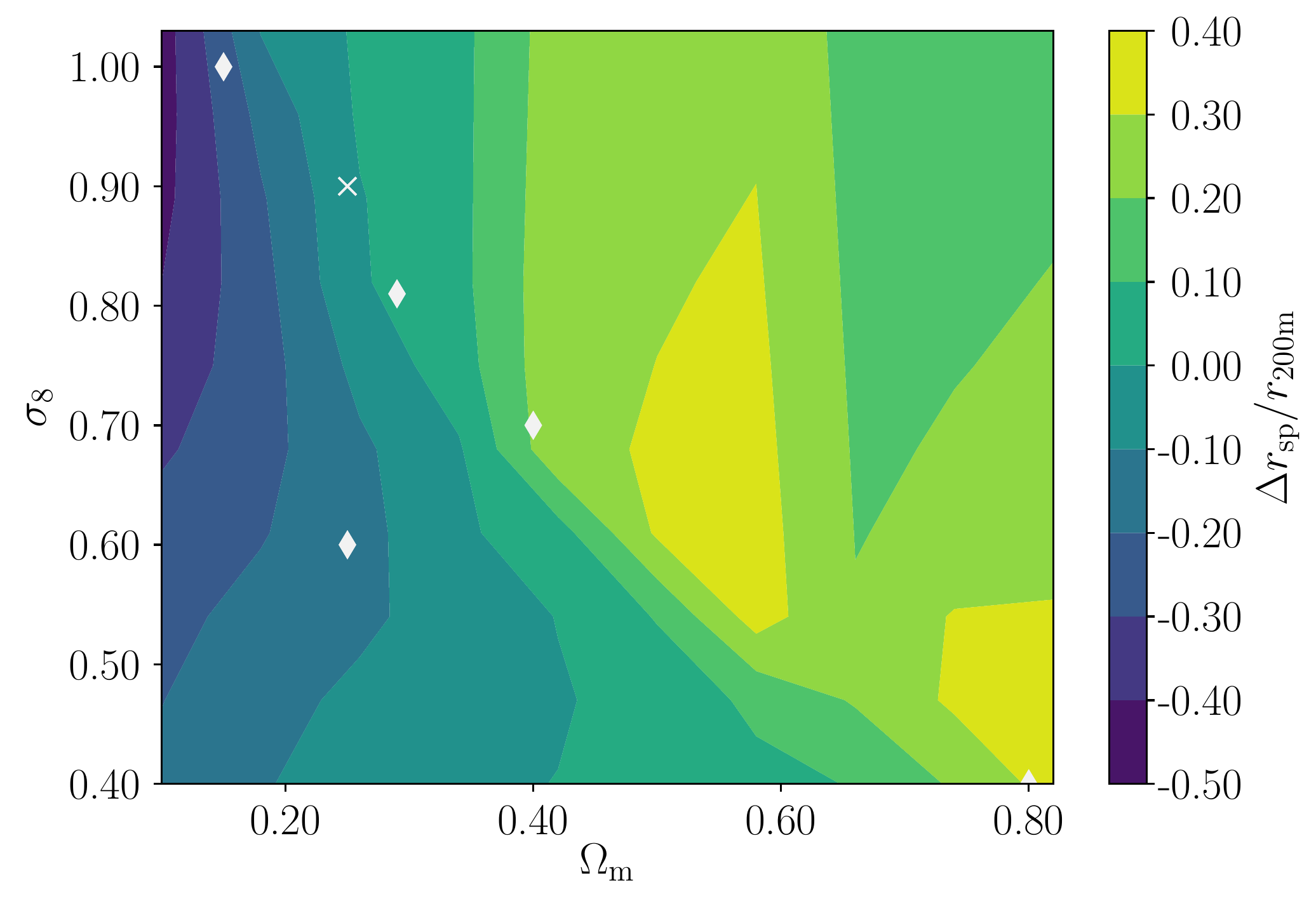}
    \caption{Predicted offset in splashback radius for matched haloes in a direct and rescaled fiducial simulation with WMAP1 parameters from the \citet{2017ApJ...843..140D} model (75th percentile).}
    \label{fig:deltaRspContours}
\end{figure}

\begin{figure}
	\includegraphics[width=\columnwidth]{\figrelpath 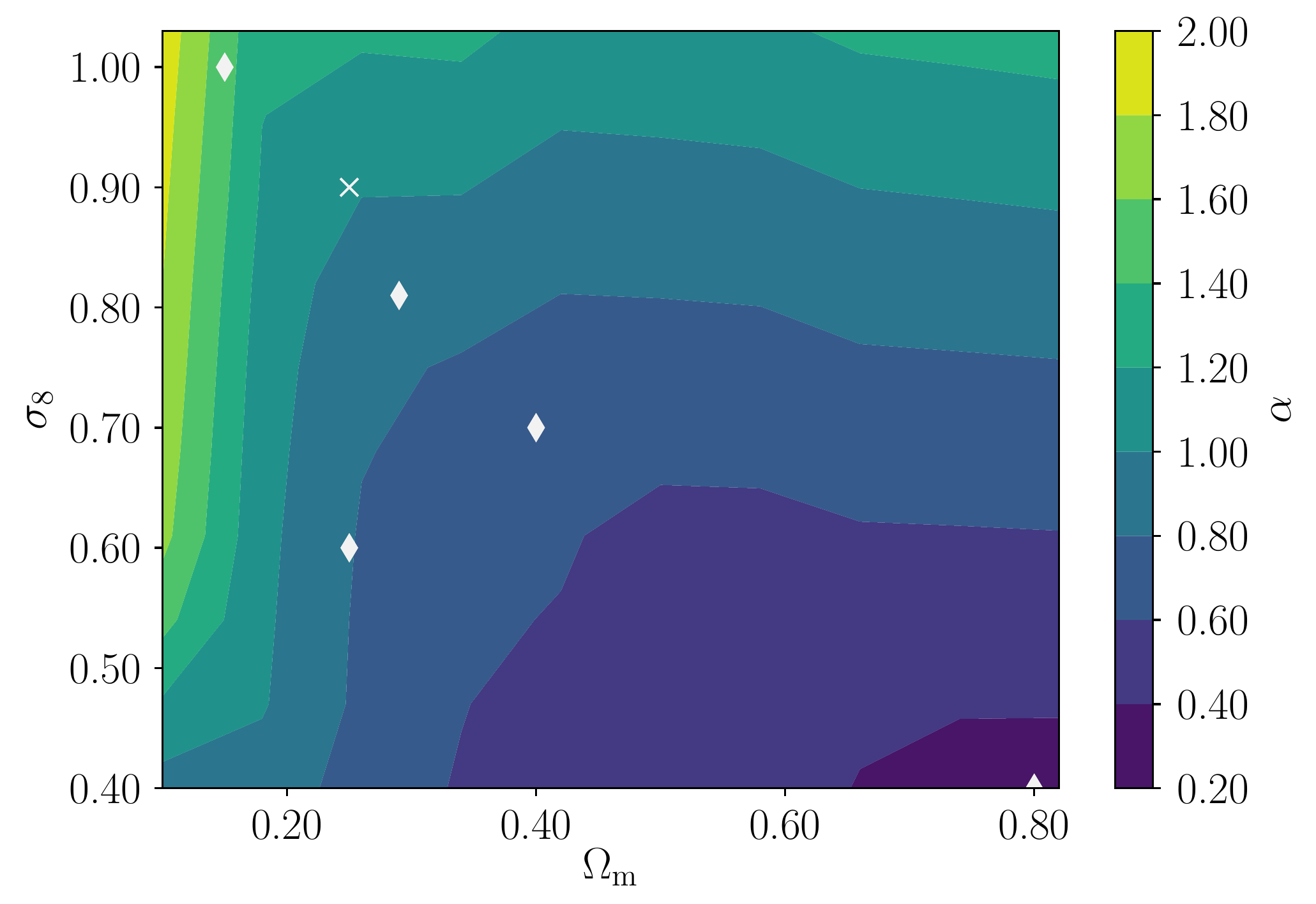}
    \caption{The length scale parameter $\alpha$ as a function of $\Delta \Omega_\text{m}$ and $\Delta \sigma_8$ w.r.t. a fiducial simulation with WMAP1 parameters.}
    \label{fig:alphaContours}
\end{figure}

\begin{figure}
	\includegraphics[width=\columnwidth]{\figrelpath 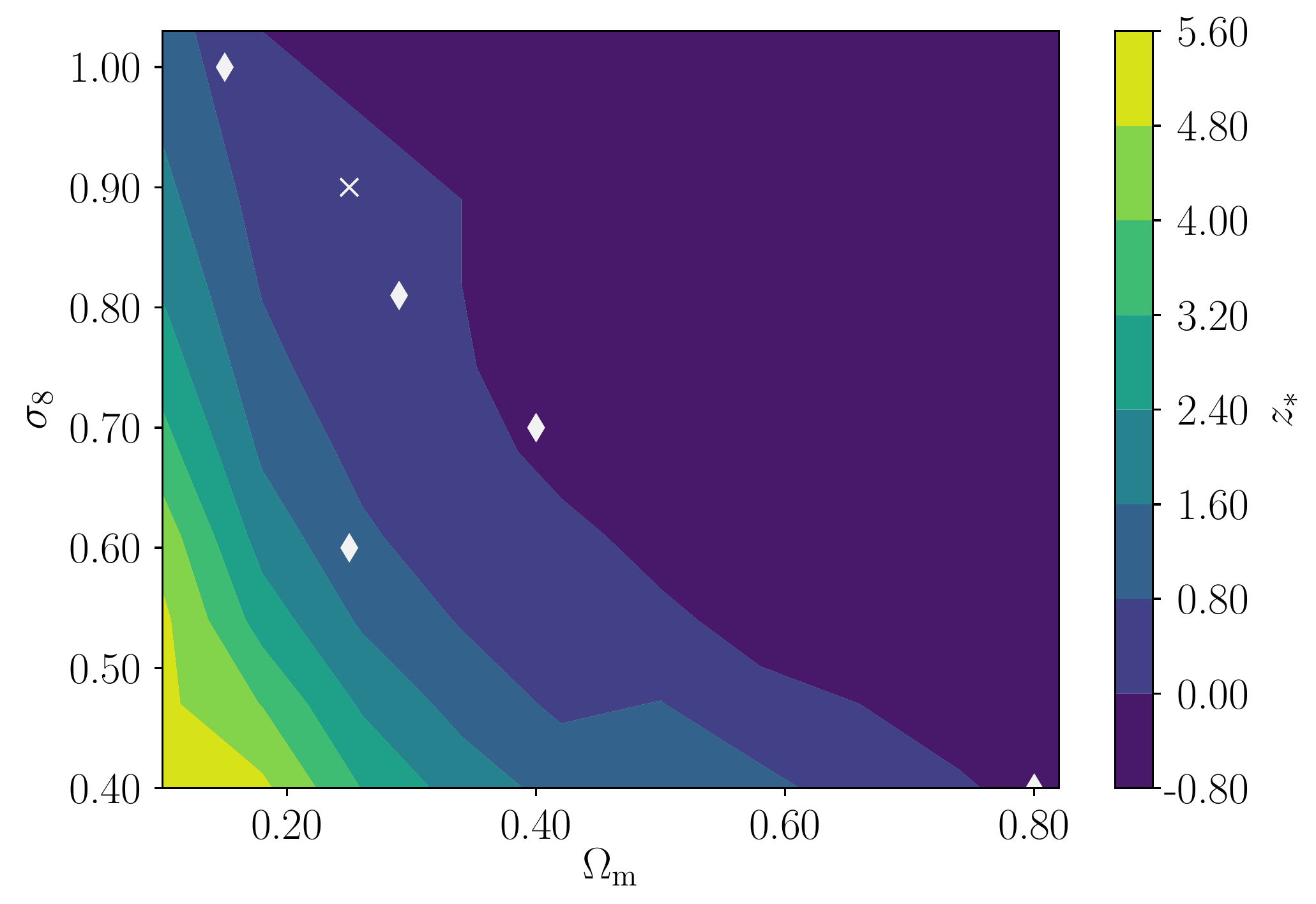}
    \caption{The time scale parameter $z_\ast$ as a function of $\Delta \Omega_\text{m}$ and $\Delta \sigma_8$.}
    \label{fig:zContours}
\end{figure}

In Fig.~\ref{fig:deltaRspContours} the predicted offsets computed with the \citet{2017ApJ...843..140D} model in the position of the splashback radius w.r.t. $r_{200\text{m}}$ for matched halo samples in different target cosmologies is shown. In large sections of the parameter plane, $\Delta r_\text{sp}/r_{200\text{m}}$ has the opposite sign as $\Delta c$ although this is not necessarily true for small changes from the fiducial run nor for the $\Delta \Omega_\text{m} > 0, \, \Delta \sigma_8 > 0$ quadrant. We plot the $(\alpha, \, z_\ast)$ pairs to emulate these different cosmologies in Figs.~\ref{fig:alphaContours} and \ref{fig:zContours}. They are smooth functions depending on $\Delta \Omega_\text{m}$ and $\Delta \sigma_8$ to the fiducial cosmology. Shrinking the simulation box is preferable to emulate a cosmology with a higher matter fraction, and expanding the box for lower matter fractions. Similarly intuitively, going to a higher redshift in the fiducial simulation could be used to match a cosmology with a lower $\sigma_8$, i.e. with a lower amplitude of the fluctuations of the matter field. This puts the Millennium simulation in a suitable position for rescaling as the WMAP1 $\sigma_8 = 0.9$ is comparably high to the current best fit matter power spectrum amplitudes.

\section{Biases for a rescaled Millennium simulation to WMAP and Planck cosmologies}
\label{sec:wmapPlanckMillennium}

\begin{figure}
	\includegraphics[width=\columnwidth]{\figrelpath 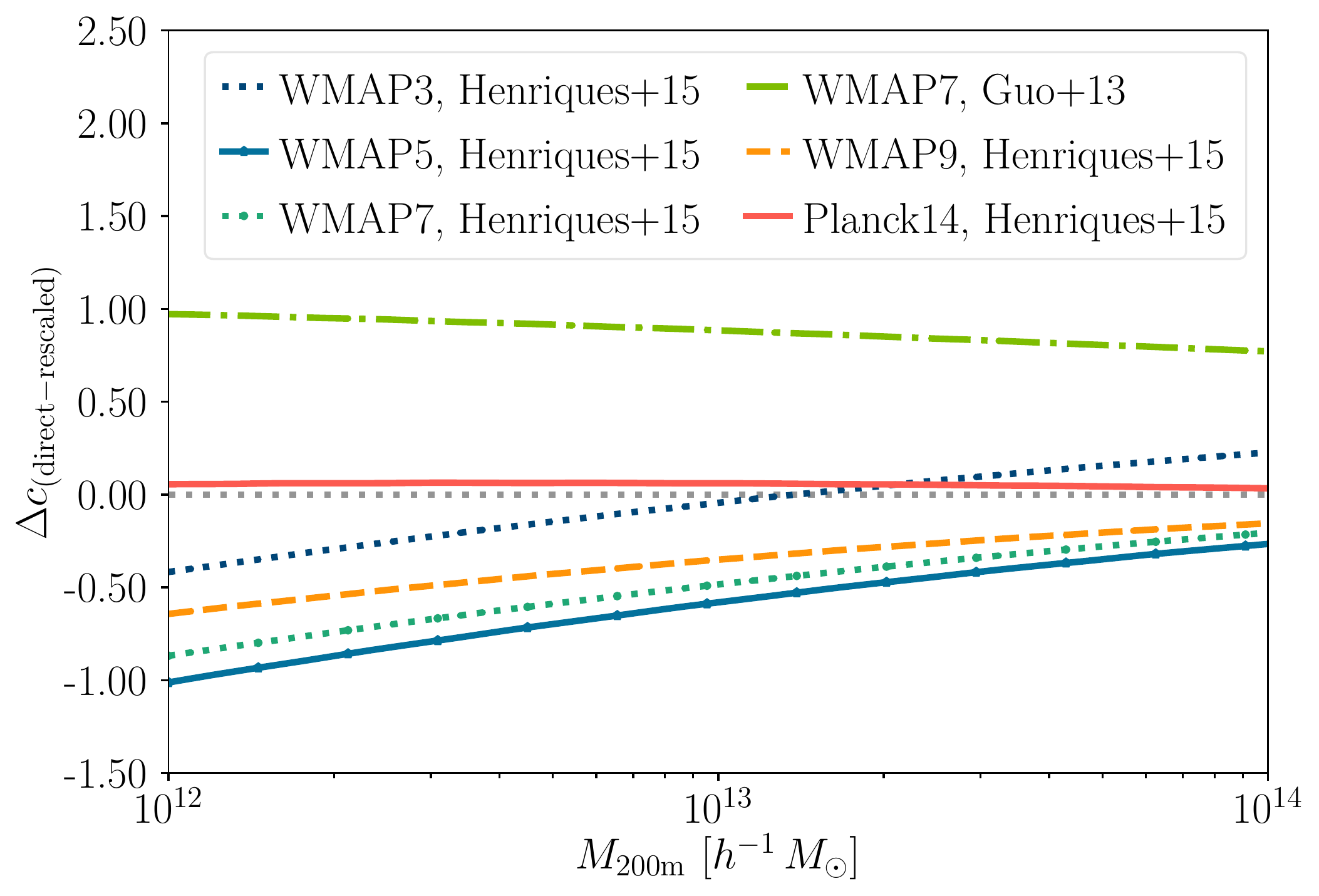}
    \caption{Predicted concentration biases for haloes rescaled using the parameters in \citet{2015MNRAS.451.2663H} and \citet{Guo:2012fy}.}
    \label{fig:samBiasesZ0}
\end{figure}

\begin{figure}
	\includegraphics[width=\columnwidth]{\figrelpath 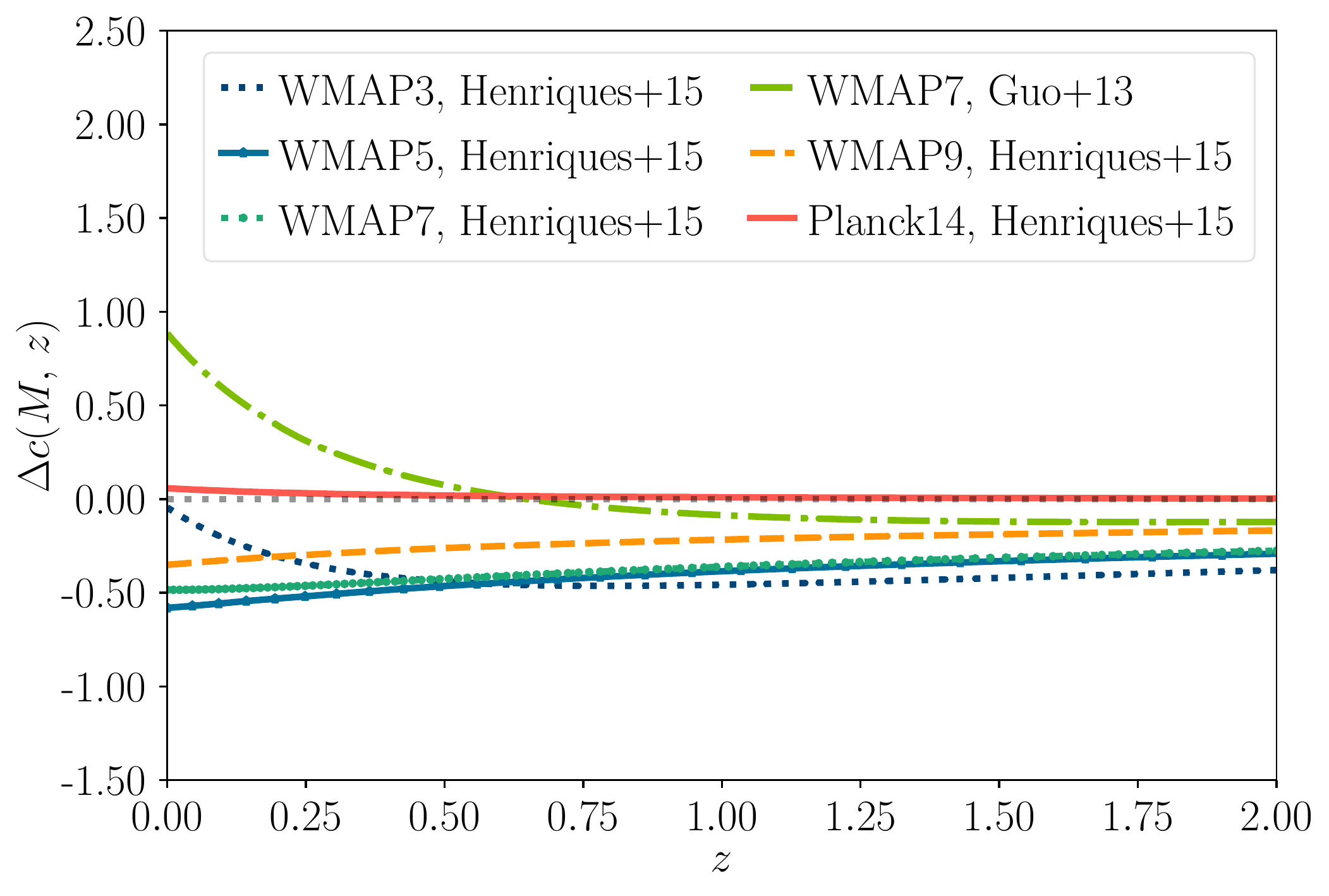}
    \caption{Redshift evolution of the concentration biases in Fig.~\ref{fig:samBiasesZ0}.}
    \label{fig:samBiasesZ}
\end{figure}

In Fig.~\ref{fig:samBiasesZ0} and Fig.~\ref{fig:samBiasesZ}, we illustrate the predicted concentration biases for the Millennium simulation \citep{Springel:2005nw} with its WMAP1 parameters \citep{2003ApJS..148..175S} rescaled to a range of cosmologies (WMAP3, WMAP5, WMAP7, WMAP9, Planck 2014) \citep{2007ApJS..170..377S, 2009ApJS..180..330K, 2011ApJS..192...18K, 2013ApJS..208...19H, 2014A&A...571A..16P} at $z=0$ according to the parameters in \citet{2015MNRAS.451.2663H} and \citet{Guo:2012fy}. For the cosmologies where there is a mass evolution of the concentration bias in Fig.~\ref{fig:samBiasesZ0}, the slope decreases at higher redshift. The predicted concentration bias for haloes with a Millennium simulation rescaled to Planck 2014 with the parameters in \citet{2015MNRAS.451.2663H} is very small and decreases for higher redshifts, which is fortuitous for future lensing analyses.

%%%%%%%%%%%%%%%%%%%%%%%%%%%%%%%%%%%%%%%%%%%%%%%%%%

% Don't change these lines
\bsp	% typesetting comment
\label{lastpage}
\end{document}